\documentclass[]{article}

\usepackage{amsmath, amsthm, amssymb, amsfonts}
\usepackage{mathtools}
\usepackage{hyperref}
\usepackage{graphicx}
\usepackage{dsfont}
\usepackage{fullpage}
\usepackage{color}
\usepackage{soul} 
\usepackage[title]{appendix}
\usepackage{tikz}
\usetikzlibrary{patterns}
\usetikzlibrary{shapes.multipart}
\usetikzlibrary{arrows}

\theoremstyle{definition}

\definecolor{labelkey}{cmyk}{.4,.2,0,0}

\newcommand{\be}{\begin{equation}}
\newcommand{\ee}{\end{equation}}
\newcommand{\bea}{\begin{eqnarray}}
\newcommand{\eea}{\end{eqnarray}}
\newcommand{\nn}{\nonumber}

\usepackage{titlesec}
\titleformat{\section}{\large\bf}{\thesection}{1em}{}
\titleformat{\subsection}[runin]{\bf}{\thesubsection}{1em}{}[.]

\usepackage{authblk}
\title{This is some thing}
\author[1]{Pierre Le Doussal}
\affil[1]{\normalsize Laboratoire de Physique de l'\'Ecole Normale Sup\'erieure, ENS, Universit\'e PSL, CNRS, Sorbonne Universit\'e, Universit\'e de Paris, 75005 Paris, France}

\title{\bf \large Dynamics at the edge for independent diffusing particles}
\date{}

\begin{document}

\maketitle

\begin{abstract}
We study the dynamics of the outliers for a large number of independent
Brownian particles in one dimension. We derive the multi-time joint distribution of the position of the rightmost particle,
by two different methods. We obtain the two time joint distribution of the 
maximum and second maximum positions, and study the counting statistics at the edge. 
Finally we derive the multi-time joint distribution of the running
maximum, as well as the joint distribution of the arrival time of the
first particle at several space points.
\end{abstract}


\section{Introduction}

The maximum of a large number $N \gg 1$ of identical independent Brownian motions, started from the origin in one dimension, properly rescaled and centered, is distributed according to the Gumbel distribution,
one of the three classes of extreme value statistics \cite{Gnedenko43,Gumbel1958,Galambos00,Nagaraja2003}.
Recently, there has been renewed interest in the statistics of the particles at the edge
of a cloud of Brownian particles diffusing in a common space-time dependent random environment
\cite{BarraquandCorwinBeta,TTPLD,BarraquandSticky,GBPLDModerate,CorwinCorwinExtremeDiff23,UsCrossover2022,WarrenEdgeCloud,das2023kpz,CorwinCorwinFPT23}. For a large number of particles, it was shown that on top of the Gumbel fluctuations, 
there is a random, environment-dependent shift in the position of 
the rightmost particle. Furthermore, the statistics of this shift was found to be related to those of some solutions of the Kardar-Parisi-Zhang equation.
Tracer diffusion experiments, involving colloid or photons, are presently aiming to test that prediction
\cite{PrivateCorwin2}.
\\

This prediction is about the position of the maximum at a given time, a one-time observable. It would be interesting to extend it to multi-time observables. With this longstanding aim in mind, one can start by asking about a much simpler problem, namely multi-time observables for identical independent Brownian motions, in the absence of a background environment. That should be useful, e.g. as a benchmark in the analysis of such experiments.
\\

In the present paper we study  the dynamics of the outliers for a large number of independent 
Brownian particles on the line, all starting at the origin. We derive the multi-time joint distribution of the position of the rightmost particle, i.e. of the maximum of all the particle positions, 
by two elementary methods, which lead to different, though equivalent, formulae.
The first method is standard for extreme value statistics, the second uses the diffusion equation.
We obtain explicit formula for the cumulants of the fluctuations of the maximum at different times.
We then extend these results to the multi-time joint distribution of the 
maximum and of the second maximum particle positions. In parallel, we obtain some
predictions about the counting statistics at the edge of the cloud, which 
describes the outliers.
Next, we study some continuum time observables of the 
rescaled maximum process. We obtain the probability that it remains
below some level during some time interval. We study 
the running maximum, that is the
maximum over all the particles and up to some fixed time, and obtain its multi-time joint
distribution. Finally, we obtain the joint
distribution of the arrival times of the first
particle (i.e. the first detection times) at 
different locations in space, and the 
distribution of the time delays between these detection events. 
\\

It must be noted that this class of problems is related to the so-called
multivariate extreme value theory and 
max-stable processes, which has a long history, 
starting with the Brown-Resnick process \cite{RescaledResnik77,Resnik2} and
the Husler-Reiss (HR) distributions \cite{HuslerReiss1989,HuslerBivariate}.
Since these seminal papers there were a number of extensions
\cite{FalkHuslerReissBook1990,deHaanMaxStable,Huser2013,BouyeFinance2002,Kabluchko2009,Engelke,HashorvaConvergence2016,Tang2021},
though apparently mainly in the fields of probability and statistics.
Our modest aim here is to instead study the problem with simple heuristic 
methods of statistical physics. In the course of our work
we will encounter some known objects, such as the HR distributions, 
in sometimes different forms, but 
we will also study more general outlier properties, counting statistics,
the running maximum problem, and arrival time statistics. 
One must also note an independent work in preparation 
on these topics \cite{SatyaGreg}, following upon
the recent work \cite{DeanEffusion23} which studies counting
statistics for stochastic processes with an extended initial condition. 
\\

The outline of the paper is as follows. In Section \ref{sec:onetime} 
we focus on one-time observables (maximum, order and counting statistics
of outliers) and recall the standard
results obtained for these quantities from extreme value statistics of i.i.d random
variables. In Section \ref{sec:first} (and Appendix \ref{app:derivationPDF})
we give a first derivation of the multi-time distribution of the maximum.
In Section \ref{sec:correlations} we give more explicit formula
for some marginals, moments and two and three time 
correlations of the maximum. The relevant calculations
are detailed in Appendices \ref{app:integrals} and \ref{app:n2}.
In Section \ref{sec:heat} 
we present a second derivation of the multi-time distribution of the maximum
based on the heat equation. It naturally allows to obtain these
distributions recursively. In Section \ref{sec:secondmax} we
study some multi-time properties of the outliers, e.g.
we obtain the multi-time joint distribution for the maximum and
second maximum, either directly, 
in Appendix \ref{app:combinatorics}, or 
by studying counting statistics near the edge of the cloud.
In Appendix \ref{app:outliers} it is indicated
how to extend these results to any rank at number of times.
Finally, in Section \ref{sec:continuum} and
Appendix \ref{app:continuous} we
study continuum time
observables of the rescaled maximum process, as well as
the multi-time statistics of the running maximum
and of the arrival times of the first particle.

\section{Outliers at a given time} \label{sec:onetime} 

Let us start by recalling standard results of extreme value statistics \cite{Gnedenko43,Gumbel1958,Galambos00,Nagaraja2003},
applied to one time observables for diffusing particles. 

\subsection{Maximum at a given time} 

Let us consider a particle whose position $x(t)$ evolves according 
to a random process. Let us denote $p_t(x)$ the one time probability distribution function (PDF)
of the position at time $t$. The example on which we will 
focus will be a Brownian particle, $x(t)=\sqrt{2 D} B(t)$,
started at $x=0$ at time $t=0$. In that case
$p_t(x)=\frac{1}{\sqrt{4 D t}} e^{- x^2/(4 D t)}$
is the standard diffusion kernel. Below we set $D=1$ since any value of $D$ can be restored by a rescaling of time.

Let us now consider $N$ identical copies
of that particle evolving independently. In our canonical
example all the Brownian particles have the same 
diffusion coefficient and all start from $x=0$ at time $t=0$.
One now defines
the position of the rightmost particle, i.e. the maximum $X(t)= \max_{i=1,\dots,N} x_i(t)$.
We are interested in the case where $N$ is large. 
By definition the cumulative distribution function (CDF) of the maximum is given by
\be 
Q(X)= {\rm Prob}(X(t)<X) = P_{<,t}(X)^N = e^{N \log(1- P_{>,t}(X))} 
\underset{N \gg 1}{\simeq}
 e^{- N P_{>,t}(X)} 
\ee 
where we denote $P_{>,t}(X)=\int_X^{+\infty} dx p_t(x)$ and 
$P_{<,t}(X)=\int_{-\infty}^X dx p_t(x)$. As is well
known, the standard diffusion falls in the Gumbel extremal class.
For any process in that class one can simply perform the change of
variable from $X$ to $z$ defined by
\be \label{change}
N P_{>,t}(X)= e^{-z} 
\ee 
In the case of diffusion it gives $\frac{N \sqrt{t}}{\sqrt{2 \pi} X} e^{- X^2/(2 t)} = e^{-z}$
leading to 
\be \label{onetime} 
X \simeq \sqrt{2 t} ( \sqrt{\log N} + \frac{z+c_N}{2 \sqrt{\log N}} ) 
\quad , \quad c_N = - \frac{1}{2} \log( 4 \pi  \log N ) 
\ee 
In this new variable the CDF is simply 
the Gumbel distribution, $e^{- e^{-z}}$, i.e. one has 
\be 
Q(X)= {\rm Prob}(X(t)<X) \simeq
e^{- e^{-z}}  \label{Gumbel} 
\ee
Note that the change of variable \eqref{change} (hence also 
\eqref{onetime} to leading order) is such that one has exactly
\be \label{differential} 
N p_t(X) dX = e^{-z} dz 
\ee 
In the following we will often (abusively) also consider $z$ as the random
variable defined by \eqref{onetime} with $X=X(t)$,
i.e. as the rescaled position of the maximum. 

\subsection{Order statistics of outliers} 
\label{sec:outliers} 

It is well known how to extend this to the $k$ first particles \cite{Gumbel1958,Galambos00,Nagaraja2003}. Let us denote
$X^{(j)}(t)$ the same set of particles, but ordered by their rank, i.e.
$X^{(1)}(t)>X^{(2)}(t)> \dots> X^{(N)}(t)$, 
so that $X(t)=X^{(1)}(t)$ is the position of the maximum, $X^{(2)}(t)$ of the second
maximum and so on. 

The joint PDF of the $k$ first outliers is 
\be 
N(N-1)\dots(N-k+1) p_t(X^{(1)}) p_t(X^{(2)}) \dots p_t(X^{(k)}) \theta_{X^{(1)}>\dots>X^{(k)}}
P_{<,t}(X^{(k)})^{N-k} dX^{(1)}  \dots dX^{(k)} 
\ee 
For large $N$ it becomes 
\be 
\simeq \left( \prod_{j=1}^k (N p_t(X^{(j)}) dX^{(j)}) \right)  \theta_{X^{(1)}>\dots>X^{(k)}}
e^{- N P_{>,t}(X^{(k)})} 
\ee
Thus the same change of variable 
\be 
N P_{>,t}(X^{(j)})= e^{-z^{(j)}} 
\ee 
which in the case of diffusion reads 
\be 
X^{(j)} \simeq \sqrt{2 t} ( \sqrt{\log N} + \frac{z^{(j)} +c_N}{2 \sqrt{\log N}} ) \label{changeoutliers} 
\ee 
allows to put the large $N$ asymptotic joint PDF of the position of the $k$ first particles in the
well known form
\be 
q(z^{(1)},\dots,z^{(k)}) = \theta_{z^{(1)}>z^{(2)}>\dots>z^{(k)}} \, e^{-z^{(1)}-z^{(2)}-\dots - z^{(k)}} e^{-e^{-z^{(k)}}} 
\ee 
or equivalently as 
\be \label{independent} 
q(z^{(1)},\dots,z^{(k)})  = \theta_{z^{(1)}>z^{(2)}>\dots>z^{(k)}}  \, \prod_{\ell=1}^{k-1} \ell e^{- \ell (z^{(\ell)}-z^{(\ell+1)})}
\times \frac{1}{(k-1)!} e^{- k z^{(k)} - e^{-z^{(k)}}} \;.
\ee 
Hence to generate the $k$ largest points, one first chooses $z_k$ and then the successive 
gaps as independent exponentially distributed variables, with distinct parameters.

\subsection{Counting statistics of outliers} \label{subsec:counting} 
Another standard way to characterize the outliers
is the counting statistics. Let us define $n_X$ as the 
number of particles at a given time $t$ with $x_i(t)>X$.
Since the particles are identical and independent the 
probability that $n_X=n$ 
follows the binomial distribution
\be 
P_X(n) = \frac{N!}{n! (N-n)!} P_{<,t}(X)^{N-n} P_{>,t}(X)^n
\ee 
In the edge regime for large $X$, i.e. for $N  P_{>,t}(X) = O(1)$ this becomes a Poisson
distribution
\be 
P_X(n) \simeq \frac{1}{n!}  (N P_{>,t}(X))^n e^{- N P_{>,t}(X)} = \frac{1}{n!} e^{-n z} e^{- e^{-z}} 
\ee  
where in the last equality we have used the change of variable \eqref{change}. In the case
of independent Brownian motions, all starting from the origin, the variables 
$X$ and $z$ are related through \eqref{onetime}. From it one recovers
the Gumbel CDF of the maximum
\be 
P_X(n=0)= {\rm Prob}(X(t)<X) \simeq e^{- e^{-z}} 
\ee 
Note that the other probabilities have also some interpretations in terms of order statistics,
i.e. $P_X(1) = {\rm Prob}(X(t)>X,X^{(2)}(t) <X)$, where $X(t)=X^{(1)}(t)$ and
$X^{(2)}(t)$ are the maximum and second maximum respectively, and so on.
Furthermore one sees that $X^{(k)} < X$ is equivalent to $n_X \in \{0,1,\dots,k-1\}$,
hence 
\be 
{\rm Prob}(X^{(k)} < X) \simeq (\sum_{n=0}^{k-1} \frac{1}{n!} e^{-n z} ) e^{- e^{-z}} 
\ee 
One can check that taking $\partial_z$ of the r.h.s. one recovers the PDF of $z=z^{(k)}$,
i.e. $q(z^{(k)})=\frac{1}{(k-1)!} e^{- k z^{(k)} - e^{-z^{(k)}}}$.

How does one recover the joint PDF of $X(t)$ and $X^{(2)}(t)$ ?
For that one needs the joint PDF of $n_{X_1}$ and $n_{X_2}$ with $X_1>X_2$
where we recall that $n_X$ the number of particles with $x_i(t)>X$. To
obtain it we split the line into three disjoint intervals (a) $x>X_1$, (b) $X_2<x<X_1$
and (c) $x<X_2$ and write the product of sums of probabilities of these events
\be 
1= (P(x>X_1) +  P(X_2<x<X_1) + P(x<X_2))^N \label{prod} 
\ee 
Here for convenience we adopt the shorthand notations, e.g.
 $P(X_2<x<X_1)= \int_{X_2}^{X_1} dx p_t(x)$ and so on.
 Expanding \eqref{prod}, we see that the probability $P_{X_1,X_2}(n_a,n_b,n_c)$
 that there are $n_a,n_b,n_c$ particles
 in each of these intervals, is given by the multinomial distribution
\be 
P_{X_1,X_2}(n_a,n_b,n_c)= \frac{N!}{n_a! n_b! n_c!} \delta_{N,n_a+n_b+n_c} 
P(x>X_1)^{n_a} P(X_2<x<X_1)^{n_b} 
P(x<X_2)^{n_c} 
\ee 
In the large $N$ limit and choosing $X_1$ and $X_2$ near the edge so
that typically $n_b,n_c=O(1)$ while $n_a \simeq N$, one obtains by similar manipulations as
above that the probability of $n_a,n_b$ is a multiple independent Poisson distribution
\be 
P_{X_1,X_2}(n_a,n_b) \simeq \frac{1}{n_a! n_b!} (e^{-z_2}-e^{-z_1})^{n_b} 
e^{- n_a z_1} e^{-e^{-z_2}} \label{Poisson2} 
\ee 
which is correctly normalized to unity. This form applies to
any problem of i.i.d random variables in the Gumbel class through the 
change of variable \eqref{change}, and for our purpose here $X_1$ and $z_1$ 
are related through \eqref{onetime} and so are $X_2$ and $z_2$.
\\

Several observables can be obtained from \eqref{Poisson2}. For instance
the joint PDF of $X(t)$ and $X^{(2)}(t)$ can be retrieved from taking
$- \partial_{X_1} \partial_{X_2}$ of the following "CDF" 
\be 
{\rm Prob}(X^{(2)}(t)  < X_2 , X^{(1)}(t)  > X_1) = P_{X_1,X_2}(n_{a}=1, n_b=0) 
\simeq e^{- z_1} e^{-e^{-z_2}} 
\ee 
and one can check that it is indeed equal to 
$ \int_{y_1>z_1} \int_{y_2<z_2} e^{-y_1 - y_2} e^{- e^{-y_2}}$. 
\\

Another interesting observable is the joint PDF of the couple $(n_{X_1},n_{X_2})$.
Since the intervals $[X_2,+\infty[$ and $[X_1,+\infty[$ overlap, the two variables are
correlated. One can write $(n_{X_1},n_{X_2})= (n_a,n_a+n_b)$, where $n_a$, $n_b$ 
are independent Poisson variables. Hence
\be 
{\rm Prob}(n_{X_1}=n_1, n_{X_2}=n_2) = \theta_{n_2 \geq n_1} \frac{\lambda_a^{n_1}}{n_1!} 
\frac{\lambda_b^{n_2-n_1}}{(n_2-n_1)!} e^{-\lambda_a-\lambda_b}
\ee 
with $\lambda_a=e^{-z_1}$ and $\lambda_b=e^{-z_2}-e^{-z_1}$ are
the mean parameters of the distribution \eqref{Poisson2}.

\section{Multi-time joint CDF for the maximum: first method} 
\label{sec:first} 

Let us now consider the dynamics of the maximum $X(t)$. Its one time CDF is given 
by \eqref{onetime}-\eqref{Gumbel}. What is the multi-time joint CDF of $X(t_1), X(t_2), \dots X(t_n)$ ?
\\

Let us first determine on what time scale these variables remain correlated. 
The first simple consideration is as follows. As recalled in 
Section \ref{sec:outliers}, at any fixed time the gap between the 
maximum $z^{(1)}$ and the second maximum $z^{(2)}$ (in the variable $z$
seen as random variables) 
is $z^{(1)}-z^{(2)}=O(1)$. Hence at a given time $t=t_1$, from
\eqref{changeoutliers}, the first gap is 
$X(t) - X^{(2)}(t) = O(\sqrt{t_1/\log N})$. These two rightmost particles undergo diffusion, hence it 
takes typically a time $t_2-t_1 \sim (X(t) - X^{(2)}(t) )^2 \sim t_1/\log N$ for them to meet.
This gives the scale of the time
difference at which the order of the first few particles at the edge reshuffles, 
and correlations start decaying. For time differences 
much larger $t_2-t_1 \gg t_1/\log N$ we expect that $X(t_1)$ and $X(t_2)$ become
uncorrelated, each described, under the appropriate scaling, by Gumbel distributions.
\\

\begin{figure}[t]
  \centerline{\includegraphics[width=0.6\columnwidth,height=8cm,keepaspectratio]{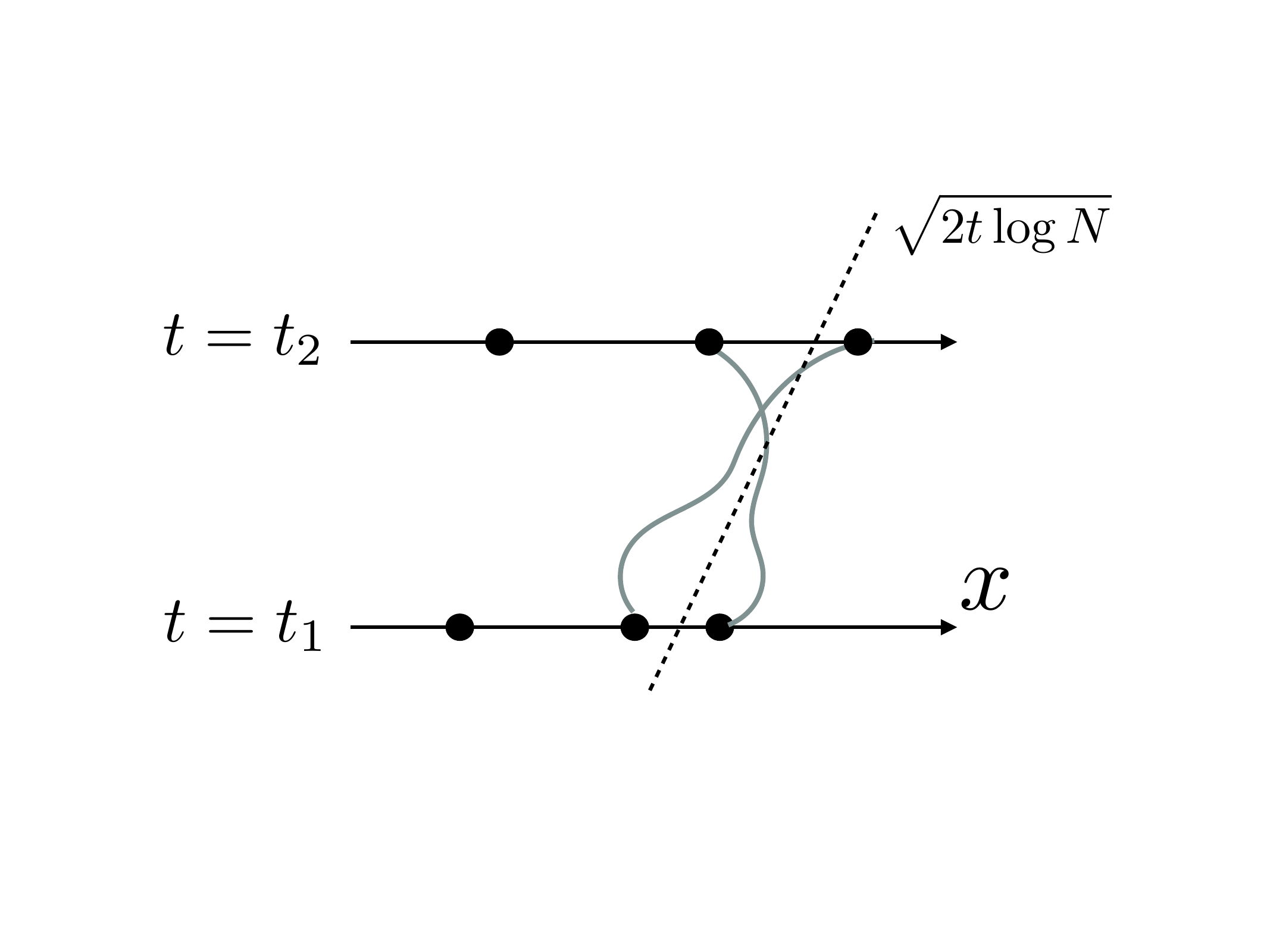}}
    \vspace{-2cm}
      \caption{Schematic view of the positions of the three rightmost particles
   at two times, with $t_2-t_1=t_1 (1+ \frac{\tau}{\log N})$. The dotted line is
   the average position of the edge, which moves to the right with unit velocity 
   in the scaled space-time variables $z,\tau$. The scaled maximum process is stationary 
   in the frame moving with the edge. However the particle of maximal position 
   at $t_1$ undergoes symmetric diffusion, and as $\tau$ increases 
   is overtaken by the other particles in a time $\tau=O(1)$, leading to
   exponential decay of correlations. }
  \label{fig:fig1}
\end{figure}

Let us give the result for the joint CDF ${\rm Prob}(X(t_1)<X_1,\dots,X(t_n)<X_n)$. 
In view of the previous paragraph we define the dimensionless rescaled time differences
$\tau_i$ as 
\be 
t_j = t_1 (1+ \frac{\tau_{j,1}}{\log N}) 
\ee 
with the notation $\tau_{i,j}=\tau_i-\tau_j$ and $\tau_1=0$. As above one performs the change of variable
\be \label{Xj} 
X_j = \sqrt{2 t_j} \sqrt{\log N} (1 + \frac{z_j +c_N}{2 \log N} ) \simeq
\sqrt{2 t_1} \sqrt{\log N} (1 + \frac{z_j + \tau_{j,1} +c_N}{2 \log N} )  
\ee 
Then, at large $N$  
we obtain (by similar manipulations as in the previous section, see Appendix \ref{app:derivationPDF}) that the joint CDF takes the form 
\bea  \label{mainresult1} 
&& {\rm Prob}(X(t_1)<X_1,\dots,X(t_n)<X_n) \simeq \exp \left( - \Phi(z_1,\dots,z_n;\tau_{2,1},\dots,\tau_{n,n-1}) \right) \\
&& 
\Phi(z_1,\dots,z_n;\tau_{2,1},\dots,\tau_{n,n-1}) 
= \int_{y_1,\dots,y_n} ( 1 - \prod_{i=1}^n \theta_{y_i<z_i}  ) 
 e^{- y_1 }
G(y_{2,1},\tau_{2,1}) \dots G(y_{n,n-1},\tau_{n,n-1}) \label{mainresult2} 
\eea 
where here and below we often use the notations $y_{i,j}=y_i-y_j$ and the shorthand $\int_{y}=\int dy=\int_{- \infty}^{+\infty} dy$
as well as $\theta_{y<z}=\theta_{z>y}=\theta(z-y)$ where $\theta(x)$ is the Heaviside function, and
\be 
G(y,\tau) = \frac{1}{\sqrt{4 \pi \tau}} e^{- \frac{(y+\tau)^2}{4 \tau} } \label{freediff} 
\ee 
is simply the free diffusion kernel, however {\it with a negative unit drift} (and with $D=2$). This drift originates
from the fact that the position of the maximum increases with time,
as can be seen e.g. in \eqref{Xj}. Hence the front of the cloud of particles moves to the right and
with respect to this front the diffusion of a single particle, which is symmetric, has a negative drift.
This is why, as we will see below, the correlations decay exponentially at large time: after
a scaled time difference $\tau=O(1)$ the cloud of particles overtakes the particle which was the rightmost 
at time $t_1$. This is illustrated in Fig. \ref{fig:fig1}.
Finally, the factor $e^{-y_1}$ reminds that the Gumbel distribution is a (non-normalizable) stationary distribution
of the rescaled maximum process in the $z$ variable (see below).
\\

Although exact, the above formula is delicate. Indeed each of the two 
terms in $( 1 - \prod_{i=1}^n \theta_{y_i<z_i} )$ is a divergent integral (for $y_j \to y_j +y$ with $y \to - \infty$), and only
the combination is finite and the terms cannot be separated.
\\

Let us give an equivalent formula for $n=2$. One recombine as 
\be
1 - \theta_{y_1<z_1}  \theta_{y_2<z_2}   = 
1 - (1- \theta_{z_1<y_1} )  (1- \theta_{z_2<y_2} ) 
=
\theta_{z_1<y_1} + \theta_{z_2<y_2}
- \theta_{z_1<y_1}  \theta_{z_2<y_2}  \label{decomposition} 
\ee 
We use the important property that $e^{-z_1} dz_1$ is a (non-normalizable) stationary measure 
of the free diffusion with a negative unit drift, i.e. it satisfies
\be \label{stationarity} 
\int dz_1 e^{- z_1 }
G(z_{2,1},\tau_{2,1}) = e^{-z_2} 
\ee 
for any real $z_1,z_2$ and $\tau_{2,1}>0$, where we here and below use the notation $z_{2,1}=z_2-z_1$. The other important property is the normalization condition
\be 
\int dy \, G(y,\tau) = 1 
\ee 
Using these two properties, inserting \eqref{decomposition} into \eqref{mainresult2}
one finds
\be 
\Phi(z_1,z_2;\tau_{2,1}) = e^{-z_1} + e^{-z_2} - \int_{z_1<y_1,z_2<y_2} 
 e^{- y_1 } G(y_{2,1},\tau_{2,1})  \label{Phin2} 
\ee 
This function admits an explicit expression in terms of error functions,
it is given below in \eqref{defphitau} and \eqref{phitauexplicit}
and in \eqref{Phiexplicit} in the Appendix. 
\\


Let us now generalize this formula to arbitrary $n$. Starting from $n=3$ one must also use 
a third property, the convolution identity
\be \label{reproducibility} 
\int dy \, G(z,y,\tau') G(y,x,\tau) = G(z,x,\tau+\tau') 
\ee 
It is then easy to see, using these three properties, that one has
\bea 
&& \Phi(z_1,z_2;\tau_{2,1})  = e^{-z_1} + e^{-z_2}  - g(z_1,z_2;\tau_{2,1}) \label{Phi2} \\
&& \Phi(z_1,z_2,z_3;\tau_{2,1},\tau_{3,2})  = e^{-z_1} + e^{-z_2} + e^{-z_3}  \label{Phi3}  \\
&& 
- g(z_1,z_2;\tau_{2,1}) - g(z_2,z_3;\tau_{3,2}) - g(z_1,z_3;\tau_{31})  
+ g_3(z_1,z_2,z_3;\tau_{2,1},\tau_{3,2}) \nn
\eea 
and so on, where we have defined
\bea 
&& g(z_1,z_2;\tau_{2,1})  = \int_{z_1<y_1,z_2<y_2} 
 e^{- y_1 } G(y_{2,1},\tau_{2,1}) = e^{-z_1} \int_{0<y_1,0<y_2} 
 e^{- y_1 } G(y_{2,1}+z_{2,1},\tau_{2,1}) \label{g_2def} 
  \\
 && g_3(z_1,z_2,z_3;\tau_{2,1},\tau_{3,2}) = \int_{z_1<y_1,z_2<y_2,z_3<y_3} 
 e^{- y_1 } G(y_{2,1},\tau_{2,1})   G(y_{3,2},\tau_{3,2}) \label{defg3}   \\
 && = e^{-z_1} \int_{0<y_1,0<y_2,0<y_3} 
 e^{- y_1 } G(y_{2,1}+ z_{2,1},\tau_{2,1})   G(y_{3,2}+ z_{3,2},\tau_{3,2}) \label{fact3} 
\eea 
and so on, with more generally, for $a_1 < \dots < a_k$, $k \geq 1$
\be 
g_k(z_{a_1},\dots,z_{a_k};\tau_{a_2,a_1},\dots,\tau_{a_k,a_{k-1}})  = \int_{z_{a_1}<y_{a_1},\dots,z_{a_k}<y_{a_k}}
e^{- y_{a_1} } G(y_{a_2,a_1} , \tau_{a_2,a_1}) \dots G(y_{a_k,a_{k-1}} , \tau_{a_k,a_{k-1}})
\ee 
with $g_2=g$ and $g_1(z)=e^{-z}$. It is easy to see from the above
properties \eqref{reproducibility} and \eqref{stationarity} 
that if any argument $z_{a_j}$ for some $j$ in $g_k$ is set
to $z_{a_j}=-\infty$ then $g_k \to g_{k-1}$ and $z_{a_j}$ and $\tau_j$ are dropped from the list of
arguments. Also, any $g_k$ obviously vanishes when any $z_{a_j}$ for some $j$  argument is taken to $+\infty$. 

One can see on \eqref{g_2def} and \eqref{fact3} that one can always factor out the term $e^{-z_1}$,
the rest being only function of differences of $z_j$'s. This important factorization property is true for any $n$, 
i.e. one can always write
\be \label{factorization} 
\Phi(z_1,\dots,z_n;\tau_{2,1},\dots,\tau_{n,n-1}) = e^{-z_1} \phi_{\tau_{2,1},\dots,\tau_{n,n-1}}(z_{2,1},\dots,z_{n,n-1}) 
\ee 
The functions $\phi$ will be further studied below.

The above result provides explicit, although lengthy, expressions for the multi-time joint CDF of the 
maximum. As we will see below, using a second method based on the heat equation, the functions $\Phi$ can also be computed
recursively with $n$. This leads to more compact expressions. 
Before explaining that, let us give some explicit results for marginals, 
cumulants and correlation functions for $n=2$ and $n=3$. 

\section{Multi-time marginals, moments and correlations for the maximum} 
\label{sec:correlations}

Let us analyze in more details the joint PDF obtained in the previous section.
We will derive very explicit results for $n=2$, and give some general
formula for $n=3$.

\subsection{Two time correlations}

Let us start with $n=2$. 

\subsubsection{Two time joint CDF}

One recalls that at large $N$ with 
$t_2-t_1= t_1 \frac{\tau_{2,1}}{\log N}$, with $\tau_{2,1}= \tau = O(1)$,
the joint CDF for the position of the maximum takes the form
\be  \label{mainresult4} 
 {\rm Prob}(X(t_1)<X_1,X(t_2)<X_2) \simeq Q_{<,<}(z_1,z_2) = \exp \left( - \Phi(z_1,z_2;\tau)  \right) 
\ee
where we denote $Q_{< <}(z_1,z_2)$ the two-time CDF of the process in the variables $z_j$
(implicitely at times $t_j$), and 
$\Phi(z_1,z_2;\tau)$
was defined in \eqref{Phin2}. 
Through the change of variable $y_j \to y_j + z_j$ it can also be written as
\be 
\Phi(z_1,z_2;\tau) = e^{-z_1} \phi_\tau(z_{2,1}) \quad , \quad 
\phi_\tau(z) = 1+ e^{-z} -  \int_{y_1>0, y_2>0} 
 e^{- y_1 } G(y_{2,1}+z,\tau) \label{defphitau} 
\ee
The integral can be computed, see Appendix \ref{app:integrals}.
This leads to the explicit form
\be 
\phi_\tau(z) = \frac{1}{2}
   \left(\text{erf}\left(\frac{\tau
   +z}{2 \sqrt{\tau }}\right)+e^{-z}
   \text{erfc}\left(\frac{z-\tau }{2
   \sqrt{\tau }}\right)+1\right) \label{phitauexplicit} 
\ee 
which obeys the important symmetry (see Appendix \ref{app:integrals})
\be 
\phi_\tau(z) e^z = \phi_\tau(-z) \label{symm0} 
\ee 
This symmetry is equivalent to the fact that $\Phi(z_1,z_2;\tau)$ is
symmetric in $z_1,z_2$ as is visible on its explicit form 
\eqref{Phiexplicit} in the Appendix. 

In summary the two time CDF of the maximum, in the rescaled variables, has the form
\be \label{Q<<} 
Q_{<<}(z_1,z_2) \simeq e^{ - e^{-z_1} \phi_\tau(z_{2,1}) } 
\ee 
where the function $\phi_\tau(z)$ is given in \eqref{phitauexplicit}.
As mentioned in the introduction, this distribution appeared 
before \cite{HuslerReiss1989,HuslerBivariate} and is
known as the bivariate HR distribution. 
\\

The function $\phi_\tau(z)$ has the following asymptotic behaviors for large argument
\bea \label{asymptoticsphitau} 
&& \phi_\tau(z) \simeq e^{-z} + \psi_\tau(z) \quad , \quad  z \to - \infty \\
&& \phi_\tau(z) \simeq 1 + \psi_\tau(z)  \quad , \quad z \to + \infty \\
&& \psi_\tau(z)  = e^{- \frac{ (z+ \tau)^2}{4 \tau}}  \frac{2 \tau^{3/2}}{z^2 \sqrt{\pi}} 
(1 + \frac{\tau (\tau -6)}{z^2} + \frac{\tau^2 (60+\tau(\tau-20)}{z^4} +
O(z^{-6})) \quad , \quad |z| \to + \infty 
\eea 
Note that $\psi_\tau(-z)=e^z \psi_\tau(z)$. These asymptotics guarantee that one recovers the one time Gumbel CDF
for $z_1 \to +\infty$ or $z_2 \to +\infty$, i.e. one has
\be 
Q_{<<}(z,+\infty) = Q_{<<}(+\infty,z) = e^{- e^{-z}} 
\ee
In addition, at large $\tau$, since the free diffusion kernel with drift 
$G(y_{2,1}+z,\tau) \to 0$, one sees from \eqref{defphitau} that
\be 
\phi_\tau(z) \simeq 1 + e^{-z} \quad , \quad \Phi(z_1,z_2;\tau) \simeq e^{-z_1} + e^{-z_2} \quad , \quad 
\tau \to +\infty
\ee 
which corresponds to two uncorrelated Gumbel variables. 

\subsubsection{Exponential moments of $z_1$ and $z_2$}

Through \eqref{Xj} with $X_j=X(t_j)$ one can (abusively)
also think of $z_1$ and $z_2$ as random variables. 
Furthermore, from now on it is useful to consider $z_1$ and $z_{2,1}=z_2-z_1$
as the two random variables of interest.

A first result, see Appendix \ref{app:n2}, is an explicit integral expression for
the joint moment generating function, which reads
\be \label{gener} 
 \langle e^{- s z_1 - b z_{2,1}} \rangle =
\Gamma(1+s) \int dz e^{-b z} \partial_{z} \left(  (1 + \frac{ \phi_\tau'(z)}{\phi_\tau(z) }) 
\frac{1}{\phi_\tau(z)^{s}} \right)
\ee
where here and below $\langle \dots \rangle$ denotes an expectation value. 
A simpler expression also exists for \eqref{gener},
see \eqref{simplerexp}, but it is less convenient for 
the analysis of the moments. For $b=0$ the integral in \eqref{gener} can be performed and the
boundary term is $1$ at $z=+\infty$ and vanish at $z=-\infty$ (see Appendix \ref{app:n2}), recovering the 
generating function of the one-time Gumbel distribution
\be 
\langle e^{- s z_1 } \rangle = \Gamma(1+s) 
\ee 
One also recovers the same result for $z_2$, by setting $b=s$, although the
algebra is slightly more involved, see Appendix \ref{app:n2}. 

\subsubsection{PDF of $z_2-z_1$}

\begin{figure}[t]
     \includegraphics[width=0.5\columnwidth]{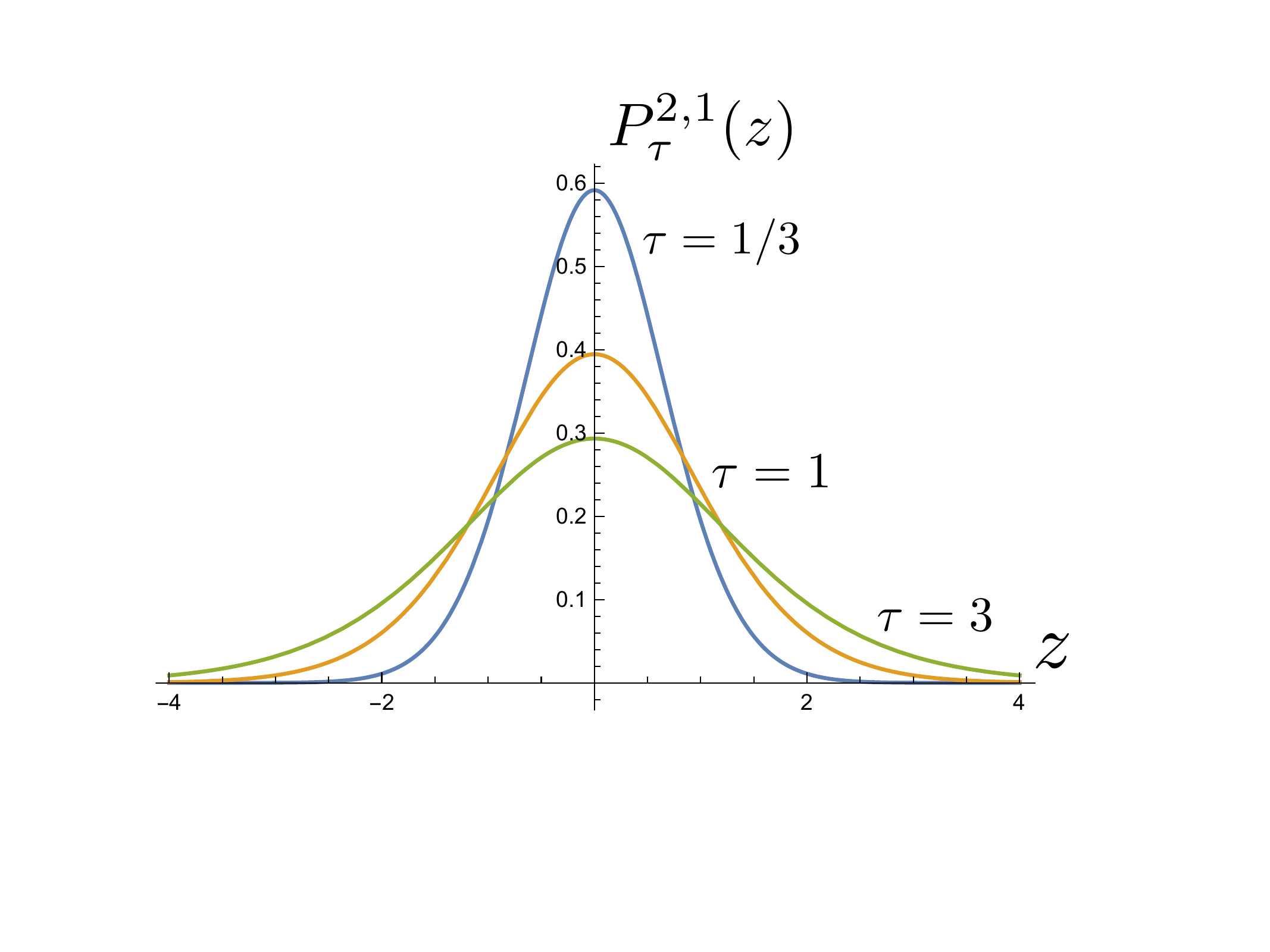}
      \includegraphics[width=0.5\columnwidth]{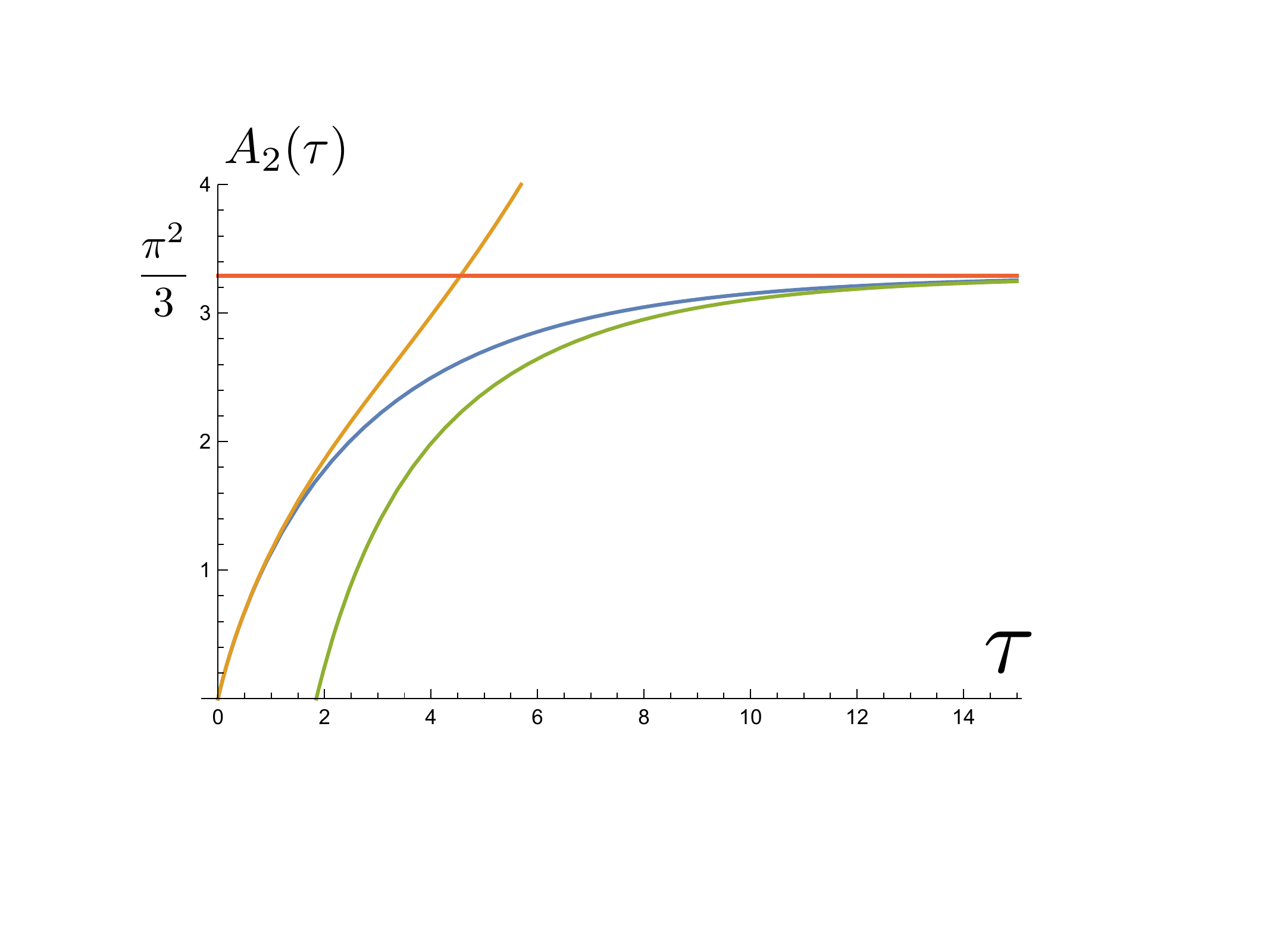}
     \vspace{-1.6cm}
   \caption{Left: marginal PDF of $z=z_{2,1}=z_2-z_1$, the scaled distance
   traveled by the maximum, with 
   $X(t_2)-X(t_1) \simeq \sqrt{\frac{t_1}{2 \log N}} (\tau+z)$. It is
   plotted for $\tau=1/3,1,3$. Right: its second moment $\langle z_{2,1}^2 \rangle=A_2(\tau)$, plotted
   versus $\tau$. The first three terms in the small $\tau$ asymptotics 
   \eqref{asymptoticsA2small}, and the first two terms in the 
   large $\tau$ asymptotics \eqref{asymptoticsA2} are also plotted, together with
   the limiting value $\pi^2/3$. Recall that $\pi^2/3-A_2(\tau)$, i.e. the curve reflected 
   versus $\pi^2/3$ describes the two-time covariance of the maximum, see Eq. 
   \eqref{cov00}. }
  \label{fig:fig2}
\end{figure}

It is possible to obtain explicitly the PDF of $X(t_2)-X(t_1)$ the distance over which
the maximum has moved. One has from \eqref{Xj} that 
$X(t_2)-X(t_1) = \sqrt{\frac{t_1}{2 \log N}} (\tau + z_{2,1})$, where we consider
$z_{2,1}=z_2-z_1$ as a random variable. Its distribution is
easily obtained. Indeed, \eqref{gener} for $s=0$ implies the following expression for the generating function
of the moments of $z_{2,1}$
\be
 \langle e^{- b z_{2,1}} \rangle = \int dz e^{-b z} \partial_{z} 
 (1 + \frac{ \phi_\tau'(z)}{\phi_\tau(z) }) \label{genersmall} 
\ee
This, in turn, implies the following expressions for the PDF (denoted $P^{(2,1)}_\tau(z)$)
and the CDF of the variable $z_{2,1}$ 
\be  \label{PPP21} 
P^{(2,1)}_\tau(z)= \partial_{z}  \frac{ \phi_\tau'(z)}{\phi_\tau(z) } = \partial_z^2 \log \phi_\tau(z)
\quad , \quad {\rm Prob}(z_{2,1} < z) = 1+ \partial_z \log \phi_\tau(z) 
\ee 
where $\phi_\tau(z)$ is given explicitly in \eqref{phitauexplicit}.
Thanks to the symmetry \eqref{symm0} we see that the PDF is an even function of $z$
\be 
P^{2,1}_\tau(-z)=P^{2,1}_\tau(z) 
\ee 
hence all odd moments of $z_{2,1}$ vanish. The PDF $P^{(2,1)}_\tau(z)$
is plotted in Fig. \ref{fig:fig2}. Its behavior for large $|z|$ at fixed $\tau$ is 
\be 
P^{2,1}_\tau(z) \simeq \frac{1}{\sqrt{4 \pi \tau}} e^{- \frac{(|z|+\tau)^2}{4 \tau}} 
(1 + \frac{2 \tau}{|z|} + \frac{2 \tau^2}{z^2} + O(|z|^{-3}) 
\ee 
hence for large argument is becomes equal to the free diffusion kernel with unit drift.
Around $z=0$ it is analytic and behaves as 
\be
P^{2,1}_\tau(z)  = 
\frac{1}{4} + \frac{e^{-\tau /4}}{\sqrt{\pi
   } \sqrt{\tau } \left(2
   \text{erf}\left(\frac{\sqrt{\tau
   }}{2}\right)+2\right)}
   -\frac{z^2}{16} 
   \left(1 + \frac{e^{-\tau /2}
   \left(\sqrt{\pi } e^{\tau /4} (5
   \tau +2)
   \left(\text{erf}\left(\frac{\sqrt
   {\tau }}{2}\right)+1\right)+6
   \sqrt{\tau }\right)}{\pi  \tau
   ^{3/2}
   \left(\text{erf}\left(\frac{\sqrt
   {\tau
   }}{2}\right)+1\right)^2}\right)
   +O(z^4)
\ee
In the limit of large $\tau$ and fixed $z$ it converges to a finite limit
\be 
P^{2,1}_\tau(z) = \frac{1}{4 \cosh^2(\frac{z}{2})} + O( e^{-\tau/4} \tau^{-1/2}) \label{largetau} 
\ee 
which is simply the PDF of the difference of two uncorrelated Gumbel variables,
as it should. Indeed \eqref{largetau} implies 
\be 
\lim_{\tau \to +\infty} \langle e^{- b z_{2,1}} \rangle = \frac{\pi b}{\sin(\pi b)} 
= \Gamma(1+b) \Gamma(1-b) 
\label{largetauexp} 
\ee 

One can study how that limit is reached. One has, at large $\tau \gg 1$
\bea  \label{largetauphi} 
\phi_\tau(z) = 1 + e^{-z} - e^{- \frac{(z+\tau)^2}{4 \tau}} \chi_\tau(z) \quad , \quad 
\chi_\tau(z) = \frac{2}{\sqrt{\pi \tau}} (1 - \frac{2}{\tau} + \frac{12+z^2}{\tau^2} + O(1/\tau^3)) 
\eea 
Inserting into \eqref{PPP21}, expanding the logarithm to first order in $\chi_\tau(z)$,
writing the exponential as
$e^{- \frac{\tau}{4} - \frac{z}{2}} \times e^{- \frac{z^2}{4 \tau}}$ and expanding
the last factor in powers of $1/\tau$, one obtains the large $\tau$ expansion
\bea 
P^{2,1}_\tau(z)   = \frac{1}{4 \cosh^2(\frac{z}{2})} +  
\frac{2 e^{-\frac{\tau}{4}}  }{\sqrt{\pi \tau} } 
\bigg( 
\frac{1}{16} \frac{3-\cosh (z)}{\cosh^3(\frac{z}{2})} 
+\frac{ -3
   z^2+\left(z^2+16\right) \cosh
   (z)-8 z \sinh (z)-16}{64 \cosh^3(\frac{z}{2})
   \tau}+O(\tau^{-2})
   \bigg) \label{Plargetau} 
\eea

\subsubsection{Moments of $z_2-z_1$}

We recall that the odd moments of $z_{2,1}=z_2-z_1$ vanish.
One can further obtain formulae for the the even moments of $z_{2,1}$ by integration by part (boundary terms vanish),
as, with $n \geq 1$ 
\be 
\langle z_{2,1}^{2n} \rangle =  A_{2n}(\tau) = 2 \int_0^{+\infty} dz z^{2n} \partial_z^2 \log \phi_\tau(z)
= 4 n (2 n -1) \int_0^{+\infty} dz z^{2n-2}  \log \phi_\tau(z)   \label{defA2n} 
\ee 
which are easy to evaluate numerically. 
The function $A_2(\tau)$ is plotted in Fig. \ref{fig:fig2}.
Using \eqref{largetauexp} one finds that at large time
the moments tend to the moments of the difference of two
uncorrelated Gumbel variables, namely
\be 
\lim_{\tau \to +\infty}  \langle z_{2,1}^{2n} \rangle = A_{2n}(+\infty) =
(-1)^{n+1} (2^{2 n}-2) \pi^{2n} B_{2n} \label{Bernoulli} 
\ee 
where the $B_k$ are the Bernoulli numbers. 
The details asymptotics of $A_{2n}(\tau)$ at large $\tau$ is obtained as follows. From
\eqref{largetauphi} one has
\be 
\log \phi_\tau(z) = \log(1 + e^{-z})  - e^{- \frac{(z+\tau)^2}{4 \tau}} \frac{\chi_\tau(z)}{1+ e^{-z}} 
+ O(e^{- 2 \frac{(z+\tau)^2}{4 \tau}} ) 
\ee 
and one performs similar manipulations as above.
Using the last identity in \eqref{defA2n}, the integral on $z$ converges term by term
in the expansion,
leading to
\bea \label{asymptoticsA2}
&&\!\!\!  \!\!\!  A_2(\tau) = \frac{\pi^2}{3} -\frac{4 \sqrt{\pi } e^{-\tau/4}}{\sqrt{\tau }}
\left( 
1- \frac{8 + \pi^2}{\tau} + 
\frac{12+\frac{3 \pi ^2}{2}+\frac{5 \pi^4}{32}}{\tau^2} 
- \frac{120+15 \pi ^2+\frac{25 \pi^4}{16}+\frac{61 \pi ^6}{384}}{\tau^3} + O(\tau^{-4}) \right) \\
&& \!\!\!  \!\!\!  A_4(\tau) 
= \frac{7 \pi^4}{15} - \frac{ 24 \pi ^{5/2} e^{-\tau /4}}{\sqrt{\tau}} 
   \left( 
1 - \frac{2+\frac{5 \pi ^2}{4}}{\tau} + \frac{12+\frac{15 \pi ^2}{2}+\frac{61 \pi
   ^4}{32}}{\tau^2} -
   \frac{120+75 \pi ^2+\frac{305 \pi
   ^4}{16}+\frac{1385 \pi ^6}{384}}{\tau^3} +O(\tau^{-4})  
\right) \nn
\eea

%
%

At short time difference $\tau \ll 1$ the variable $z_{21}$ is of order 
$\sqrt{\tau}$. Defining the $O(1)$ random variable $w$ such that 
$z_{21}= w \sqrt{\tau}$ one finds that its PDF $p_\tau(w)$ admits a small
$\tau$ expansion
\be 
p_\tau(w)= \frac{e^{-\frac{w^2}{4}}}{2
   \sqrt{\pi }}-\frac{\sqrt{\tau} 
   \left(\sqrt{\pi }
   e^{-\frac{w^2}{4}} w \,
   \text{erf}\left(\frac{w}{2}\right
   )+\pi 
   \left(\text{erf}\left(\frac{w}{2}
   \right)^2-1\right)+2
   e^{-\frac{w^2}{2}}\right)}{4 \pi
   }+O\left(\tau\right)
\ee 
So not surprisingly $z_{2,1}$ undergoes free diffusion at short time, but there are 
corrections as $\tau$ increases. The short time expansion of the lowest
moments and cumulants, as well as of the kurtosis, reads
\bea \label{asymptoticsA2small}
&& \langle z_{2,1}^2 \rangle = \langle z_{2,1}^2 \rangle^c = A_2(\tau) = 
   2 \tau -\frac{4}{3} \sqrt{\frac{2}{\pi }}
   \tau^{3/2} + 0.217996 \tau^2 - 0.0129398 \tau^{5/2} + O(\tau^3)  \\
&& \langle z_{2,1}^4 \rangle = A_4(\tau)  = 12 \tau^2 -\frac{72}{5} \sqrt{\frac{2}{\pi
   }} \tau^{5/2} + O(\tau^3)  ~,~ \langle z_{2,1}^4 \rangle^c = \frac{8}{5} \sqrt{\frac{2}{\pi }}
   \tau^{5/2} +  O(\tau) 
   ~,~ {\rm Ku}= \frac{2}{5} \sqrt{\frac{2}{\pi }}
   \sqrt{\tau } + O(\tau) \nn \\
&& \langle z_{2,1}^6 \rangle = A_6(\tau)  = 120 \tau^3 -\frac{1380}{7}
   \sqrt{\frac{2}{\pi }} \tau^{7/2} + O(\tau^4) \nn 
\eea 
The small and large $\tau$ asymptotics are compared
with the numerical calculation of $A_2(\tau)$ in Fig. \ref{fig:fig2}.

\subsubsection{Covariance of $X(t_1)$ and $X(t_2)$} 

Let us translate some of these results in the original variables. Let us recall the relations
\bea 
&& X(t_1)= X_1 \simeq \sqrt{2 t_1} (\sqrt{\log N} + \frac{z_1}{2 \sqrt{\log N}}) \\
&& X(t_2)= X_2 \simeq \sqrt{2 t_1} (\sqrt{\log N} + \frac{\tau + z_2}{2 \sqrt{\log N}}) 
\eea 
Hence one has 
\be
X_2- X_1  \simeq \sqrt{\frac{t_1}{2 \log N} }  ( \tau + z_{2,1} )  \quad , \quad 
\tau=\frac{t_2-t_1}{t_1} \log N = O(1)
\ee 
The results of the previous subsection thus imply
for $\tau = O(1)$ and $p \geq 1$
\bea && \langle X_2-X_1 \rangle \simeq \sqrt{\frac{t_1}{2 \log N} }  \, \tau , \quad , \quad 
 \langle \left( X_2-X_1 -  \langle X_2-X_1 \rangle \right)^{2p+1} \rangle  = o( (\frac{t_1}{2 \log N})^{2p+1}) \\
&& {\rm Var} X_1 = {\rm Var} X_2 \simeq  \frac{ t_1 }{2 \log N} \frac{\pi^2}{6} 
\quad , \quad  {\rm Var}(X_2-X_1) \simeq \frac{ t_1 }{2 \log N} A_2(\tau) \\
&& {\rm Cov}(X_1,X_2)  = \frac{1}{2} ( {\rm Var}(X_1) + {\rm Var}(X_2)  - {\rm Var}(X_2-X_1)) \\
&& = \frac{ t_1 }{4 \log N}  ( \frac{\pi^2}{3} - A_2(\tau)) \quad , \quad 
 A_2(\tau) = 4 \int_0^{+\infty}  dz  \log \phi_\tau(z) \label{cov00} 
\eea
where we have used the values for the first two moments
of the Gumbel distribution, 
$\langle z_1^2 \rangle = \langle z_2^2 \rangle= \frac{\pi^2}{6}+ \gamma_E^2$
and $\langle z_1 \rangle = \langle z_2 \rangle= \gamma_E$.
The covariance is illustrated in Fig. \ref{fig:fig2}.
Using the asymptotics \eqref{asymptoticsA2}, we see that at large
time difference $\tau \gg 1$ the covariance of the maximum at two
different times decay as
\be \label{covlarge} 
 {\rm Cov}(X(t_1),X(t_2))  \simeq \frac{  t_1 }{4 \log N}  \frac{4 \sqrt{\pi } e^{-\tau/4}}{\sqrt{\tau }}
\ee 
As discussed in the previous section, the large time decay is exponential
since the particle which is the rightmost at time $t_1$ undergoes symmetric
diffusion, while the front of the other particles advances, with an effective unit
drift, on the time scales $\tau = O(1)$. 

\subsubsection{Correlations between $X(t_1)$ and $X(t_2)-X(t_1)$} 

There are correlations between the variable $z_1$ and $z_{2,1}$.
For instance from \eqref{gener} one obtains 
\bea 
&& \langle z_1 e^{- b z_{2,1}} \rangle 
- \langle z_1 \rangle \langle e^{- b z_{2,1}}  \rangle  = - \int dz e^{- b z} \partial_z \left( (1 + \frac{ \phi_\tau'(z)}{\phi_\tau(z) } ) \log \phi_\tau(z) \right)
\eea 
The lowest cross moment is trivial
\be 
\langle z_1 z_{2,1} \rangle  = - \frac{1}{2} \langle z_{2,1}^2 \rangle
\ee 
The first non trivial cross cumulant is
\bea 
&& \langle z_1 z_{2,1}^2 \rangle^c =  \langle z_1 z_{2,1}^2 \rangle 
- \langle z_1 \rangle \langle  z_{2,1}^2  \rangle  = - \int dz z^2 \partial_z \left( (1 + \frac{ \phi_\tau'(z)}{\phi_\tau(z) } ) \log \phi_\tau(z) \right)
\eea 
which upon rescaling yields the corresponding cross cumulant for the maximum $X(t_1)$
and its variation $X(t_2)-X(t_1)$.

\subsubsection{Conditional probability of $z_1$ given $z_2-z_1$} 
\label{subsec:conditioned} 

It is also interesting to compute the PDF of $z_1$ conditioned on a given value of $z_{2,1}=z$,
which turns out to have a simple form.
One finds
\be \label{qconditioned} 
q_\tau(z_1 | z_{2,1}=z) = 
\frac{\partial_{z_1} \partial_{z_2} e^{- e^{-z_1} \phi_\tau(z_{2,1})}|_{z_{2,1}=z} }{\partial_z^2 
\log \phi_\tau(z)} = \phi_\tau(z) 
(B_\tau(z) e^{-z_1} + \phi_\tau(z) (1- B_\tau(z)) 
e^{-2 z_1} ) ) e^{- e^{-z_1} \phi_\tau(z)} 
\ee 
with $B_\tau=\phi_\tau(\phi_\tau'+\phi_\tau'')/(\phi_\tau \phi_\tau''-(\phi'_\tau)^2 )$.
Upon the deterministic shift
\be 
z_1 = \tilde z_1 + \log \phi_\tau(z) 
\ee 
we see that $\tilde z_1$ has the same distribution that the one of a random variable which 
(i) with probability $B_\tau(z)$ is Gumbel (ii) 
with probability $1-B_\tau(z)$ has the PDF of the second maximum
$e^{-2 \tilde z_1} e^{- e^{- \tilde z_1}}$. Note that $0<B_\tau(z)<1$
is an even function of $z$ which reaches its strictly positive $\tau$-dependent 
minimum at $z=0$, and
with $B_\tau(\pm \infty)=1$. 

\subsection{Three time correlations} \label{3timecorr} 

For three times we will again consider $z_1$, $z_{2,1}=z_2-z_1$ and $z_{3,2}=z_3-z_2$
as the random variables of interest. Note that we use the notation $z_{21}$ and $z_{32}$
for the corresponding arguments of functions, or for real integration variables.
We use the same notation for $z_1,z_2,z_3$ as random variables and function arguments.

Anticipating a bit, in the next section, Section \ref{sec:heat}, we will obtain the simplest form for the three
time function $\Phi$. As noted in \eqref{factorization}, once again the dependence in $z_1$ can be singled out as
\be 
\Phi(z_1,z_2,z_3;\tau_{3,2},\tau_{2,1}) = e^{- z_1} \phi_{\tau_{2,1},\tau_{3,2}}(z_{2,1},z_{3,2}) 
\ee 
a property which extends to any number of time $n$. 
For $n=3$ one has
the tedious but explicit formula 
\bea  
&& \phi_{\tau_{2,1},\tau_{3,2}}(z_{21},z_{32}) = \int dy \, G(z_{32}+z_{21}-y,\tau_{3,2}) 
\phi_{\tau_{2,1}}(\min(z_{21},y)) \label{phitauexplicit3} \\
&&
= \frac{1}{2} \int \frac{dy}{\sqrt{4 \pi \tau_{32}} } e^{- \frac{(z_{32}+z_{21} - y + \tau_{3,2})^2}{4 \tau_{3,2}} }
   \left(\text{erf}\left(\frac{\tau_{2,1}+\min(z_{21},y)}{2 \sqrt{\tau_{2,1} }}\right)+e^{-\min(z_{21},y)}   \text{erfc}\left(\frac{\min(z_{21},y)-\tau_{2,1} }{2
   \sqrt{\tau_{2,1} }}\right)+1 \right) \nn
\eea 
Although we could not find perform this integral in closed form, it
can be easily evaluated numerically.
Let us now put this result to use.

One can generalize the manipulations in the previous section 
(see Appendix \ref{app:n2}) and obtain again the exponential moments as
\bea \label{gener3time} 
 \langle e^{- s z_1 - b z_{2,1} - c z_{3,2}} \rangle &=&
\Gamma(1+s) \int dz_{32} dz_{21} e^{- s z_1 - b z_{21} - c z_{32}} \\
& \times &
\partial_{z_{32}} (\partial_{z_{21}}- \partial_{z_{32}} ) 
 \left( (  1 + \frac{ \partial_{z_{21}} \phi_{\tau_{2,1},\tau_{3,2}}(z_{21},z_{32})}{\phi_{\tau_{21},\tau_{32}}(z_{21},z_{32})} )
 \phi_{\tau_{2,1},\tau_{3,2}}(z_{21},z_{32})^{-s}  \right) \nn
\eea
where now $z_{21}$ and $z_{32}$ denote two real integration variables.
Hence the joint PDF $P_{\tau_{2,1},\tau_{3,2}}(z_{21},z_{32})$ of the
random variables $z_{2,1}=z_2-z_1$ and $z_{3,2}=z_3-z_2$ is obtained as
\bea
P_{\tau_{2,1},\tau_{3,2}}(z_{21},z_{32})=
\partial_{z_{32}} (\partial_{z_{21}}- \partial_{z_{32}} ) \partial_{z_{21}}  \log 
 \phi_{\tau_{2,1},\tau_{3,2}}(z_{21},z_{32}) \label{PDF3} 
\eea
where $\phi_{\tau_{2,1},\tau_{3,2}}(z_{21},z_{32})$ is given explicitly in 
\eqref{phitauexplicit3}. Upon rescaling and shifting, \eqref{PDF3} gives the joint PDF
of the variations $X(t_3)-X(t_2)$ and $X(t_2)-X(t_1)$.

From \eqref{PDF3} one can evaluate the corresponding joint moments
through a double integral. 
The lowest non trivial such moments are of third order, i.e. $\langle z_{2,1}^2 z_{3,2} \rangle$
and $\langle z_{2,1} z_{3,2}^2 \rangle$. Indeed the order two correlation 
is simply related to two time moments
\be 
\langle z_{21} z_{32} \rangle = \langle z_{21} z_{32} \rangle_c
=\frac{1}{2} (A(\tau_{31})- A(\tau_{21}) - A(\tau_{32}) ) 
\ee

\section{Multi-time joint CDF for the maximum, from the heat equation} 
\label{sec:heat} 

We now give another calculation of the multi-time joint CDF for the maximum,
using the diffusion equation. It is slightly less controlled technically,
but quite intuitive physically.

Let us start with the one-time CDF. For a single particle $x(t)$ undergoing free 
diffusion, the CDF, which we denote here for convenience $P_<(x,t)={\rm Prob}(x(t)<x)$, satisfies
\be 
\partial_t P_< = \frac{1}{2} \partial_x^2 P_<  \label{heat} 
\ee 
with $P_<(-\infty,t)=0$ and $P_<(+\infty,t)=1$. Here the initial condition $P_<(x,t=0)=\theta(x)$,
although one can consider more general ones. 
Writing $P_<(x,t)=e^{-f(x,t)}$, the field $f(x,t)$ satisfies
\be 
\partial_t f = \frac{1}{2} \partial_x^2 f - \frac{1}{2} (\partial_x f)^2  
\ee 
Considering now $N$ identical copies, the 
CDF of the maximum is given by ${\rm Prob}(X(t)<x)=P_<(x,t)^N= e^{-N f(x,t)}=e^{-F(x,t)}$
where we defined $F(x,t)=N f(x,t)$. Hence the field $F(x,t)$ satisfies 
\be
\partial_t F = \frac{1}{2} \partial_x^2 F - \frac{1}{2 N} (\partial_x F)^2 \label{eqH} 
\ee 
with $F(-\infty,t)=+\infty$ and $F(+\infty,t)=0$. 

Let us now use diffusive
scaling, i.e. we define $F(x,t)=\tilde F(y,t)$ with
$y=x/\sqrt{2 t}$. This leads to 
\be 
2 t \partial_t \tilde F = y \partial_y \tilde F +  \frac{1}{2} \partial_y^2 \tilde F - \frac{1}{2 N} (\partial_y \tilde F)^2
\ee 
To anticipate the known result for $N$ large, we now make the further change of variable
\be 
\tilde F(y,t) = \hat F(z,t) \quad , \quad y = \sqrt{\log N} (1 + \frac{z + c_N}{2 \log N}) \quad , \quad 
dy = \frac{dz}{2 \sqrt{\log N}} 
\ee 
It gives
\be 
\frac{t}{\log N} \partial_t \hat F =  \partial_z \hat F +   \partial_z^2 \hat F
+ \frac{z + c_N}{2 \log N}  \partial_z \hat F 
- \frac{1}{N} (\partial_z \hat F)^2  \label{eqrescaled} 
\ee 
which, until now, is exact for any $N$. 
\\

Let us now consider time $t=O(1)$ and large $N \gg 1$. If we are looking for typical events, i.e. $\hat F=O(1)$,
and the equation formally becomes stationary
\be 
\partial_z \hat F +   \partial_z^2 \hat F \simeq  0
\ee 
Hence one sees that the Gumbel distribution, which corresponds to $\hat F(z)=e^{-z}$ 
is indeed a stationary distribution. Note that studying the finite $N$ corrections, and convergence
to stationarity \cite{footnoteconvergence}, would require to examine the various regimes in $z$ and their
matching \eqref{eqrescaled}, but we do not need it here.
\\

It turns out that it is relatively simple to obtain also the multi-time joint CDF from \eqref{eqrescaled},
i.e. the dynamics of the maximum. For this one needs to rescale the time, more precisely
rescale the relative time from a fixed reference time. To this aim we
fix some time $t=t_1$ and consider times very close to $t_1$, $t-t_1=t_1 \frac{\tau}{\log N}$.
Denoting for convenience $\hat F(z,t_1(1+\frac{\tau}{\log N})) \to \hat F(z,\tau)$, the
l.h.s of \eqref{eqrescaled} becomes
\be 
\frac{t}{\log N} \partial_t \hat F \simeq \frac{t_1}{\log N} \partial_t \hat F
=  \partial_\tau \hat F 
\ee 
Thus dropping the terms which are subdominant at large $N$ in the region $z=O(1)$, $\hat F=O(1)$,
we obtain
\bea
\partial_\tau \hat F = \partial_z \hat F +   \partial_z^2 \hat F  \label{freedrift} 
\eea 
which is precisely the diffusion with negative unit drift encountered in the previous sections,
of associated Green's function $G(z,\tau)$ in \eqref{freediff}. 
\\

Let us now apply this method to solve the two time problem $n=2$. The main point is that one can
use the same representation (with $t_2>t_1$)
\be 
{\rm Prob}(X(t_1)< x_1 , X(t_2)< x_2) = {\rm Prob}(x(t_1)< x_1 , x(t_2)< x_2)^N 
= e^{ - F(x_2,t_2; x_1,t_1) } 
\ee 
The single particle joint CDF, $Q_{t_1,t_2}(x_1,x_2)={\rm Prob}(x(t_1)< x_1 , x(t_2)< x_2)$, satisfies the same heat equation \eqref{heat}
as a function of $x_2$ and $t_2$, but with "initial" condition 
$Q_{t_1,t_1}(x_1,x_2)=P_{<}(\min(x_1,x_2),t_1)$. Hence 
$F(x_2,t_2; x_1,t_1)$ satisfies the same equation \eqref{eqH} as a function of $x_2$ and $t_2$, but with
initial condition $F(x_2,t_1;x_1,t_1)=F(\min(x_1,x_2),t_1)$, and
where $F(x,t)$ is the function studied above.
In the large $N$ limit and in the variables $z_1,z_2,\tau$ one thus finds that
$F(x_2,t_2;x_1,t_1) = \hat F(z_2,\tau;z_1)$ satisfies the negative unit drift diffusion equation \eqref{freedrift} with 
initial condition
\be 
\hat F(z_2,\tau=0;z_1) = e^{- \min(z_1,z_2)} 
\ee 
This implies that 
\bea 
 \hat F(z_2,\tau;z_1) &=& \int_{-\infty}^{z_1} dy_2 e^{-y_2} G(z_2-y_2,\tau) 
+ e^{-z_1} \int^{+\infty}_{z_1} dy_2 G(z_2-y_2,\tau)  \label{Hfinal2} \\
& =& e^{-z_1} \left( 
\int_{-\infty}^{0} dy \, e^{-y} G(z_{2,1} -y,\tau)  
+  \int^{+\infty}_{0} dy \, G(z_{2,1}-y,\tau) \nn
\right) 
\eea 
Remarkably, although the calculations are quite different, 
this gives exactly the same result as the other method, i.e. one finds
by explicit calculation of the integrals in \eqref{Hfinal2} 
\be 
\hat F(z_2,\tau;z_1) = \Phi(z_1,z_2;\tau) = e^{-z_1} \phi_\tau(z_{2,1}) 
\quad , \quad \phi_\tau(z) = \frac{1}{2}
   \left(\text{erf}\left(\frac{\tau
   +z}{2 \sqrt{\tau }}\right)+e^{-z}
   \text{erfc}\left(\frac{z-\tau }{2
   \sqrt{\tau }}\right)+1\right) \label{phitauexplicit2} 
\ee 
\\

This method can be iterated to obtain the result for $n=3$. One must again solve the negative unit drift diffusion equation \eqref{freedrift} in the variables $z_3$ and $\tau_{3,2}$ with the "initial" condition
\be 
\hat F(z_3,\tau_{3,2};z_2,\tau_{2,1}|z_1) |_{\tau_{3,2}=0} = \hat F(\min(z_2,z_3),\tau_{2,1};z_1) 
\ee
This gives
\be
 \hat F(z_3,\tau_{3,2};z_2,\tau_{2,1};z_1) = 
\int dy_3 G(z_3-y_3,\tau_{3,2})  \hat F(\min(z_2,y_3),\tau_{2,1};z_1)  
\ee 
%
%
%
Expanding into various integrals we obtain an expression which, 
remarkably again, although being apparently quite different, numerically
coincides with the one obtained by the first method.
Indeed we have checked numerically that 
\be 
\hat F(z_3,\tau_{3,2};z_2,\tau_{2,1};z_1) = \Phi(z_1,z_2,z_3;\tau_{2,1},\tau_{3,2}) 
= e^{-z_1} \phi_{\tau_{21},\tau_{32}}(z_{2,1},z_{3,2}) 
\ee 
where $\Phi$ and $\phi$ are defined in \eqref{Phi3} and \eqref{factorization}. 

Hence this second method gives a recursive way to construct the multi-time functions $\Phi$.
Specifically we obtain the recursion, for any $n \geq 2$
\be 
\Phi(z_1,\dots,z_n;\tau_{2,1},\dots,\tau_{n,n-1}) = 
\int dy \, G(z_n-y,\tau_{n,n-1})  \Phi(z_1,\dots,z_{n-2},\min(z_{n-1},y);\tau_{2,1},\dots,\tau_{n-1,n-2}) 
\ee 
Equivalently, for the functions $\phi$ defined in \eqref{factorization} the recursion
reads
\be 
\phi_{\tau_{2,1},\dots,\tau_{n,n-1}}(z_{2,1},\dots,z_{n,n-1}) 
= \int dy \, G(z_{n,1}-y,\tau_{n,n-1})   
\phi_{\tau_{2,1},\dots,\tau_{n-1,n-2}}(z_{2,1},\dots,z_{n-1,n-2},\min(z_{n,n-1},y))
\ee 
where $z_{n,1}=z_{2,1}+z_{3,1}+ \dots + z_{n,n-1}$. 
This recursive construction was used above in Section \eqref{3timecorr}.

\section{Multi-time observables for outliers}
\label{sec:secondmax} 

In this section we study some multi-time observables for the outliers
in the case of $N$ independent Brownian motions all starting from the origin.
In the text we focus on the maximum $X^{(1)}(t)$ and second maximum $X^{(2)}(t)$
and obtain their joint two-time distribution. This is achieved by
studying the two-time counting statistics with several intervals.
In the Appendix \ref{app:outliers} it is indicated
how to extend these results to a secondary maxima of any rank, and to any number of times.

\subsection{Two-time distribution of maximum and second maximum} 

In this section for convenience we will adopt slightly different notations from the rest of the paper, so we denote
$t$ and $t'$ the two different times, use $1$ and $2$ for first and second maximum,
and prime quantities denote the quantities at time $t'>t$. To study large $N$ we perform again the change of variable
\be 
N P_{>,t}(X_i)= e^{-z_i} \quad , \quad N P_{>,t'}(X_i')= e^{-z_i'} \quad , \quad t'-t = t \frac{\tau}{\log N} 
\ee 
and for the problem at hand this leads to the change of variable (which we use
interchangeably for the random process as well
as for the real variables) 
\bea 
&& X^{(1)}(t) \equiv X_1 \simeq \sqrt{2 t} ( \sqrt{\log N} + \frac{z_1+c_N}{2 \sqrt{\log N}} ) \\
&& X^{(2)}(t) \equiv X_2 \simeq \sqrt{2 t} ( \sqrt{\log N} + \frac{z_2+c_N}{2 \sqrt{\log N}} ) \\
&& X^{(1)}(t') \equiv X'_1 \simeq \sqrt{2 t'} ( \sqrt{\log N} + \frac{z'_1+c_N}{2 \sqrt{\log N}} ) \\
&& X^{(2)}(t') \equiv X'_2 \simeq \sqrt{2 t'} ( \sqrt{\log N} + \frac{z'_2+c_N}{2 \sqrt{\log N}} )
\eea 
In these variables our main result is that for $\tau=O(1)$ the joint PDF  
$q(z_1,z_2,z'_1,z'_2;\tau)$ of the random variables $z_1,z_2,z'_1,z'_2$
reads, for $z_1>z_2$, $z'_1>z'_2$, 
\bea \label{q4} 
&& q(z_1,z_2,z'_1,z'_2;\tau) =  \partial_{z_1} \partial_{z'_1} \partial_{z_2} \partial_{z'_2} 
\bigg( ( \Phi(z_1, z'_2;\tau) \Phi(z_2,z'_1;\tau) - \Phi(z_1,z'_1;\tau)  )  e^{- \Phi(z_2,z'_2;\tau)} \bigg) \\
&&  = \partial_{z_1} \partial_{z'_1} \partial_{z_2} \partial_{z'_2}  \bigg( 
\left( e^{-z_1-z_2} \phi_\tau(z'_2-z_1) \phi_\tau(z'_1-z_2) - e^{-z_1} \phi_\tau(z'_1-z_1) 
 \right) e^{- e^{-z_2} \phi_\tau(z'_2-z_2)} \bigg)
\eea 
and is zero otherwise. This formula is completely explicit if one uses the
expression of $\phi_\tau(z)$ given in \eqref{phitauexplicit}.
\\

We obtain this result by two different methods. The first one is direct but tedious
and given in the Appendix \ref{app:combinatorics}. The second one uses the counting statistics, which is interesting in it's own sake and which we 
now describe. Note that this method also yields the two-time joint CDF of the second maximum, see
formula 
\eqref{CDF2timesecond}.
\\

\subsection{Two-time counting statistics} 
\label{sec:multicounting} 

Let us generalize the considerations about counting statistics of Section 
\ref{subsec:counting}. Given $X_1,X_2,X'_1,X'_2$ with $X_1>X_2$ and 
$X'_1>X'_2$, we again split the line into three disjoint intervals (a) $x>X_1$, (b) $X_2<x<X_1$
and (c) $x<X_2$ at time $t$, and into three disjoint intervals (a') $x'>X'_1$, (b') $X'_2<x'<X'_1$
and (c') $x'<X'_2$ at time $t'$. Each particle has some probability of being in one of these intervals at $t$ and another one at $t'$. We denote these probabilities as follows, e.g.
\be
P_{aa'} = P(x>X_1,x'>X'_1) = \langle \theta_{x>X_1} \theta_{x'>X'_1} \rangle \quad , \quad P_{ab'} = P(x>X_1,X'_2<x'<X'_1)
\ee
and so on, where we recall that $\langle \dots \rangle$ denote expectation values.
Since the events are mutually exclusive one has, for each particle
\bea
P_{aa'} + P_{ab'} + P_{ac'} + P_{ba'} + P_{bb'} + P_{bc'} + 
P_{ca'} + P_{cb'} + P_{cc'} = 1 
\eea 
Raising to the power $N$ and expanding we can read off the joint probabilities
that there are $\{ n_{i,i'} \}=\{n_{aa'},n_{ab'},\dots,n_{cc'}\}$ particles which are respectively in the interval
$i=a,b,c$ at $t$, and in the interval $i'=a',b',c'$ at $t'$. It is simply the multinomial distribution
\be \label{PoissonMultiple} 
P(\{ n_{ii'} \}) = \frac{N!}{\prod_{i=a,b,c} \prod_{i'=a',b',c'} n_{ii'} !} 
\prod_{i=a,b,c} \prod_{i'=a',b',c'}P_{ii'}^{n_{ii'}} 
\ee 
In the large $N$ limit and at the edge, all the occupation numbers $n_{ii'}=O(1)$, except $n_{cc'}$ which
is a macroscopic number, $n_{cc'} \simeq N$, and one has the asymptotics, e.g. from 
\eqref{mainresult1} for $n=2$
\be 
P_{cc'}^{n_{cc'}} \simeq P(x < X_1,x' < X_1')^N \simeq e^{- \Phi(z_1,z_1';\tau) }
\ee
where $\Phi$ is defined in \eqref{Phin2}, and its explicit expression is given in 
\eqref{defphitau} and \eqref{phitauexplicit}, or also 
in \eqref{Phiexplicit}. The multinomial coefficient in \eqref{PoissonMultiple} 
provides one power of $N$ for each of the remaining $P_{ii'}$.
To evaluate these probabilities at large $N$ one first recall that 
\be \label{NPaa} 
N P_{aa'} = N P(x>X_1,x'>X_1') \simeq g(z_1,z_1';\tau) = e^{-z_1} + e^{-z_1'} - \Phi(z_1,z_1';\tau) 
\ee
where the function $g$ is defined in \eqref{g_2def} and its explicit expression is given in 
\eqref{gexplicit}.
Next one expresses
all the other probabilities in terms of this one, and of the single time probabilities. For
instance one has 
\bea 
&& N P_{ab'} = N P(x>X_1,X'_2<x'<X'_1) = N \langle \theta_{x>X_1} (\theta_{x'>X'_2} - 
\theta_{x'>X'_1} ) \rangle \simeq g(z_1,z_2';\tau) - g(z_1,z_1';\tau) \nn \\
&& N P_{ac'} = N P(x>X_1,x'<X'_2) = N \langle \theta_{x>X_1} (1-\theta_{x'>X'_2}) \rangle
\simeq e^{-z_1} - g(z_1,z_2';\tau)  
\eea
and so on.  This leads to the following multiple independent Poisson distribution 
\be \label{bigPoisson2} 
 {\rm Prob}(n_{aa'},n_{ab'},n_{ac'},n_{ba'},n_{bb'},n_{bc'},n_{ca'},n_{cb'}) 
= \frac{\lambda_{aa'}^{n_{aa'}} \, \lambda_{ab'}^{n_{ab'}} \, \lambda_{ac'}^{n_{ac'}} \lambda_{ba'}^{n_{ba'}} 
\lambda_{bb'}^{n_{bb'}} \, \lambda_{bc'}^{n_{bc'}} \lambda_{ca'}^{n_{ca'}} \, \lambda_{cb'}^{n_{cb'}}   }
{n_{aa'}! n_{ab'}! n_{ac'} ! n_{ba'} ! n_{bb'} ! n_{bc'} ! n_{ca'} ! n_{cb'} ! } e^{- \Phi(z_2,z'_2,\tau)}  
\ee
where the mean parameters, i.e. such that $\langle n_{ii'} \rangle = \lambda_{ii'}$, are given by
\bea \label{alllambda} 
&& \lambda_{aa'} = g(z_1,z'_1;\tau) \quad , \quad 
 \lambda_{ab'} = g(z_1,z'_2;\tau)-g(z_1,z'_1;\tau) \quad , \quad 
 \lambda_{ac'} = e^{-z_1}- g(z_1,z'_2;\tau) \\
&& \lambda_{ba'} = g(z_2,z'_1;\tau)-g(z_1,z'_1;\tau) \quad , \quad 
 \lambda_{bb'} = g(z_2,z'_2;\tau)-g(z_1,z'_2;\tau) - g(z_2,z'_1;\tau) + g(z_1,z'_1;\tau) \nn \\
&& \lambda_{bc'} = e^{-z_2}-e^{-z_1} - g(z_2,z_2';\tau) + g(z_1,z_2';\tau) 
\quad , \quad   \lambda_{ca'} = e^{-z'_1}- g(z_2,z'_1;\tau) \nn \\
&& \lambda_{cb'} = e^{-z_2'}-e^{-z_1'} - g(z_2,z_2';\tau) + g(z_2,z_1';\tau) \nn 
\eea 
By summing over the $\{n_{ii'} \}$ one can check that this distribution is correctly normalized to unity.
That is, the sum of all the $\lambda_{ii'}$ equals exactly $\Phi(z_2,z'_2;\tau)$
(using the second relation in \eqref{NPaa}). 

One can also check that the one time result \eqref{Poisson2} is recovered.
Indeed one has $n_a=n_{aa'}+n_{ab'}+n_{ac'}$ is the sum of three independent
Poisson variables, and the same for $n_b=n_{ba'}+n_{bb'}+n_{bc'}$, 
independent of $n_a$. The
mean parameters simply add up and one can check that 
\bea 
&& \lambda_a = \lambda_{aa'} + \lambda_{ab'} + \lambda_{ac'} = e^{-z_1} \\
&& \lambda_b = \lambda_{ba'} + \lambda_{bb'} + \lambda_{bc'} = e^{-z_2} - e^{-z_1} 
\eea 
in agreement with \eqref{Poisson2}. The same check can be performed
for $\lambda_{a'}$ and $\lambda_{b'}$ with $z_1,z_2$ replaced by $z'_1,z'_2$. 

From the general result \eqref{bigPoisson2} the probability of various events can be 
computed. For instance one obtains the joint "CDF" of the maximum and second maximum
at two times. Indeed one can check that the event where one has simultaneously
\be 
X^{(1)}(t) >X_1, X^{(2)}(t)<X_2, X^{(1)}(t')>X'_1, X^{(2)}(t') <X'_2 
\ee 
is equivalent to the event 
\bea \label{cases} 
&& n_{aa'}=1 \, \, \text{and} \, \,  n_{ab'}=n_{ac'}=n_{ba'}=n_{bb'}=n_{bc'}=n_{ca'}=n_{cb'}=0 \\
&& \text{OR} \,\,\, 
n_{ac'}=n_{ca'}=1 \, \, \text{and} \, \,  n_{aa'}=n_{ab'}=n_{ba'}=n_{bb'}=n_{bc'}=n_{cb'}=0 \nn
\eea 

\begin{figure}[t]
     \includegraphics[width=0.5\columnwidth]{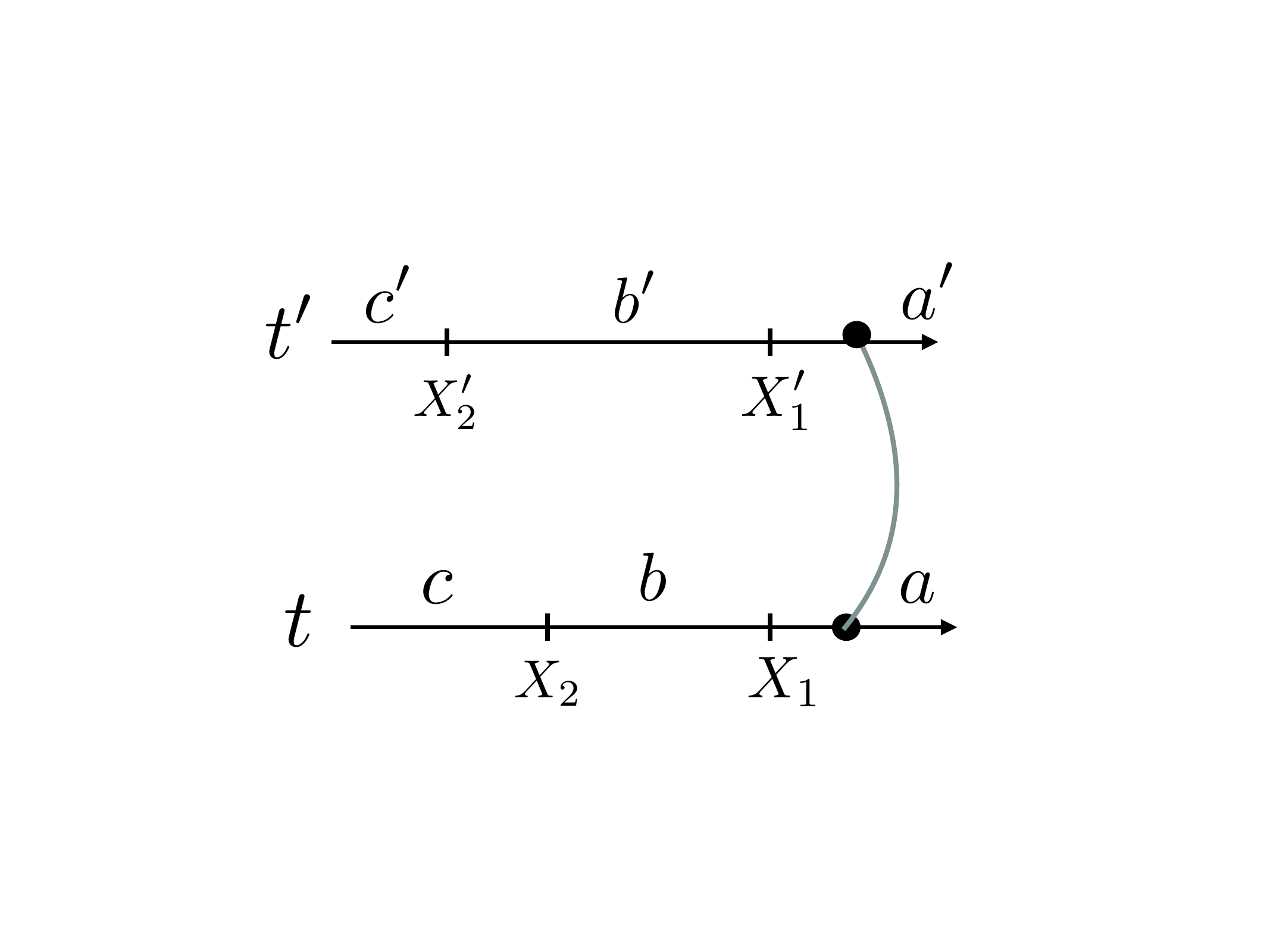}
      \includegraphics[width=0.5\columnwidth]{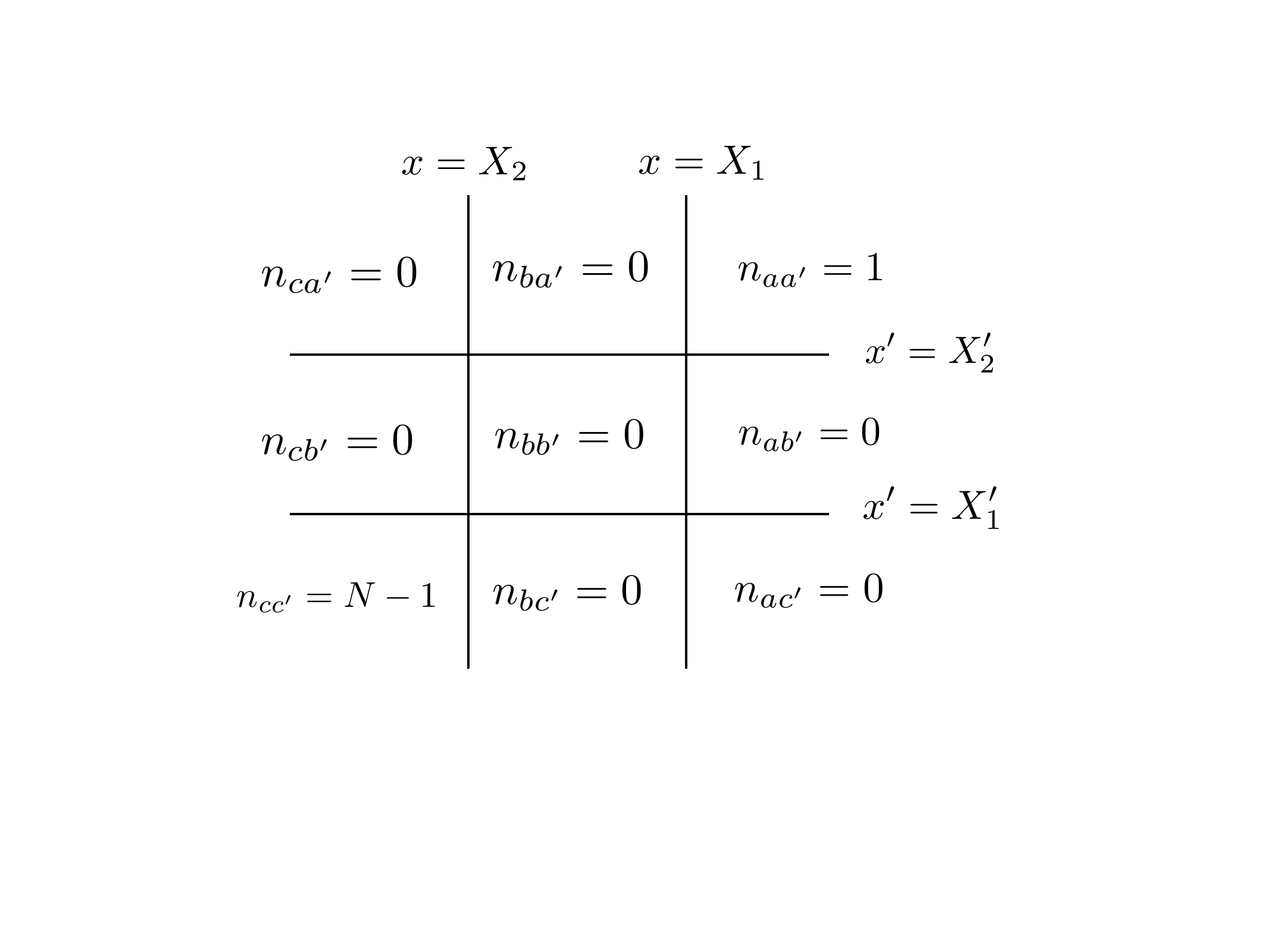}
       \includegraphics[width=0.5\columnwidth]{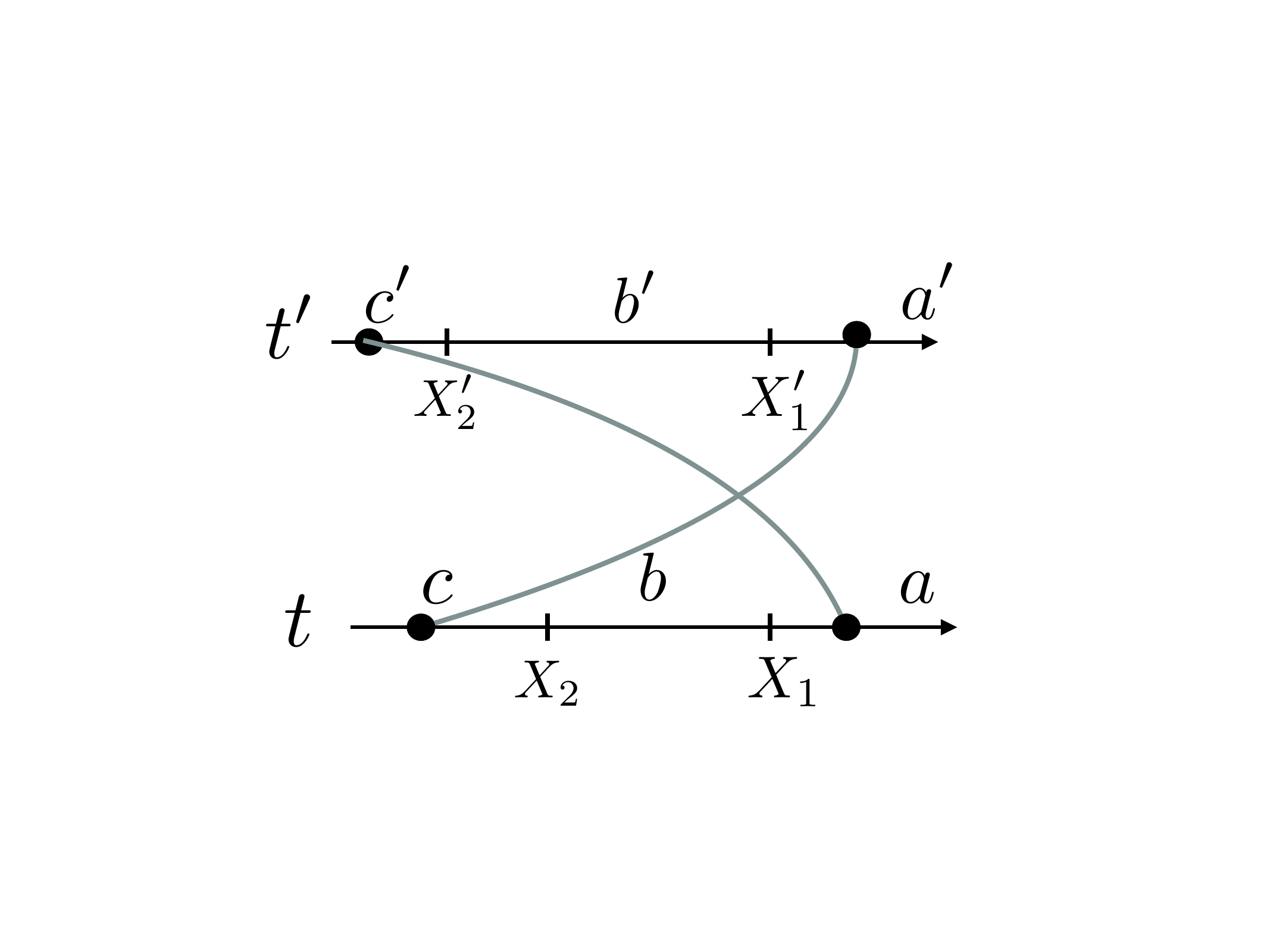}
      \includegraphics[width=0.5\columnwidth]{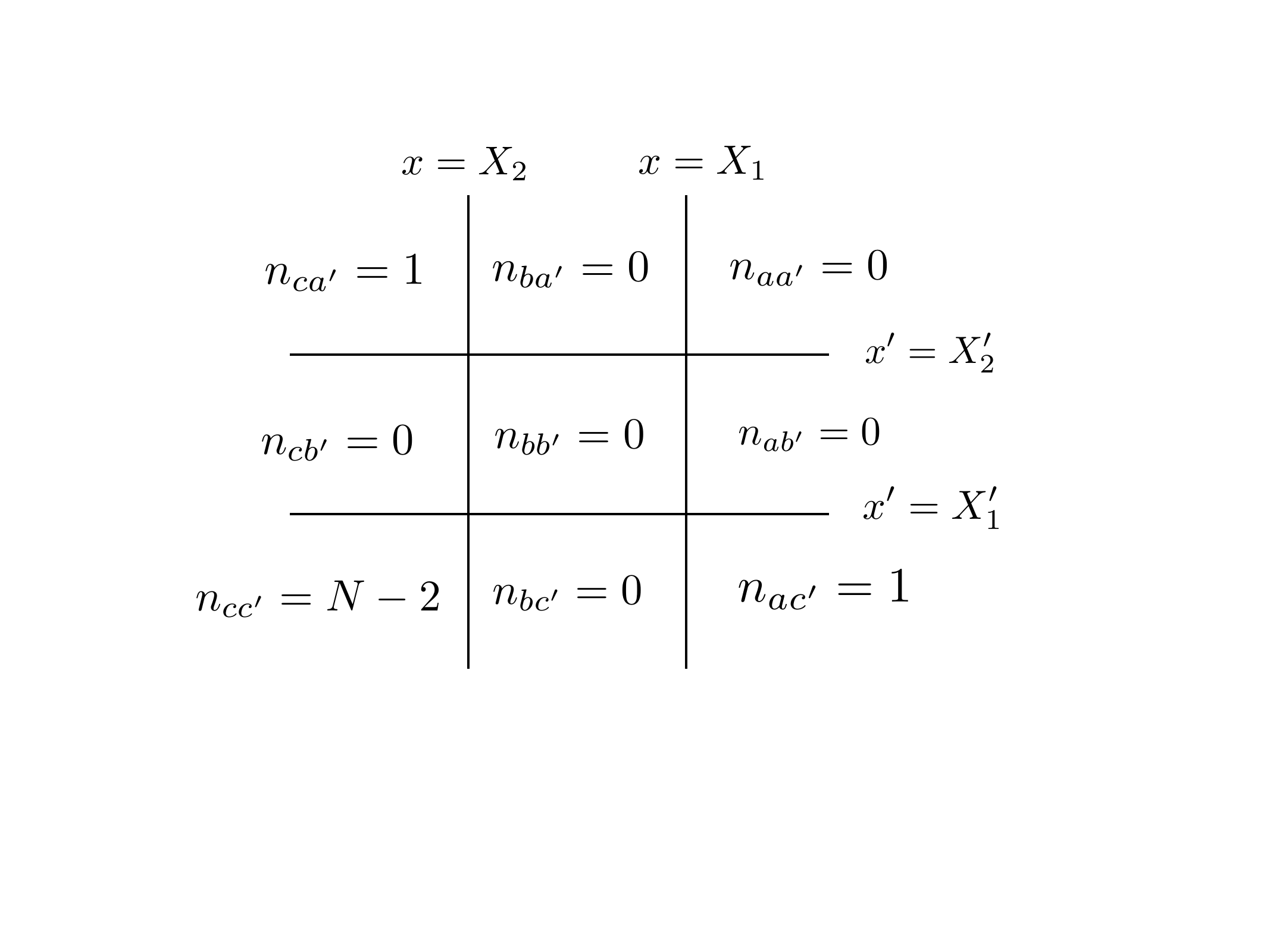}
         \vspace{-2cm}
   \caption{Illustration of the two cases in \eqref{cases} which occur in the
   calculation of the two time maximum and second maximum CDF in \eqref{cum12max}.
   Top: first case. Left: the rightmost particle at $t$ remains so at $t'$. Right: the values of the occupation
   numbers defined in the text, corresponding to this event. Since they
   are independent Poisson distributed, see Eq. \eqref{bigPoisson2},
   this case leads to the term $\lambda_{aa'}$ in \eqref{cum12max}.
   Bottom: second case. Left: the rightmost particle at $t$ is different
   from the rightmost at $t'$. Since by definition of the CDF in \eqref{cum12max}
   the middle interval is always empty (in both cases), these
   particles come from $c,c'$ as illustrated. Right: the 
   values of the occupation
   numbers corresponding to this event. This case
   leads to the term $\lambda_{ac'} \lambda_{ca'}$ in \eqref{cum12max}.}
  \label{fig:fig3}
\end{figure}

Thus from \eqref{bigPoisson2} we obtain
\bea \label{cum12max} 
&& {\rm Prob}( X^{(1)}(t) >X_1, X^{(2)}(t)<X_2, X^{(1)}(t')>X'_1, X^{(2)}(t') <X'_2 ) \\
&& \simeq ( \lambda_{aa'} + \lambda_{ac'} \lambda_{ca'} ) e^{- \Phi(z_2,z'_2;\tau)} \nn 
\\
&& =  \left( g(z_1,z'_1;\tau) + (e^{-z_1}- g(z_1,z'_2;\tau))
(e^{-z'_1}- g(z_2,z'_1;\tau))
 \right) e^{- \Phi(z_2,z'_2;\tau)}  \nn 
\eea 
Note that the first term $\lambda_{aa}$ correspond to an event where the same particle
is rightmost both at $t$ and $t'$, while the second one $\lambda_{ac'} \lambda_{ca'}$
corresponds to an event where the righmost particle has changed, and
in both cases the middle interval ($b$ and $b'$) has remained empty. 
This is illustrated in Fig. \ref{fig:fig3}.

Using the second relation in \eqref{NPaa} the r.h.s of \eqref{cum12max} can be rewritten as
 \be \label{equiv12} 
 \left(  e^{-z_1} + e^{-z'_1} - \Phi(z_1,z'_1;\tau) + 
 ( \Phi(z_1,z'_2;\tau) - e^{- z'_2})  ( \Phi(z_2,z'_1;\tau) - e^{- z_2}) ) \right) e^{- \Phi(z_2,z'_2;\tau)} 
\ee
Now taking four derivatives we obtain the joint PDF 
of the max and second max at two times, $q(z_1,z_2,z'_1,z'_2)$ in the variables $z_i,z'_i$,
%
%
\be
q(z_1,z_2,z'_1,z'_2;\tau) = \partial_{z_1} \partial_{z'_1} \partial_{z_2} \partial_{z'_2}   \text{Eq. \eqref{equiv12}} 
= \partial_{z_1} \partial_{z'_1} \partial_{z_2} \partial_{z'_2} 
\bigg( ( \Phi(z_1, z'_2;\tau) \Phi(z_2,z'_1;\tau) - \Phi(z_1,z'_1;\tau)  )  e^{- \Phi(z_2,z'_2;\tau)} \bigg) 
\ee 
as we see that the contributions of the additional exponentials vanish.
This is the result \eqref{q4} announced above. This derivation
is different, but equivalent to the one given in the Appendix
\ref{app:combinatorics}.
\\


Another observable of interest, related to the two-time maximum, is the joint PDF of 
$n_{X_1}$ and $n'_{X'_1}$, where $n_{X}$ is the number of particules with $x_i(t)>X$ and 
$n'_{X'}$ is the number of particules with $x'_i(t)>X'$. One has 
\bea \label{dec1} 
&& n_{X_1}=n_{aa'} + n_{ab'} + n_{ac'} = n_{aa'} + m \\
&& n'_{X_1}=n_{aa'} + n_{ba'} + n_{ca'} = n_{aa'} + m'
\eea 
where $m$, $m'$ are two independent Poisson variables, independent of $n_{aa'}$, and of mean 
$\lambda=\lambda_{ab'}+\lambda_{ac'}$ 
and $\lambda'=\lambda_{ba'}+\lambda_{ca'}$, respectively.
The couple $n_{X_1},n'_{X'_1}$ thus obeys a bivariate Poisson distribution. 
Note that bivariate Poisson distributions also appear in
the two-time counting statistics in the bulk, as discussed in \cite{DeanEffusion23}. 
Its distribution is 
\bea 
 {\rm Prob}(n_{X_1}=n_1,n'_{X'_1}=n'_1) &=&
\sum_{n_{aa}=0}^{\min(n_1,n'_1)} 
\frac{\lambda^{n_1-n_{aa}}}{(n_1-n_{aa})!} 
\frac{(\lambda')^{n'_1-n_{aa}}}{(n'_1-n_{aa})!} 
\frac{\lambda_{aa}^{n_{aa}}}{n_{aa}!} e^{- \lambda - \lambda' - \lambda_{aa}} \\
& = & \frac{(-1)^{n_1}}{n_1!} \frac{1}{n'_1!} \lambda_{aa}^{n_1} (\lambda')^{n_1'-n_1} 
U(-n_1, 1-n_1+n'_1, - \frac{\lambda \lambda'}{\lambda_{aa}}) e^{- \lambda - \lambda' - \lambda_{aa}} 
\eea 
where $U$ is the confluent hypergeometric function and
\be 
\lambda= e^{-z_1} - g(z_1,z'_1;\tau)  \quad , \quad \lambda'= e^{-z'_1} - g(z_1,z'_1;\tau) 
\quad , \quad \lambda_{aa} = g(z_1,z'_1;\tau) 
\ee 
with $\lambda+\lambda'+\lambda_{aa}=\Phi(z_1,z'_1;\tau)=e^{-z_1} \phi_\tau(z_1'-z_1)$. 
The characteristic function is
\be 
\langle e^{u_1 n_{X_1} + u'_1 n_{X'_1} } \rangle 
= e^{\lambda_{aa} (e^{u_1+u'_1}-1) + \lambda (e^{u_1}-1) + 
\lambda' (e^{u'_1}-1)} 
\ee 
One obtains in particular the two time covariance of the number of particles 
\be 
{\rm Cov}(n_{X_1},n'_{X'_1}) = \lambda_{aa} = g(z_1,z'_1;\tau) 
= e^{-z_1} + e^{-z'_1}  - \Phi(z_1,z'_1;\tau)  
\ee 
where we recall the asymptotics of the function $g$ (see Appendix \ref{app:integrals})
\bea 
&& g(z_1,z'_1;\tau)  \simeq e^{- \max(z_1,z'_1)} \quad ,\quad \tau \to 0 \\
&& g(z_1,z'_1;\tau)  = e^{-z_1} e^{- (z'_1-z_1+\tau)^2/(4 \tau)} \chi_\tau(z'_1-z_1) 
\quad ,\quad \tau \to +\infty
\eea 
where the large $\tau$ behavior of $\chi_\tau(z)$ is given in \eqref{largetauphi}.
\\

One can further study the joint PDF of $n_{X_1}, n_{X_2},n'_{X'_1},n'_{X'_2}$.
It is a multivariate Poisson distribution, which can 
be obtained from \eqref{bigPoisson2}, with in addition to \eqref{dec1}
\bea
&& n_{X_2}= n_{X_1} + n_{ba'} + n_{bb'} + n_{bc'} \\
&& n'_{X_2}= n'_{X'_1} + n_{ab'} + n_{bb'} + n_{cb'} 
\eea 
Its characteristic function can be easily written as above, but we will not pursue this here. 
One can simply give the two-time covariance of the number of
particles in $[X_2,X_1]$ at $t$ and in $[X'_2,X'_1]$ at $t'$, obtained as
\bea 
{\rm Cov} \left(  (n_{X_2}-n_{X_1} ) (n'_{X'_2}-n'_{X'_1} ) \right) 
& \simeq & {\rm Var} \left( n_{bb'} \right) = \lambda_{bb'} \\
&
= & g(z_2,z'_2;\tau)-g(z_1,z'_2;\tau) - g(z_2,z'_1;\tau) + g(z_1,z'_1;\tau)
\eea 
using \eqref{alllambda}. 
\\

{\bf Subcase with only two disjoint intervals, and two-time joint CDF of the second maximum}. A subcase of the above result \eqref{bigPoisson2} is obtained
by taking $X_1,X'_1 \to + \infty$. One can then denote $X_2=X$ and $X'_2=X'$.  
This amounts to divide the line into two disjoint intervals (b) $x>X$, (c) $x<X$
at time $t$, and into two disjoint intervals (b) $x>X$, (c) $x<X$ at time $t'$.
In the large $N$ limit, denoting $z,z'$ the rescaled coordinates, one
obtains the PDF
\be \label{smallPoisson2} 
 {\rm Prob}(n_{bb'},n_{bc'},n_{cb'}) 
= \frac{\lambda_{bb'}^{n_{bb'}} \, \lambda_{bc'}^{n_{bc'}} \lambda_{cb'}^{n_{cb'}}   }
{n_{bb'}! n_{bc'}! n_{cb'} ! } e^{- \Phi(z,z',\tau)}  
\ee
with
\be  \label{smalllambda} 
\lambda_{bb'} =  g(z,z';\tau) \quad , \quad 
\lambda_{bc'} = e^{-z} -g(z,z';\tau)  \quad , \quad 
\lambda_{cb'} = e^{-z'} -g(z,z';\tau) 
\ee 
e.g. from \eqref{alllambda} taking $z_1,z'_1 \to +\infty$,
since the $g$ function vanishes for any positive infinite argument
(the other $\lambda_{ii'}$ vanish and the corresponding $n_{ii'}$
are frozen to $0$). Note that the counting statistics discussed above 
for $n_{aa}$, $m$ and $m'$ in \eqref{dec1} is also recovered setting $(z,z')=(z_1,z'_1)$,
$bb'=aa'$, $\lambda_{bc'} =\lambda$ and $\lambda_{cb'} =\lambda'$.

The two-interval counting statistics \eqref{smallPoisson2},
\eqref{smalllambda} allows to obtain some two-time distributions.
First, of course, the two time joint CDF of the maximum is recovered
as ${\rm Prob}(n_{bb'}=0,n_{bc'}=0,n_{cb'}=0)$, 
setting $(z,z')=(z_1,z'_1)$. The two time joint CDF of the second
maximum can also be obtained. Recall that in Section
\eqref{subsec:counting} we noted that for a single time
the event $X^{(2)}(t) < X$ is equivalent to $n_X=0,1$. 
Here one has $n_X=n_{bb'}+n_{bc'}$ and $n'_{X'}=n_{bb'}+n_{cb'}$.
Hence the event $X^{(2)}(t) < X$ and $X^{(2)}(t') < X'$
corresponds to $n_X=0,1$ and $n'_{X'}=0,1$, 
which corresponds to the union of events
$(n_{bb'},n_{bc'},n_{cb'}) \in \{(0,0,0),(1,0,0),(0,1,0),(0,0,1),(0,1,1)\}$. 
This leads to the two time joint CDF of the second
maximum
\bea \label{CDF2timesecond} 
&& {\rm Prob}(X^{(2)}(t) < X, X^{(2)}(t') < X' ) 
= 1 + \lambda_{bb'} + \lambda_{bc'} + \lambda_{cb'} + \lambda_{bc'} \lambda_{cb'} \\
&& = \left( 1 + e^{-z} + e^{-z'} - g(z,z';\tau) + (e^{-z} -g(z,z';\tau))(e^{-z'} -g(z,z';\tau)) \right)
e^{- \Phi(z,z';\tau)} \nn \\
&& = \left( 1 + \Phi(z,z';\tau) +  (\Phi(z,z';\tau) - e^{-z'})(\Phi(z,z';\tau) - e^{-z}) \right) 
e^{- \Phi(z,z';\tau)} \nn
\eea 

Note that one can also find the two-time joint PDF of the second maximum
from \eqref{cum12max}, as 
\bea 
&& q(z_2,z'_2;\tau) = \int_{z_2}^{+\infty} dz_1 
\int_{z'_2}^{+\infty} dz'_1 
\partial_{z_1} \partial_{z'_1} {\cal Q}(z_1,z_2,z'_1,z'_2) = {\cal Q}(z_2,z_2,z'_2,z'_2) \\
&& {\cal Q}(z_1, z_2, z'_1, z'_2)= \partial_{z_2} \partial_{z'_2}   \text{Eq. \eqref{equiv12}} 
\eea
since the boundary terms at infinity do not contribute
as the function $g$ vanish when any $z$ argument goes to $+\infty$.
(beware that $z_1=z_2$ and $z'_1=z'_2$ should be taken only after
taking the two derivatives $\partial_{z_2} \partial_{z'_2}$). We have checked
that this calculation gives the same result as taking 
$\partial_{z_2} \partial_{z'_2}$ on the CDF \eqref{CDF2timesecond}.

The above calculation can be extended to obtain the two time joint CDF of the $k$-th
maximum for any $k$. One must simply enumerate all the values of the triplet
$(n_{bb'},n_{bc'},n_{cb'})$ such that 
$n_X \in \{0,1,\dots,k-1\}$ and $n'_{X'} \in \{0,1,\dots,k-1\}$.
A similar method yields the joint PDF of second maximum at $t$ and
main maximum at $t'$, or any other combination.


Finally, all the calculations in this Section can be extended to any number of times, any rank
order and any number of intervals, although it quickly becomes tedious.
This is sketched in Appendix \ref{app:outliers}
where we give explicit formula e.g. for the joint PDF of the
two-time three first maxima, and the three time first two maxima.

\section{Continuous time observables and rescaled process} 
\label{sec:continuum} 

\subsection{Probability that the maximum remains below some curve for $t \in [t_1,t_2]$} 

The multi-time joint PDF formula \eqref{mainresult1} is asking for
a "path integral" generalization. Given again the maximum process 
$X(t)=X_N(t)= \max_{i=1,\dots,N} x_i(t)$, an interesting observable in
that respect, which we study in this subsection, is
\be \label{obscont} 
{\rm Prob} (X(t) < M(t) ~,~ \forall t \in [t_1,t_2]) 
\ee 
upon proper rescaling of $X,M$ and $t_2-t_1$. 
\\

Let us make the following preliminary observations. Consider $n=2$ and the factorized
form \eqref{defphitau}. Considering $z_1$ and $z_2$ as random
variables whose CDF is $Q_{<<}$ in \eqref{Q<<}, the factorization
implies that for any real $a,z$ 
\be  \label{interpret} 
{\rm Prob}( \max(z_1, z_2 - a) < z ) = e^{- e^{-z} \phi_\tau(a)} 
\ee 
Hence the random variable $\max(z_1, z_2 - a)$ for fixed $a$ is itself a Gumbel random variable,
but shifted by $\log \phi_\tau(a)$ (a deterministic quantity). 
The equation \eqref{interpret} gives another nice interpretation to the function $\phi_\tau(a)$.
Recall from Section \ref{subsec:conditioned} that the factorized form \eqref{defphitau} 
{\it does not imply} that the PDF of $z_1$ conditioned to a given value of $z_{2,1}$ 
is Gumbel (its precise form is  a bit different, see \eqref{qconditioned}).
\\

For the original maximum process, the property \eqref{interpret} implies that 
the random variable
$\max(X(t_1), X(t_2) - M)$, when properly scaled
(and where $M$ 
and $t_2-t_1$ are properly scaled) is
also a shifted Gumbel random variable. 
\\

Clearly the property \eqref{interpret} extends to any number of times
and for any $n \geq 2$
\be \label{equalinlaw} 
\max(z_1, z_2 - a_2,\dots,z_n-a_n)  \quad \text{equal in law to} \quad 
G + \log \phi_{\tau_{2,1},\dots,\tau_{n,n-1}}(a_2,a_{3,2},\dots,a_{n,n-1})
\ee 
where $G$ is a Gumbel random variable.
\\

To take the continuum limit one must consider the rescaled process. 
For this one must fix one time $t_1$, and 
defines the rescaled maximum process (or extremal process) $z(\tau)$
(a function of the rescaled time $\tau$, which lives in the vicinity of time $t_1$) 
by the equivalence at large $N$  
\be 
X(t) \simeq \sqrt{2 t \log N} (1 + \frac{z(\tau) + c_N}{2 \log N} ) \quad , \quad t-t_1 = t_1 \frac{\tau}{\log N} 
\ee 
or, more properly as the process
\be 
z(\tau) = \lim_{N \to +\infty} \left( (2 \log N ) \left( \frac{X(t)}{\sqrt{2 t \log N} }|_{t= t_1 (1+ \frac{\tau}{\log N})}
- 1 \right) - c_N \right) 
\ee 
Such limits (and more general max stable processes) 
were considered rigorously in the statistics and probability literature,
starting with the seminal work \cite{RescaledResnik77}. 
The one-time distribution of the process $z(\tau)$ is the Gumbel distribution. 
In the previous part of the paper we have studied the $n$ time CDF's, 
$e^{- \Phi}$, of the 
process $z(\tau)$. 

Let us return to the observable \eqref{obscont} and choose to scale
\be 
M(t) = X_1 + \sqrt{ \frac{t_1}{2 \log N}} (m(\tau) + \tau) \quad , \quad 
X_1= \sqrt{2 t_1 \log N} (1 + \frac{z_1 + c_N}{2 \log N} )
\quad , \quad 
\tau= \frac{t-t_1}{t_1} \log N 
\ee 
where $m(\tau)$ is any function with $m(0)=0$. Then one has
at large $N$

\be 
{\rm Prob} (X(t) < M(t) ~,~ \forall t \in [t_1,t_2]) \simeq 
{\rm Prob}(z(\tau) < z_1 + m(\tau) ~,~ \forall \tau \in [0,\tau_{2,1}]) 
\ee 

Now we can guess the continuum limit from Eq. \eqref{mainresult2}.
Indeed $\Phi$ is the expectation of
$( 1 - \prod_{i=1}^n \theta_{y_i<z_i}  )$ over a Brownian
with diffusion coefficient $D=2$ and drift $-1$, started
at $y_1$ which is distributed with $e^{-y_1}$. 
The conjecture is thus
\be 
{\rm Prob}(z(\tau) < z_1 + m(\tau) ~,~ \forall \tau \in [0,\tau_{2,1}])
= e^{ - \Psi(z_1;\tau_{2,1};m(\tau)) } \label{inf} 
\ee 
with 
\be 
\Psi(z_1;\tau_{2,1};m(\tau)) = 
\int dy e^{-y_1} \left( 1- \theta(z_1-y_1)  
{\rm Prob}(y_1+ \sqrt{2} B(\tau) - \tau < z_1 + m(\tau) ~,~ \forall \tau \in [0,\tau_{2,1}]) \right)
\ee
where $B(\tau)$ is a standard Brownian.
So it is expressed in terms of the probability that the above 
mentioned Brownian does not hit the moving point $z_1+m(\tau)$.
One sees that, shifting $y_1 \to y_1 + z_1$, one has
\bea 
&& \Psi(z_1,\tau_{2,1};m(\tau)) = e^{-z_1} \Psi(\tau_{2,1};m(\tau)) \\
&& \Psi(\tau_{2,1};m(\tau)) =
\int dy e^{-y_1} \left( 1- \theta(-y_1)  
{\rm Prob}(y_1+ \sqrt{2} B(\tau) - \tau < m(\tau) ~,~ \forall \tau \in [0,\tau_{2,1}]) \right)
\eea
Hence we can rewrite \eqref{inf} as
\be 
{\rm Prob}(z(\tau) - m(\tau) < z_1  ~,~ \forall \tau \in [0,\tau_{2,1}])
= e^{ - e^{-z_1} \Psi(\tau_{2,1};m(\tau)) } \label{inf2} 
\ee 
which implies that the random variable
\be 
\max_{\tau \in [0,\tau_{2,1}]} (z(\tau) - m(\tau)) 
\quad \text{equal in law to} \quad 
G + \log \Psi(\tau_{2,1};m(\tau))  
\ee 
where $G$ is a Gumbel random variable.
This is the continuum limit of \eqref{equalinlaw}.
A scaled version can then be deduced for $X(t)-M(t)$. 

One can derive this formula in the case where $m(\tau)=(w-1) \tau$ is
a linear function of $\tau$. This is done in Appendix 
\eqref{app:checks}. In that case we have, with $y_1<0$
 \be
{\rm Prob}(y_1+ \sqrt{2} B(\tau) - \tau < (w-1) \tau ~,~ \forall \tau \in [0,\tau_{2,1}]) 
= {\rm Prob}( {\sf T}^{-w}_{-y_1}>\tau_{2,1}) 
\ee 
where ${\sf T}^{-w}_{z}$ is the first passage time at level $z>0$
for a Brownian starting at the origin 
with drift $-w$ and diffusion coefficient $D=2$ (see Appendix \ref{app:FPT}).
This leads to
\be 
{\rm Prob}(z(\tau) - (w-1) \tau < z_1  ~,~ \forall \tau \in [0,\tau_{2,1}])
= e^{ - e^{-z_1} \Psi_w(\tau_{2,1}) } \label{inf2} 
\ee 
with (changing variable to $y=-y_1$)
\be \label{resultT} 
\Psi_w(\tau) = \int dy e^{y} (1- \theta(y)  {\rm Prob}({\sf T}^{-w}_{y} > \tau)) 
\ee 
This function can be computed explicitly for any $w$, the result is 
given in \eqref{explicit} in Appendix \eqref{app:checks}. Here we only display the result for $w=1$, i.e.
for $m(\tau)=0$,
which reads
\be \label{Psi1} 
\Psi_1(\tau) =
\frac{1}{2} (\tau +2)
   \text{erfc}\left(\frac{\sqrt{\tau
   }}{2}\right)-\tau -\frac{e^{-\tau
   /4} \sqrt{\tau }}{\sqrt{\pi }}
\ee
which implies that $\max_{\tau \in [0,\tau_{2,1}]} z(\tau)$
is a Gumbel variable shifted by $\log \Psi_1(\tau_{2,1})$.

\subsection{Running maximum and arrival time of first particle: one-time distributions}

Let us first consider a single standard Brownian $x(t)$, with $r(t)=\max_{0 \leq t' \leq t} x(t')$
its running maximum, and the CDF for $R_1>0$
\be \label{P1def} 
{\rm Prob}(r(t_1)<R_1)
= {\rm Prob}( T_{R_1} > t_1) = {\cal P}_1
\ee 
where $T_R$ is the first passage time of the standard Brownian at level $R \geq 0$.
It is given by (see Appendix \ref{app:FPT})
\be \label{P1} 
{\cal P}_1 = {\rm Prob}( T_{R_1} > t_1  ) 
= \int_{-\infty}^{R_1} dx_1  
(\frac{e^{- \frac{x_1^2}{2 t_1}}}{\sqrt{2 \pi t_1}} - 
\frac{e^{- \frac{(x_1-2 R_1)^2}{2 t_1}}}{\sqrt{2 \pi t_1}} )
\ee 
which is the probability that the Brownian with an absorbing wall
at $R_1$  has survived up to $t_1$. 
\\

Let us now consider the {\it running maximum $R(t)$} for $N$ identical standard Brownian motions
starting from the origin
\be 
R(t) = \max_{0 \leq t' \leq t} X(t) = \max_i r_i(t)  \quad , \quad r_i(t) = \max_{0 \leq t' \leq t} x_i(t') 
\ee 
Let us first study the one-time CDF of the running maximum (which is a standard calculation
but which sets the stage for the multi-time generalization given below). It is given by
\be 
{\rm Prob}(R(t_1)<R_1) = {\rm Prob}(r(t_1)<R_1)^N 
= {\rm Prob}( T_{R_1} > t_1)^N = {\rm Prob}( T^{\rm min}_{R_1} > t_1) \label{bothways} 
\ee 
where the last term is equal to the probability that 
the minimum of the first passage times $T^{\rm min}_{R_1}  =
\min_i T^i_{R_1}$ at $R_1$ of 
$N$ identical copies is larger than $t_1$. This is also the
{\it arrival time at $R_1$ of the first particle}, an important quantity.

At large $N$, we will scale $R_1$ 
as usual as $R_1 = \sqrt{2 t_1 \log N} (1+ \frac{z_1 + c_N}{2 \log N})$ so that 
$1- {\cal P}_1=O(1/N)$ and
\be 
{\rm Prob}(R(t_1)<R_1) 
= {\rm Prob}( T_{R_1} > t_1)^N \simeq e^{- N (1- {\cal P}_1)} 
\ee 

Let us now estimate, from \eqref{P1}, using similar manipulations as in Appendix \ref{app:derivationPDF} 
\bea 
&& N(1- {\cal P}_1) =N  \int dx_1   \frac{e^{- \frac{x_1^2}{2 t_1}}}{\sqrt{2 \pi t_1}} 
\left( 1 - \theta(R_1-x_1) \left( 1- e^{- \frac{2 R_1(R_1-x_1)}{t_1} }\right) 
 \right) \\
&& \simeq
\int dy_1 e^{-y_1} 
(1- \theta(z_1-y_1) (1- e^{-  2 (z_1-y_1)} ) )\nn \\
&& = e^{-z_1} \int dy e^{y} (1 - \theta(y) (1- e^{-2 y}) ) 
= e^{-z_1} (\int_{y>0} e^{y} + \int_{y<0} e^{- y} )= 2 e^{-z_1} \label{166} 
\eea 
where we have changed variables denoting $x_1 = \sqrt{2 t_1 \log N} (1+ \frac{y_1 + c_N}{2 \log N})$,
followed by $y_1=z_1-y$. Note that the term $e^{-y_1} - e^{2 z_1-y_1}$ upon expanding the 
middle line can be interpreted as the stationary measure of the diffusion with
negative drift in presence of a hard wall at $z_1$. 
At the end, not surprisingly, at large $N$ the running maximum
has a Gumbel distribution
\be \label{running1} 
{\rm Prob}(R(t_1)<R_1) \simeq e^{ - 2 e^{-z_1}  } \simeq {\rm Prob}(X(t_1)<R_1)^2 
\ee 
with, however, a shift as compared to the instantaneous maximum, i.e. $z_1 = G + \log 2$. 

Since one can read \eqref{bothways} both ways (i.e. for the running maximum
or for the arrival time of the first particle), the above result also implies that 
\be 
{\rm Prob}( T^{\rm min}_{R_1} > t_1) \simeq e^{ - 2 e^{-z_1}  }  \label{Tmin1} 
\ee 
where the arrival time of the first particle (see Appendix \eqref{app:detailsrunning1} 
for details)
\be
T^{\rm \min}_{R_1} = t_1 = \frac{R_1^2}{2 \log N} ( 1 - \frac{z_1 + c_N}{\log N}) ) 
\ee
where from \eqref{Tmin1} $z_1 = G + \log 2$ and $G$ is Gumbel distributed. 

\subsection{Running maximum: two-time distribution}.

We can now ask about the two-time joint CDF of the running maximum, 
at two given times, $t_2>t_1$ 
\be 
{\rm Prob}(R(t_1)<R_1, R(t_2) < R_2) = {\rm Prob}(r(t_1)<R_1, r(t_2) < R_2)^N = 
{\rm Prob}( T_{R_1} > t_1 , T_{R_2} > t_2 )^N
\ee 
with $R_2>R_1$, 
which now involves the two-time joint "CDF" of the first passage times of a single Brownian
at $R_1$ and $R_2$. The latter is given by
 \be \label{calP0} 
{\cal P} = {\rm Prob}( T_{R_1} > t_1 , T_{R_2} > t_2 ) 
= \int_{-\infty}^{R_1} dx_1 
 \int_{-\infty}^{R_2} dx_2 
(\frac{e^{- \frac{x_1^2}{2 t_1}}}{\sqrt{2 \pi t_1}} - 
\frac{e^{- \frac{(x_1-2 R_1)^2}{2 t_1}}}{\sqrt{2 \pi t_1}} )
 ( \frac{e^{- \frac{(x_2-x_1)^2}{2 (t_2-t_1)}}}{\sqrt{2 \pi (t_2-t_1)}} -  
 \frac{e^{- \frac{(x_2+x_1-2 R_2)^2}{2 (t_2-t_1)}}}{\sqrt{2 \pi (t_2-t_1)}} )
\ee 
which is the probability that the Brownian with an absorbing wall
at $R_1$ for $t \in [0,t_1]$ and an absorbing wall
at $R_2$ for $t \in [t_1,t_2]$ has survived up to $t_2$. 

Note that for $N=1$, i.e. for a single Brownian $x(t)$, the two time
PDF of $R(t_1)=R_1$ and $R(t_2)=R_2$ was obtained in 
\cite{RunningMaxBenichou2016} (their Eq. (6)). 
This PDF vanishes for $R_2<R_1$ since the running maximum can only increase
with time, but there is however a $\delta(R_2-R_1)$ component in the PDF.
Its weight corresponds to the probability that $x(t)$ reaches $R(t_1)$ at
some time before $t_1$, but never crosses again the level $R(t_1)$ 
for $t \in ]t_1,t_2]$, so that $R(t_2)=R(t_1)$. As we will see
below there is a similar feature for $N>1$. 
\\

In the large $N$ limit we will scale as usual 
\be \label{sc} 
R_i = \sqrt{2 t_i \log N} (1+ \frac{z_i + c_N}{2 \log N}) \quad , \quad t_2-t_1=t_1 \frac{\tau}{\log N}
\ee 
and insert in \eqref{calP0}. 
The calculation is sketched in the Appendix \ref{app:2timerunning}. One finds
that at large $N$ the two time CDF of the running maximum takes the form, 
for $R_1 \leq R_2$, which corresponds to $z_{2,1}=z_2-z_1 \geq -\tau$,
\be
{\rm Prob}(R(t_1)<R_1, R(t_2) < R_2) \simeq e^{- \Gamma(z_1,z_2;\tau) } 
= e^{ - e^{-z_1} \gamma_\tau(z_{21}) }  \label{2running} 
\ee
where for $z \geq -\tau$
\bea 
\gamma_\tau(z) = \int dy_1 \int dy_2 \, e^{y_1} \frac{e^{- \frac{(z-y_{2,1}+\tau)^2}{4 \tau}}}{\sqrt{4 \pi \tau}}
(1 - \theta(y_1) \theta(y_2) 
(1- e^{-  2 y_1} ) (1- e^{- \frac{(z+\tau+y_1)y_2}{\tau}} )) \label{gammatau0} 
\eea 
a generalization of \eqref{166}. It turns out that this integral can be computed explicitly
(see Appendix \ref{app:2timerunning})
and one finds 
\be 
\gamma_\tau(z) = 2 \text{erf}\left(\frac{\tau +z}{2
   \sqrt{\tau }}\right)+e^{-z}
   \text{erfc}\left(\frac{z-\tau }{2
   \sqrt{\tau }}\right)+e^{2 \tau
   +z} \text{erfc}\left(\frac{3 \tau
   +z}{2 \sqrt{\tau }}\right) \label{gammaexplicit} \quad , \quad z \geq - \tau
\ee 
Since the running maximum always increases, 
for $R_2 \leq R_1$ one has
\be
{\rm Prob}(R(t_1)<R_1, R(t_2) < R_2) = {\rm Prob}(R(t_2) < R_2) \simeq e^{-2 e^{-z_2}} \label{R2R1} 
\ee 
The boundary case $R_2=R_1$ corresponds to $z_1=z_2+\tau$, i.e. $z_{2,1}=-\tau$.
Then we see that \eqref{R2R1} is consistent with the boundary value $\gamma_\tau(z=-\tau) = 2 e^\tau$
which one obtains from \eqref{gammaexplicit}. Thus one can extend \eqref{2running}
to any $z_{2,1}$ if one defines 
\be 
\gamma_\tau(z) = 2 e^{- z} \quad \text{for} \quad z \leq -\tau \label{gamma0} 
\ee

One can check that 
that $\gamma_\tau(z)$ is a decreasing
function of $z$, with $\gamma_\tau(z)  > \phi_\tau(z)$, 
which is consistent with the fact that the running maximum is always larger than the instantaneous 
maximum.

Let us study the asymptotic behaviors of $\gamma_\tau(z)$.
The large $z$ behavior is
\bea 
\gamma_\tau(z) = 2 + e^{- \frac{(z + \tau)^2}{4 \tau}} \frac{16 \tau^{5/2}}{\sqrt{\pi } z^3} 
(1-\frac{3 \tau }{z}+\frac{2 \tau  (5
   \tau
   -6)}{z^2}+O(z^{-3}) ) \quad , \quad z \to + \infty
   \eea 
The limit $\gamma_\tau(+\infty)=2$ is consistent with the one time result 
\eqref{running1}. The large $\tau$ limit is
\bea \label{gammalargetau} 
\gamma_\tau(z) = 2 (1 + e^{-z}) - e^{- \frac{\tau}{4} - \frac{z}{2}} \frac{16}{3 \sqrt{\pi \tau}} 
( 1 - \frac{80 + 3 z (4 + 3 z)}{36 \tau} 
   +O(\tau^{-2}) ) \quad , \quad \tau \to + \infty
   \eea 
The asymptotic value $2 (1 + e^{-z})$ yields $e^{-e^{-z_1} \gamma_\tau(z_{21})}
\to e^{- 2 e^{-z_1}} e^{- 2 e^{-z_2}}$, i.e. $R(t_1)$ and $R(t_2)$ become statistically
independent and one recovers the product of the one-time distributions.
\\

Consider now $z_1$ and $z_{2,1}=z_2-z_1$, i.e the scaled positions of the running
maximum from \eqref{sc}, as random variables. Their 
exponential moments can be computed as in \eqref{final}, replacing
$\phi_\tau(z)$ by $\gamma_\tau(z)$. Similarly on has
$1/\gamma_\tau(z)^{a+b} \to 2^{-(a+b)}$ as as $z \to +\infty$. Hence
\eqref{gener}, as well as 
\eqref{genersmall}, hold with the replacement $\phi_\tau(z)$ by $\gamma_\tau(z)$.
One notes from \eqref{gamma0} that the combination which appears in these
formula
\be 
1 + \frac{ \gamma_\tau'(z)}{\gamma_\tau(z)} = 0 \quad \text{for} \quad z < -\tau
\ee 
vanishes for $z<-\tau$,
and converges to $1$ exponentially fast as $z \to +\infty$. However 
this combination vanishes discontinuously at $z=-\tau$. Indeed one finds
\be 
1 + \frac{ \gamma_\tau'(z)}{\gamma_\tau(z)}|_{z=-\tau^+} = {\rm Erfc}(\sqrt{\tau}) \label{jump} 
\ee 
Hence the marginal PDF of $z=z_{2,1}=z_2-z_1$, which we denote 
by $\tilde P^{(2,1)}_\tau(z)$, now acquires a delta function part
and is given by, for $z \geq -\tau$ 
\be 
\tilde P^{(2,1)}_\tau(z)= q_\tau \delta(z+\tau)  
+ \partial_{z}^2 \log \gamma_\tau(z) \quad , \quad q_\tau = {\rm Erfc}(\sqrt{\tau}) 
 \quad , \quad z \geq -\tau \label{Prun} 
\ee 
and vanishes for $z<-\tau$. The second term is smooth: it has a finite 
value $1- {\rm Erfc}(\sqrt{\tau})^2$ at $z=-\tau$, with positive first
derivative, so $\tilde P^{(2,1)}_\tau(z)$ has a maximum for 
some $\tau$-dependent value of $z$. As explained above for $N=1$, the weight of the delta
part in in \eqref{Prun} corresponds to the probability $q_\tau$ that $R(t_2)=R(t_1)$. Specifically
it corresponds to events such that the running maximum $R(t_1)$ was achieved
by one particle at some time before $t_1$ and that all particles have 
remained below that level for $t \in [t_1,t_2]$. Note that when $\tau \to 0$
the delta part in \eqref{Prun} dominate the smooth part. 
Finally the CDF of $z_{2,1}$ is given by
\be 
 {\rm Prob}(z_{2,1} < z) = 1+ \partial_z \log \phi_\tau(z) 
\ee 
and exhibits a jump $q_\tau$ at $z=-\tau$. 

\begin{figure}[t]
     \includegraphics[width=0.5\columnwidth]{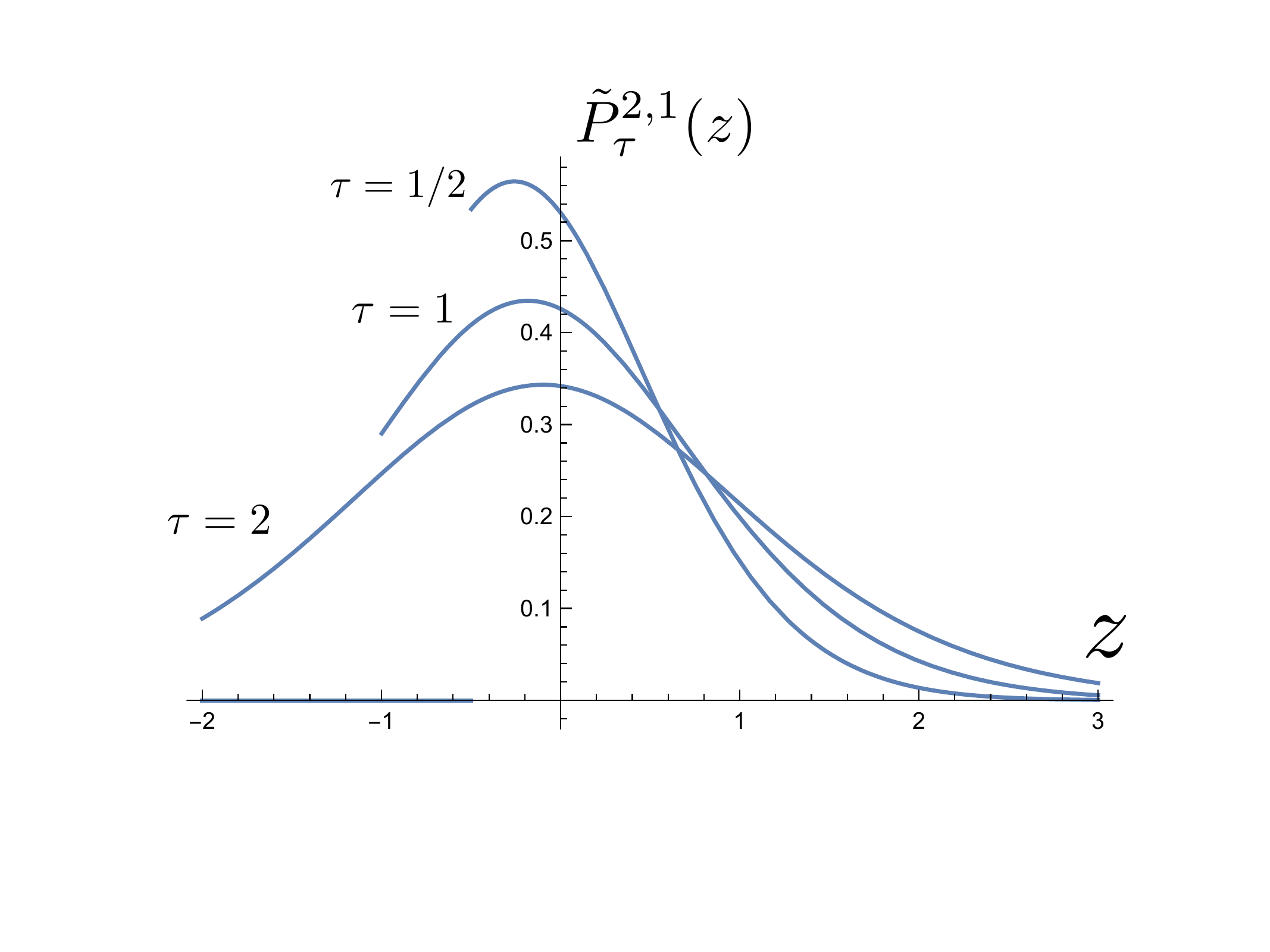}
      \includegraphics[width=0.5\columnwidth]{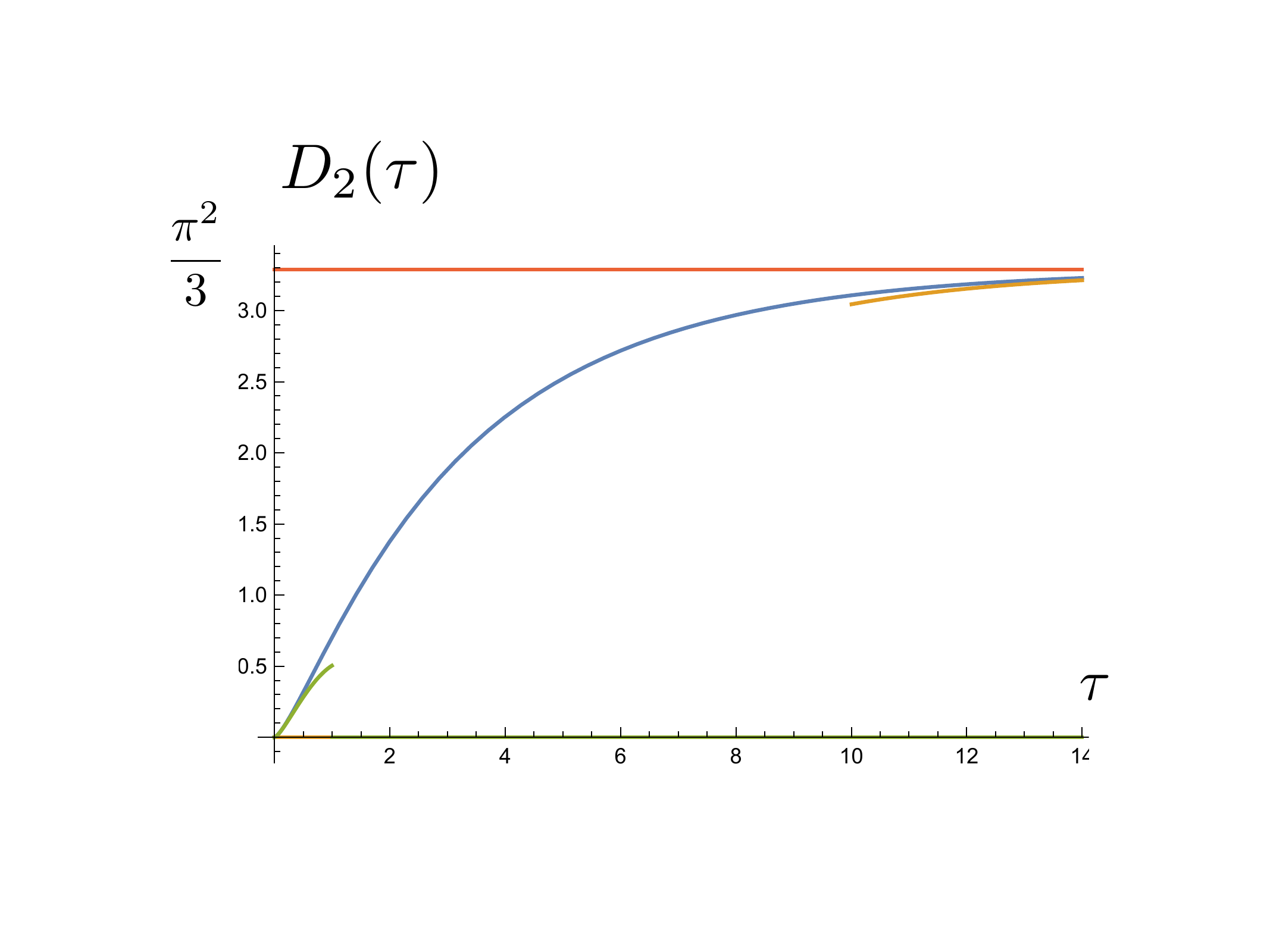}
     \vspace{-1.6cm}
   \caption{Left: marginal PDF of $z=z_{2,1}=z_2-z_1$, the scaled distance
   traveled by the running maximum, with 
   $R(t_2)-R(t_1) \simeq \sqrt{\frac{t_1}{2 \log N}} (\tau+z)$. It is
   plotted for $\tau=1/2,1,2$ and vanishes discontinuously for $z<-\tau$. 
   In addition there is a delta function at $z=-\tau$, not shown,
   of amplitude $q_\tau= {\rm Erfc}(\sqrt{\tau})\approx0.32,0.16,0.05$, corresponding
   to the event $R(t_2)=R(t_1)$,
   see \eqref{Prun}. Right: its second moment $\langle z_{2,1}^2 \rangle=D_2(\tau)$, plotted
   versus $\tau$. The first two terms in the small $\tau$ asymptotics 
   \eqref{D2small}, and the first three terms in the 
   large $\tau$ asymptotics \eqref{D2large} are also plotted, together with
   the limiting value $\pi^2/3$. Convergence to the asymptotics 
   is slower than for $A_2(\tau)$. Recall that $\pi^2/3-D_2(\tau)$, i.e. the curve reflected 
   versus $\pi^2/3$ describes the two-time covariance of the running maximum, see Eq. 
   \eqref{runningcov}. }
  \label{fig:fig4}
\end{figure}

The PDF in \eqref{Prun} is plotted in 
Fig. \ref{fig:fig4} and has the following asymptotic behaviors. 
For fixed $\tau$ and $z \to +\infty$ it decays as 
\be 
\tilde P^{(2,1)}_\tau(z) =
 e^{- \frac{(z + \tau)^2}{4 \tau}} \frac{2 \sqrt{\tau}}{z \sqrt{\pi}} \left( 
1-\frac{\tau}{z}+\frac{\tau (5
   \tau-2)}{z^2}+O(z^{-3}) \right) \quad , \quad z \to +\infty
\ee 
while for fixed $z$ and $\tau \to +\infty$ it behaves as
\bea 
\tilde P^{2,1}_\tau(z)   =   \frac{1}{4 \cosh^2(\frac{z}{2})} + \frac{e^{-\tau}}{\sqrt{\pi \tau}} \delta(z+\tau) +
\frac{e^{- \frac{\tau}{4}  } }{\sqrt{\pi \tau} } \left( 
\frac{3-\cosh (z)}{6 \cosh^3(\frac{z}{2}) } 
   + O(\frac{1}{\tau}) \right) 
\quad , \quad \tau \to +\infty  \label{runlargetauu} 
\eea 
which should be compared with the result \eqref{Plargetau} for the instantaneous maximum.
Hence at large $\tau$, $z_{2,1}$ is distributed again as the difference of two Gumbel random variables,
the even moments have exactly the same limit as in \eqref{Bernoulli}, and the odd
ones tend to zero. 

The integer moments of the random variable $z_{2,1}$ are obtained as, for $k \geq 1$
\be
\langle z_{2,1}^{k} \rangle =  D_{k}(\tau) = \int_{-\tau}^{+\infty} dz z^k P^{2,1}_\tau(z) =
q_\tau (-\tau)^{k} + 
\int_{-\tau}^{+\infty} dz z^{k} \partial_z^2 \log \gamma_\tau(z) \label{defDk} 
\ee 
The function $D_2(\tau)$ is plotted in Fig. \ref{fig:fig4} using that formula.
An alternative expression can be obtained 
upon integration by parts 
\be 
\langle z_{2,1}^{k} \rangle =  D_{k}(\tau) = 
(1-k) (-\tau)^{k}  + k(k-1)  \int_{-\tau}^{+\infty} dz z^{k-2} 
\log (\frac{\gamma_\tau(z)}{2} ) \label{Dknew} 
\ee 
where we have used \eqref{jump} 
and $\gamma_\tau(-\tau)=2 e^\tau$, $\gamma_\tau(+\infty)=2$
and that $\gamma'_\tau(z)/\gamma_\tau(z)$ decays exponentially fast at
$z \to +\infty$. For $k=1$ this implies that the first moment vanishes, 
$\langle z_{2,1} \rangle =0$, 
as it should since $\langle z_1 \rangle = \langle z_2 \rangle
= \log 2 + \gamma_E$ from the one-time result \eqref{running1}.
\\

To obtain the large time asymptotics of the moments we use 
\eqref{runlargetauu} (to a higher order, not shown). Inserting into \eqref{defDk}
the last term of \eqref{runlargetauu} leads to
a convergent integral on $z \in ]-\infty,+\infty[$.
The term $q_\tau (-\tau)^{k} \simeq (-\tau)^k e^{-\tau}/\sqrt{\pi \tau}$
is subdominant as compared to the leading decay $\sim e^{-\tau/4}$. 
One finds $\langle z_{2,1} \rangle = D_1(\tau)=0$ and 
\bea \label{D2large} 
&& \langle z_{2,1}^2 \rangle = D_2(\tau) = \frac{\pi^2}{3}  -\frac{16 \sqrt{\pi } e^{-\tau/4}}{3 \sqrt{\tau }} 
\left( 1 - \frac{\frac{20}{9}+\frac{\pi ^2}{4}}{\tau} + \frac{\frac{364}{27}+\frac{5 \pi
   ^2}{3}+\frac{5 \pi ^4}{32}}{\tau^2} + 
O(\frac{1}{\tau^3}) \right)  \\
&& \langle z_{2,1}^3 \rangle = D_3(\tau) = \frac{16 \pi^{5/2} e^{-\tau/4}}{3 \tau^{3/2}} 
(1 -\frac{5}{12 \tau} \left(16+3 \pi^2\right) + \frac{\frac{1820}{27}+\frac{125 \pi
   ^2}{9}+\frac{61 \pi ^4}{32}}{\tau^2} + O(\frac{1}{\tau^3}) 
) \\
&& 
\langle z_{2,1}^4 \rangle = D_4(\tau) 
= \frac{7 \pi^4}{15} -\frac{32 \pi ^{5/2} e^{-\tau
   /4}}{\sqrt{\tau }}  \left( 1 - \frac{5}{36 \tau} \left(16+9 \pi
   ^2\right)
   + \frac{\frac{364}{27}+\frac{25 \pi
   ^2}{3}+\frac{61 \pi ^4}{32}}{\tau^2} 
   + O(\frac{1}{\tau^3}) \right) 
\eea 
Note that the third moment of $z_2-z_1$, which is also the third cumulant
vanishes at large $\tau$, as all the odd moments, and its leading
order is one order lower than the corrections to the even moments.
\\

Let us study now the close time asymptotics.
At short time difference $\tau \ll 1$, we can scale 
\be 
z_{2,1}= -\tau + w \sqrt{\tau} \quad , \quad w \geq 0 
\ee 
where $w$ is a $O(1)$ positive random variable. This
is a bit different from the case of the instantaneous maximum.
Upon this scaling we find that the PDF $p_\tau(w)$ of the random
variable $w$ admits the following small $\tau$ expansion
\bea \label{smallexp} 
&& p_\tau(w)= \left( 1- \frac{2 \sqrt{\tau}}{\sqrt{\pi}}(1- \frac{\tau}{3} 
+ O(\tau^2))  \right)  \delta(w) 
+ \sqrt{\tau}  \, 
\text{erfc}\left(\frac{w}{2}\right) \\
&& + \tau^{3/2} 
    \left(\frac{e^{-\frac{w^2}{4}}
   w \left(5
   \text{erfc}\left(\frac{w}{2}\right)-1\right)}{\sqrt{\pi
   }}+\frac{1}{2}
   \text{erfc}\left(\frac{w}{2}\right) \left(-\left(3 w^2+2\right)
   \text{erfc}\left(\frac{w}{2}\right)+w^2+2\right)-\frac{4
   e^{-\frac{w^2}{2}}}{\pi}\right)+O(\tau^{5/2}) \nn
\eea 
which is normalized to unity order by order in $\tau$. 
Since $\langle z_{2,1} \rangle=0$, one must have 
$\langle w \rangle = \int_0^{+\infty} dw \, w \, p_\tau(w)=\sqrt{\tau}$, which is indeed satisfied
by \eqref{smallexp} to the order displayed.
Note that in the small $\tau$ limit $z_{2,1}$ has {\it an intermittent behavior}, it is
equal to $-\tau$ with probability $1- O(\sqrt{\tau})$ (which
corresponds to the event $R(t_2)=R(t_1)$)
and is of order $O(\sqrt{\tau})$ with probability $O(\sqrt{\tau})$
(which
corresponds to the event $R(t_2)>R(t_1)$).

From \eqref{smallexp} one finds the small $\tau$ expansion of the moments of
$z_{2,1}$ of lowest order, as well as the skewness $Sk$
\bea \label{D2small} 
&&  \langle z_{2,1}^2 \rangle = \langle z_{21}^2 \rangle^c 
 = D_2(\tau)= \tau \langle w^2 \rangle - \tau^2 = 
 \tau \langle w^2 \rangle^c = 
\frac{8}{3\sqrt{\pi}} \tau^{3/2} - \tau^2 +
\frac{8 \left(4
   \sqrt{2}-5\right)}{15 \sqrt{\pi
   }} \tau^{5/2} + O(\tau^{7/2}) \nn \\
&& \langle z_{2,1}^3 \rangle = \langle z_{21}^3 \rangle^c 
 = D_3(\tau)= \tau^{3/2} \langle w^3 \rangle - 3 \tau^2 \langle w^2 \rangle + 2 \tau^3 = 
 \tau^{3/2} \langle w^3 \rangle^c \\
 && ~~~~~~~~~~~~~~~~~~~~~~~~~~~~~~= 3 \tau ^2-\frac{8 \tau
   ^{5/2}}{\sqrt{\pi }}+\frac{8 \tau
   ^3}{\pi }-\frac{8 \left(4
   \sqrt{2}-5\right) \tau ^{7/2}}{5
   \sqrt{\pi }}+O(\tau^4) \nn \\
 && Sk = \frac{D_3(\tau)}{D_2(\tau)^{3/2}} = \frac{9 \sqrt{3} \pi ^{3/4}}{16
   \sqrt{2} \tau^{1/4}} 
\bigg(    1+\frac{(27 \pi -64) \sqrt{\tau
   }}{24 \sqrt{\pi
   }}+\left(-\frac{3}{2}-\frac{6
   \sqrt{2}}{5}+\frac{8}{3 \pi
   }+\frac{135 \pi }{128}\right)
   \tau + O(\tau^{3/2}) \bigg) \nn 
\eea 
The large and small $\tau$ asymptotics are shown in
Fig. \eqref{fig:fig4}. As compared to $A_2(\tau)$, one needs larger values of $\tau$
(respectively smaller) to approximate the function by the first
few terms of the series. 
\\

Finally, let us recall that in the original variables for the 
running maximum $R(t_i)=R_i$, from \eqref{sc} one has
\be
R_2- R_1  \simeq \sqrt{\frac{t_1}{2 \log N} }  ( \tau + z_{2,1} )  \quad , \quad 
\tau=\frac{t_2-t_1}{t_1} \log N = O(1)
\ee 
The one time distributions of $z_1$ and $z_2$ are shifted Gumbel,
hence their variance are the same as for the instantaneous maximum,
i.e. one has $ {\rm Var} R_1 = {\rm Var} R_2 \simeq  \frac{ t_1 }{2 \log N} \frac{\pi^2}{6}$.
This leads to the two time covariance of the running maximum
\be \label{runningcov} 
 {\rm Cov}(R(t_1),R(t_2))  \simeq \frac{ t_1 }{4 \log N}  ( \frac{\pi^2}{3} - D_2(\tau)) 
 \simeq_{\tau \to + \infty}  \frac{  t_1 }{4 \log N}  \frac{16 \sqrt{\pi } e^{-\tau/4}}{3 \sqrt{\tau }}
\ee
which can be compared to \eqref{covlarge}

\subsection{Arrival time of the first particle: two-time distribution}

Consider now, for fixed $R_2 > R_1$, the joint distribution 
of $T^{\rm min}_{R_1}$ and $T^{\rm min}_{R_1} $,
the arrival times of the first particle respectively at $x=R_1$ et $x=R_2$.
We can introduce again the rescaled variables $z_1$ and $z_2$ as
\bea
&& T^{\rm \min}_{R_1} = t_1 = \frac{R_1^2}{2 \log N} ( 1 - \frac{z_1 + c_N}{\log N}) ) \\
&& T^{\rm \min}_{R_2} = t_2 = \frac{R_2^2}{2 \log N} ( 1 - \frac{z_2 + c_N}{\log N}) ) 
\eea
Clearly if $R_2$ and $R_1$ are sufficiently separated $z_1$ and $z_2$ (seen as 
random variables) will be two independent Gumbel variables, each with the one-time CDF 
\eqref{Tmin1}. To see how close $R_2$ and $R_1$ must be so that non-trivial
correlations exist we look at the ratio
\be 
\frac{t_2-t_1}{t_1} = \frac{R_2^2-R_1^2}{R_1^2} - \frac{z_2-z_1}{\log N} \frac{R_2^2}{R_1^2} 
+ O(\frac{1}{(\log N)^2}) \label{t21} 
\ee 
Since we want this ratio to be of order $1/\log N$ we need to choose
\be 
\frac{R_2-R_1}{R_1} = \frac{\rho}{2 \log N}   \label{defrho} 
\ee 
where $\rho=O(1)$ is a fixed number. Hence we
can approximate $\frac{R_2^2-R_1^2}{R_1^2} \simeq \frac{2 (R_2-R_1)}{R_1}$
in the first term in the r.h.s. of \eqref{t21}, and $R_2/R_1 \simeq 1$ in the second term
there,
and our usual variable $\tau$ becomes 
\be 
\tau =  \frac{t_2-t_1}{t_1}  \log N \simeq \rho - z_{2,1}  \label{deftaunew} 
\ee 
but we have to remember that $\tau$ it is now fluctuating. 
Hence the variable $\tau$ is related to the variable $z_{2,1}$
of the previous section. Note in particular that since $\tau>0$,
one must have $z_{2,1}<\rho$. 
\\

We can now use the results of the previous section and obtain 
from \eqref{2running}, the joint "CDF" of the first particules arrival times 
$T^{\rm min}_{R_1}$, $T^{\rm min}_{R_2}$, 
for fixed dimensionless distance $\rho>0$ defined in \eqref{defrho}
\bea
 {\rm Prob}( T^{\rm min}_{R_1} > t_1 , T^{\rm min}_{R_2} > t_2 ) &=&
{\rm Prob}(R(t_1)<R_1, R(t_2) < R_2) \\
&\simeq& Q_{<<}(z_1,z_2) =
 e^{ - e^{-z_1} \sigma_\rho(z_{2,1}) } \quad , \quad 
\sigma_\rho(z) = \gamma_{\tau=\rho - z}(z)  \label{2Tmin} 
\eea
where $Q_{<<}(z_1,z_2)$ is the CDF in the rescaled variables.
For large $\rho$, using \eqref{gammalargetau}, 
one finds $Q_{<<}(z_1,z_2) \to e^{-2 e^{-z_1} - 2 e^{-z_2}}$
which corresponds to two independent shifted Gumbel random variables, as expected. 
\\

In fact it is more convenient to eliminate $z_2$ and use $z_1$ and
$\tau$ as the basic random variables. It means that we write
\be \label{tauz1} 
 T^{\rm \min}_{R_1} = t_1 = \frac{R_1^2}{2 \log N} ( 1 - \frac{z_1 + c_N}{\log N})  
\quad , \quad  \frac{T^{\rm \min}_{R_2} - T^{\rm \min}_{R_1}}{T^{\rm \min}_{R_1}}
\simeq 
\frac{T^{\rm \min}_{R_2} - T^{\rm \min}_{R_1}}{(R_1^2/2 \log N)}
= 
\frac{\tau}{\log N}
\ee
hence $\tau$ has the interpretation of the (rescaled) delay time between detecting
a first particle at $R_1$ and detecting a first particle at $R_2$ (which, of course, may
not be the same particle). Note that in the second equation in \eqref{tauz1},
$T^{\rm \min}_{R_1}$ in the denominator can be approximated by its
leading order value $R_1^2/2 \log N$, to the same order at large $N$.

The random variables $z_1$ and $\tau>0$ defined in \eqref{tauz1} are distributed with the joint PDF
(using the change of variable \eqref{change2} with $\partial_{z_{21}}= -\partial_\tau$) 
\be 
q(z_1,\tau) = - \partial_{\tau} (\partial_{z_1} + \partial_{\tau} ) e^{ - e^{-z_1} \Sigma_\rho(\tau)  } \quad , \quad
\Sigma_\rho(\tau) = \gamma_{\tau}(\rho-\tau) \quad , \quad \rho>0 \quad , \quad \tau>0
\ee 
where the function $\Sigma_\rho(\tau)$ has the explicit form, from \eqref{gammaexplicit}
\be \Sigma_\rho(\tau) =  
2 \text{erf}\left(\frac{\rho }{2
   \sqrt{\tau }}\right)+e^{\tau
   -\rho }
   \text{erfc}\left(\frac{\rho -2
   \tau }{2 \sqrt{\tau
   }}\right)+e^{\rho +\tau }
   \text{erfc}\left(\frac{\rho +2
   \tau }{2 \sqrt{\tau }}\right) 
  \ee 
It obeys the identity $\Sigma_\rho(\tau)  -  \Sigma'_\rho(\tau) = 2 \,  \text{erf}\left(\frac{\rho }{2
   \sqrt{\tau }}\right)$. We recall that the parameter $\rho$ is defined in \eqref{defrho}
   and is proportional to the spatial separation of the two points $R_1,R_2$ where the
   arrival times are measured. 
 \\
 
Below we will need the small and large $\tau$ asymptotics of $\Sigma_\rho(\tau)$,
which we now study.
 For $\tau \to 0$ at fixed $\rho$ one finds
 \be 
 \Sigma_\rho(\tau) =
 2  +  \frac{16 \tau^{5/2}}{\sqrt{\pi } \rho^3}
 \left( 1-\frac{12 \tau }{\rho ^2}+\frac{4
   \left(\rho ^2+45\right) \tau
   ^2}{\rho ^4}+O\left(\tau
   ^3\right) \right) \exp
   \left(-\frac{\rho ^2}{4 \tau
   }\right)
   \ee
   so that, for $\tau \to 0$ at fixed $\rho$ one has $\frac{\Sigma'_\rho(\tau)}{\Sigma_\rho(\tau)} \to 0$
   exponentially fast. To study large $\tau$ at fixed $\rho$ it is more convenient to write
 $\Sigma_\rho(\tau)$ in the equivalent form
  \be \Sigma_\rho(\tau) =  
 2 e^{\tau-\rho} + 2 \text{erf}\left(\frac{\rho }{2
   \sqrt{\tau }}\right) + e^{\tau + \rho }
   \text{erfc}\left(\frac{2
   \tau+\rho }{2 \sqrt{\tau }}\right) - e^{\tau-\rho }
   \text{erfc}\left(\frac{2
   \tau -\rho }{2 \sqrt{\tau }}\right)
 \ee
From it one obtains the large $\tau$ asymptotics
 \be \Sigma_\rho(\tau) = 2 e^{\tau-\rho} + \frac{2 \rho}{\sqrt{\pi \tau}} \left( 
 1- \frac{\rho ^2+6}{12 \tau
   }+\frac{\rho ^4+20 \rho
   ^2+120}{160 \tau
   ^2}+O(\tau^{-3})  \right) 
 \ee 
In that limit the following combination, useful below, vanishes as 
 \be 
 1 - \frac{\Sigma'_\rho(\tau)}{\Sigma_\rho(\tau)} = e^{\rho-\tau}  \,  \text{erf}\left(\frac{\rho }{2
   \sqrt{\tau }}\right) + O(e^{-2 \tau}) = e^{\rho-\tau}  \,  
 \frac{ \rho}{\sqrt{\pi \tau}} \left(  1-\frac{\rho ^2}{12 \tau
   }+\frac{\rho ^4}{160 \tau
   ^2}+O(\tau^{-3})\right) 
 \ee 
\\

\begin{figure}[t]
 \includegraphics[width=0.5\columnwidth]{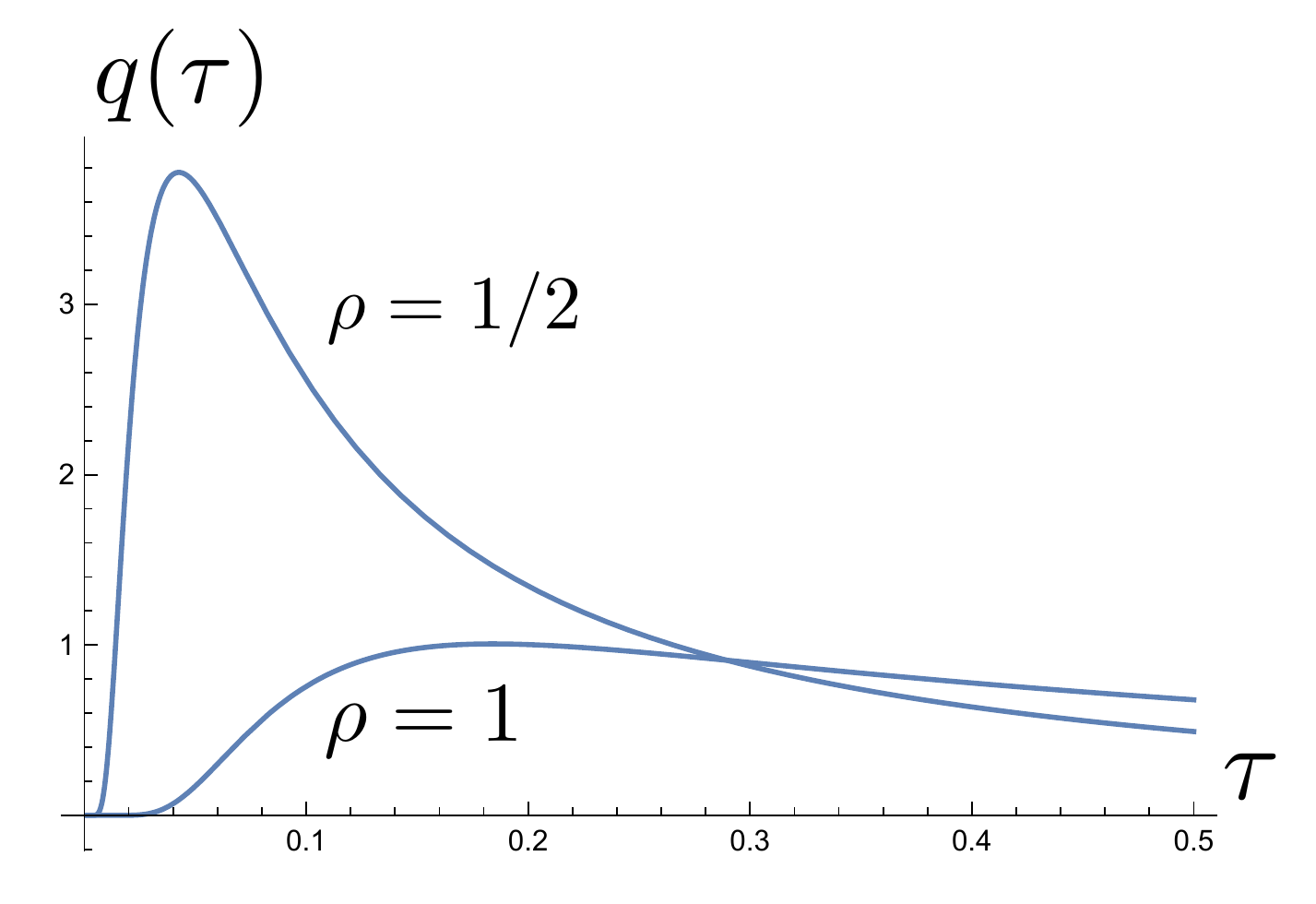}
  \includegraphics[width=0.5\columnwidth]{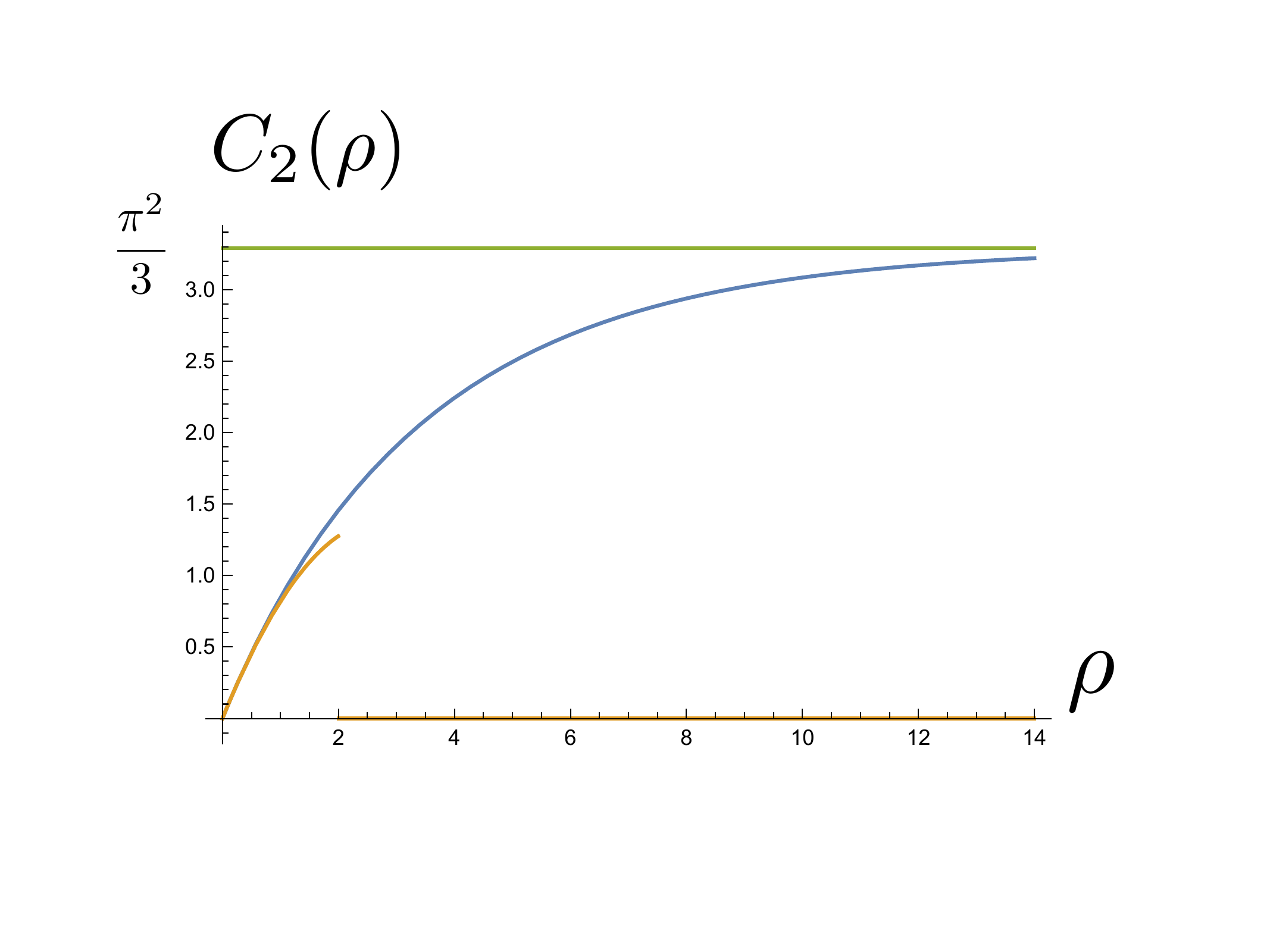}
     \vspace{0cm}
   \caption{Left: PDF $q(\tau)$ of the scaled time delay $\tau$
   between the arrival of the first particles at $R_1$ and at $R_2$,
   with $\frac{T^{\rm \min}_{R_2} - T^{\rm \min}_{R_1}}{T^{\rm \min}_{R_1}}
\simeq 
\frac{\tau}{\log N}$, for some values of the parameter $\rho$,
with $\frac{R_2-R_1}{R_1} = \frac{\rho}{2 \log N}$.
Right: its second cumulant $\langle \tau^2 \rangle_c=C_2(\rho)$, plotted
   versus $\rho$. The small $\rho$ asymptotics and the 
   the limiting value $\pi^2/3$ at large $\rho$ are also shown. }
  \label{fig:fig5}
\end{figure}

Let us now turn to the exponential moments and to the marginal PDF of 
the delay time $\tau$. One finds, by a similar calculation as in \eqref{final},
\be 
\langle e^{-s z_1 - b \tau} \rangle = - \Gamma(1+s) \int_0^{+\infty} d\tau e^{-b \tau} 
\partial_\tau \left( (1 - \frac{\Sigma'_\rho(\tau)}{\Sigma_\rho(\tau)}) \frac{1}{\Sigma_\rho(\tau)^s} \right) 
\ee 
For $b=0$, the integrand is a total derivative with 
boundary values $0$ at $\tau=+\infty$ and
$2^{-s}$ at $\tau=0$ (from the above asymptotics),
and one recovers the one time result,
$\langle e^{-s z_1 } \rangle =  \Gamma(1+s) 2^{-s}$. 
Inserting $s=0$ one finds the PDF $q(\tau)$ and the "CDF" of the (scaled) delay time, i.e. the random variable $\tau$ 
\be 
q(\tau)= - \partial_\tau  (1 - \frac{\Sigma'_\rho(\tau)}{\Sigma_\rho(\tau)}) 
= \partial_\tau^2 \log \Sigma_\rho(\tau) \quad , \quad {\rm Prob}(\tau > t) = 
1 - \partial_\tau \log \Sigma_\rho(\tau) 
\ee 
This PDF is plotted in Fig. \ref{fig:fig5}.
The PDF $q(\tau)$ vanishes exponentially fast for $\tau \to 0$,
\bea 
q(\tau) =  \frac{\rho  e^{-\frac{\rho ^2}{4
   \tau }}}{2 \sqrt{\pi } \tau ^{3/2}}
\left( 1+\frac{4 \tau ^2}{\rho ^2}-\frac{8
   \tau ^3}{\rho ^4}+\frac{16
   \left(\rho ^2+3\right) \tau
   ^4}{\rho ^6}+O\left(\tau
   ^5\right) \right) 
\eea 
It is easy to see that the leading behavior is exactly the PDF of the first passage time 
of a single (symmetric standard) Brownian (using the definitions \eqref{defrho}
and \eqref{deftaunew} of $\rho$ and $\tau$). This is because
for small $\tau$ the rightmost particle at $t_1$ is
also the rightmost particle at $t_2$. The PDF $q(\tau)$ exhibits a maximum 
for some value of $\tau$, and then decreases exponentially at
large $\tau$ as
\be \label{qsmalltau} 
q(\tau) = \frac{ \rho}{\sqrt{\pi \tau}} e^{\rho-\tau}  \left( 
1+\frac{6-\rho ^2}{12 \tau
   }+\frac{\rho ^2 \left(\rho
   ^2-20\right)}{160 \tau
   ^2}+O(\tau^{-3})\right) + O(e^{-2 \tau}) 
\ee 
Using integration by parts and the above asymptotics one finds that the 
average scaled delay time is simply
\be 
M_1(\rho) = \langle \tau \rangle = \int_0^{+\infty} d\tau \, \tau \, q(\tau) = 
- \int_0^{+\infty} d\tau \, \tau \,\partial_\tau  (1 - \frac{\Sigma'_\rho(\tau)}{\Sigma_\rho(\tau)}) = \rho
\ee 
which in the original variables means
\be 
\langle \frac{T^{\rm \min}_{R_2} - T^{\rm \min}_{R_1}}{T^{\rm \min}_{R_1}} \rangle
\simeq 2 \frac{R_2-R_1}{R_1} 
\ee 
Similarly one finds by integration by parts, the second moment and the
second cumulant
\be 
M_2(\rho) = \langle \tau^2 \rangle = \int_0^{+\infty} d\tau \, \tau^2 \, q(\tau) =
2 \int_0^{+\infty} d\tau \, \log( \frac{\Sigma_\rho(\tau)}{2 e^{\tau-\rho}}) 
\quad , \quad C_2(\rho)= \langle \tau^2 \rangle_c = M_2(\rho)-\rho^2
\ee 
As one can see in Fig. \ref{fig:fig5} the second cumulant 
reaches a finite value at large $\rho$, again equal to $C_2(+\infty)=\pi^2/3$. 
At small $\rho$ it is well approximated by $C_2(\rho) \simeq \rho - 0.181 \rho^2$. 
\\

To understand the large $\rho$ limit, one can decompose $\tau$ into its average
$\langle \tau \rangle = \rho$ and its fluctuating part $\delta$, $\tau = \rho + \delta$,
and one finds that for large $\rho$ at fixed $\delta$
\be 
\Sigma_\rho(\tau=\rho + \delta) = 2 + 2 e^{\delta} - e^{- \frac{\rho}{4} + \frac{\delta}{4} } \frac{16}{3 \sqrt{\pi \rho}} (1 + O(\rho^{-1})) 
\ee 
which leads to
\be 
q(\tau=\rho + \delta) = \frac{1}{4 \cosh^2(\delta/2)} 
+ \frac{\left(e^{\delta } \left(22-9
   e^{\delta }\right)-1\right)
   e^{\frac{\delta -\rho}{4}}}{6
   \sqrt{\pi \rho} \left(e^{\delta
   }+1\right)^3} (1 + O(\rho^{-1}))
\ee 
The leading term is again the PDF of the difference between two independent Gumbel
random variables that we encountered several times previously. The 
correction term correctly integrates to zero for $\delta \in ]-\infty,+\infty[$,
with also zero first moment. The second moment gives
\be
C_2(\rho) = \frac{\pi^2}{3} - \frac{\gamma}{6 \sqrt{\pi \rho}} e^{- \frac{\rho}{4} + \frac{\delta}{4} } (1 + O(\rho^{-1})) \quad , \quad \rho \to +\infty
\ee 
with $\gamma \approx 142.172$. We see that the second cumulant $\langle \tau^2 \rangle_c$,
which is infinite for the first passage time of a symmetric Brownian, here is finite
and dominated at large $\rho$ by the $O(1)$ vicinity of $\tau=\rho$.

\newpage

Finally, using similar methods as in this and the previous sections, one can obtain a formula
for the multi-time CDF of the running maximum and of the arrival time of the first particle,
${\rm Prob}(R(t_1)<R_1,\dots,R(t_n)<R_n)= {\rm Prob}(T^{\rm min}_{R_1} > t_1 ,\dots,T^{\rm min}_{R_n} > t_n )$ , for $n \geq 3$ times.
It is given in Appendix \ref{app:multi-running}.


\section{Conclusion}

In this paper, using simple methods of statistical physics, we have studied the dynamics of a cloud of a large number of independent identical Brownian
particles near its edge in one dimension. We have focused on the few rightmost particles, i.e.
with the largest positions (the outliers). 
To probe their dynamics we have computed the joint distribution of
the maximum position at a set of different times, 
and extended it to the maximum and second maximum,
and eventually to any finite rank, although the formula quickly become
complicated. For the maximum itself we have obtained 
distributions which appeared before in some form in probability
theory and statistics. We have found a physically appealing derivation using
the diffusion equation which naturally leads to a recursive construction
of these distributions. For the outliers, a useful tool was the counting statistics,
which, for independent particles, leads to multivariate Poisson 
distributions. In a second part we have studied other properties of
the rescaled maximum process, such as the probability that
it remains below some space-time curve. We have studied the
multi-time statistics of the running maximum of the cloud, that is the
maximum of all positions up to time $t$. Since the running maximum is intimately related
to the first passage time, we have also obtained the 
statistics of the "arrival times of the first particle", at several
locations. In particular we obtained an explicit formula
for the distribution of delay time between the first detection 
of a particle at two different neighboring locations. 

We believe that the above result will be of interest for numerics 
or experiments probing the behavior of a cloud of diffusing particles.
We have restricted here to the case of identical particles all
starting from the origin, but the study can be extended to more
general initial conditions, e.g. as considered in \cite{DeanEffusion23},
or to non-identical particles. It would be of great interest to
extend these results to diffusion in presence of a random environment,
e.g. as discussed
in the introduction. The method based on the
diffusion equation may provide a route
in that direction. 

There are possible applications to a number of
other problems. One is single-file diffusion, i.e. Brownian 
motions which do not interact except that they  
reflect on each others at each collision
\cite{SingleFileDiffusion1955,SingleFileDiffusionLargeDevSadhu2014,DeanEffusion23}.
It amounts to
considering the ordered set of positions in our problem,
$x_i(t) \to x^{(i)}(t)$, and the present results immediately apply 
to the dynamics at the edge. This would describe the dynamics
of a gas at very high temperature with hard core repulsion. 

Another example is related to the random energy model (REM). 
Consider a portfolio with $N$ stocks, each performing independent
Black-Scholes geometric Brownian motions, of total value
$Z = \sum_i e^{x_i(t)}$, where $x_i(t)$ are
the positions of the particles in our Brownian cloud
model. One can scale $t=  \tilde t \log N$ and
retain the results of the present paper since it is
just a uniform rescaling of the $t_i$'s. The
rescaled time plays the role of 
an inverse temperature $\beta=\sqrt{\tilde t}$. It is 
well known \cite{DerridaREM}
that (minus) the intensive free energy $f=\frac{1}{ \log N \sqrt{\tilde t}} \log Z(\tilde t)$ 
exhibits a REM freezing transition from a high temperature phase for $\tilde t<2$ with
$f=\frac{1}{\sqrt{\tilde t}} + \frac{\sqrt{\tilde t}}{2}$, towards
 a glass phase for $\tilde t>2$, with $f=\sqrt{2}$. In the zero temperature
limit $\tilde t \to +\infty$ one has $f=\max_i x_i(t)/(\sqrt{t \log N})$
and the free energy of the REM 
identifies with the maximum of $N$ Gaussian random variables (properly scaled),
while the extensive free energy $F=f \log N$ has $O(1)$ Gumbel distributed fluctuations.
The multi-time distribution
of the maximum discussed in this paper 
thus describes the time evolution of the
portfolio at large $\tilde t$, with correlations existing in small time windows. It would
be interesting to compute the analog of the multi-time distributions studied here, but for
finite $\tilde t$, i.e. in the finite temperature regime for the REM.

%

\bigskip

{\bf Aknowledgments}. I thank S. N. Majumdar and G. Schehr for attracting my attention 
to the problem of multi-time correlations of extremes 

\newpage
\bigskip

\appendix

\begin{Large}
{\bf Appendix} 
\end{Large} 

\vspace{-0.3cm}

\section{Derivation of the multitime joint CDF}
\label{app:derivationPDF} 

Here we provide a simple derivation of the result \eqref{mainresult1}, \eqref{mainresult2} in the text.
Since the walkers are independent one has 
\be
{\rm Prob}(X(t_1)<X_1,\dots,X(t_n)<X_n) = 
{\rm Prob}(x(t_1)<X_1,\dots,x(t_n)<X_n)^N  
= e^{ N \log {\rm Prob}(x(t_1)<X_1,\dots,x(t_n)<X_n) }
\ee
Now one has, from normalization
\bea 
&& {\rm Prob}(x(t_1)<X_1,\dots,x(t_n)<X_n) 
= \int_{x_1<X_1} \dots  \int_{x_n<X_n} p_{t_1}(x_1) p_{t_2-t_1}(x_2-x_1) \dots 
p_{t_n-t_{n-1}}(x_n-x_{n-1}) \nn \\
&& = 1 -  I(X_1,\dots,X_n;t_1,\dots t_n) \\
&& \!\!\! \!\!\! I(X_1,\dots,X_n;t_1,\dots t_n) = 
 \int dx_1 \dots dx_n (1 - \prod_{i=1}^n \theta(x_i < X_i) ) 
p_{t_1}(x_1) p_{t_2-t_1}(x_2-x_1) \dots 
p_{t_n-t_{n-1}}(x_n-x_{n-1}) \nn
\eea 
Hence one has
\be 
{\rm Prob}(X(t_1)<X_1,\dots,X(t_n)<X_n) =  
e^{ N \log (1 - I(X_1,\dots,X_n;t_1,\dots t_n) ) }
\ee 
Until now this is exact for any $N$. Let us now consider $N \gg 1$.

For large $N$ the probability remains
of order unity when $ I(X_1,\dots,X_n;t_1,\dots t_n) ) = O(1/N)$. 
The change of variable from $X_j$ to $z_j$ and from $t_j$
to $\tau_j$ will produce exactly the correct factor. 
Let us recall that 
\be 
t_j = t_1 (1+ \frac{\tau_{j,1}}{\log N}) 
\ee 
with the notation $\tau_{i,j}=\tau_i-\tau_j$ and $\tau_1=0$.
In the expression for $ I(X_1,\dots,X_n;t_1,\dots t_n)$
one also performs the change of variables
\bea \label{Xj2} 
&& X_j = \sqrt{2 t_j} \sqrt{\log N} (1 + \frac{z_j +c_N}{2 \log N} ) \simeq
\sqrt{2 t_1} \sqrt{\log N} (1 + \frac{z_j + \tau_{j,1} +c_N}{2 \log N} )  \\
&& x_j = \sqrt{2 t_j} \sqrt{\log N} (1 + \frac{y_j +c_N}{2 \log N} ) \simeq
\sqrt{2 t_1} \sqrt{\log N} (1 + \frac{y_j + \tau_{j,1} +c_N}{2 \log N} )
\eea 
Hence we have
\be 
dx_j = \frac{ \sqrt{2 t_j} }{2 \sqrt{\log N}} dy_j \simeq \frac{ \sqrt{2 t_1} }{2 \sqrt{\log N}} dy_j 
\ee 
Consider now, for $j=1,\dots,n-1$
\bea 
p_{t_{j+1}-t_{j}}(x_{j+1}-x_j) &=& \frac{1}{\sqrt{2 \pi (t_{j+1}-t_j)}} e^{- \frac{(x_{j+1}-x_j)^2}{2 (t_{j+1}-t_j)}} 
 \simeq \frac{1}{\sqrt{2 \pi t_1 \frac{\tau_{j+1,j}}{\log N}}} 
 e^{ - \frac{ ( \frac{ \sqrt{2 t_1} }{2 \sqrt{\log N}} (y_{j+1}+\tau_{j+1,1} -( y_j + \tau_{j,1})))^2}{
 \frac{\tau_{j+1,j}}{\log N} } } \nn \\
& = &\frac{\sqrt{\log N}}{\sqrt{2 \pi t_1 \tau_{j+1,j}} }
e^{ - \frac{(y_{j+1}-y_j + \tau_{j+1,j})^2}{4 \tau_{j+1,j} } } = \frac{\sqrt{2 \log N}}{\sqrt{t_1}} G(y_{j+1,j} , \tau_{j+1,j}) 
\eea 
Hence
\be 
dx_{j+1} p_{t_{j+1}-t_{j}}(x_{j+1}-x_j)  \simeq dy_{j+1} G(y_{j+1,j} , \tau_{j+1,j})  
\ee 
Finally from \eqref{differential}
\be 
N p_{t_1}(x_1) dx_1 = e^{-y_1} dy_1
\ee 
putting all the factors together we obtain
\be 
I(X_1,\dots,X_n;t_1,\dots t_n)  \simeq \frac{1}{N}  \int_{y_1,\dots,y_n} ( 1 - \prod_{i=1}^n \theta_{y_i<z_i}  ) 
 e^{- y_1 }
G(y_{2,1},\tau_{2,1}) \dots G(y_{n,n-1},\tau_{n,n-1}) 
\ee 
leading to the result \eqref{mainresult2} 
in the text.

\section{Calculations of some integrals}
\label{app:integrals} 

In this Appendix we compute the functions $g(z_1,z_2;\tau)$, $\Phi(z_1,z_2;\tau)$ 
and $\phi_\tau(z)$ defined in the text, show a symmetry property, and
discuss their limit as $\tau \to 0$.

From its definition in \eqref{g_2def} one has
\be 
g(z_1,z_2;\tau_{2,1})  = \int_{z_1<y_1,z_2<y_2} 
 e^{- y_1 } G(y_{2,1},\tau) = e^{-z_1} I(z_{2,1},\tau) \label{gapp} 
\ee 
in terms of the integral
\be
I(z,\tau) = \int_{y_1>0, y_2>0} e^{- y_1 } G(y_{2,1}+z,\tau) 
\ee 
where we recall that $z_{2,1}=z_2-z_1$ and $y_{2,1}=y_2-y_1$
and the last equality in \eqref{gapp} is obtained by the shift of
integration variables $y_i \to y_i + z_i$. One first recall that for any $b>0$
\be 
\int \frac{dk}{2 \pi}  \frac{e^{- k^2 \tau - i k x}}{i k + b  } = 
 \int \frac{dk}{2 \pi} 
 e^{- k^2 \tau - i k (y+x) - b y }  = 
\int_{y>0}  \frac{1}{\sqrt{4 \pi \tau}}  e^{- \frac{(x+y)^2}{4 \tau} - b y}   =
\frac{1}{2} e^{b^2 \tau + b x}  {\rm erfc} ( \frac{x+ 2 b \tau}{2 \sqrt{\tau}}) 
\ee 
Hence one has
 \bea
 I(z,\tau) 
&=& \int \frac{dk}{2 \pi} 
 e^{- k^2 \tau - i k (z+ \tau) } \int_{y_1>0, y_2>0} 
 e^{- y_1 - i k y_{2,1}} \\
 &=& \int \frac{dk}{2 \pi} 
 e^{- k^2 \tau - i k (z+ \tau) } \frac{1}{1 - i k} \frac{1}{i k + 0^+} = g_\tau(z+\tau) \label{Iztau} 
\eea
where we have defined
\bea 
 g_\tau(a) &:=& \int \frac{dk}{2 \pi} 
 e^{- k^2 \tau - i k a } \frac{1}{1 - i k} \frac{1}{i k + 0^+}  
 = \int \frac{dk}{2 \pi} 
 e^{- k^2 \tau - i k a } ( \frac{1}{1 - i k} + \frac{1}{i k + 0^+} ) \\
 & = & \frac{1}{2} e^{\tau-a}  {\rm erfc} ( \frac{2  \tau-a}{2 \sqrt{\tau}}) 
 + \frac{1}{2} {\rm erfc} ( \frac{a}{2 \sqrt{\tau}})  \label{gtaua} 
\eea 
We finally obtain
\be \label{gexplicit} 
 g(z_1,z_2;\tau) = e^{-z_1} g_\tau(z_{2,1}+\tau) 
 = \frac{1}{2} \left( e^{-z_1} {\rm erfc}(\frac{z_2-z_1+\tau}{2 \sqrt{\tau}}) 
+ e^{-z_2} {\rm erfc}(\frac{z_1-z_2+\tau}{2 \sqrt{\tau}})  \right) 
\ee 
Let us also recall the definitions of
the functions $\Phi$ and $\phi_\tau$, from \eqref{Phi2} and \eqref{defphitau} 
\be 
\Phi(z_1,z_2;\tau)  = e^{-z_1} + e^{-z_2}  - g(z_1,z_2;\tau) = e^{-z_1} \phi_\tau(z_{2,1}) 
\ee
This leads to 
\be \label{Phiexplicit} 
\Phi(z_1,z_2;\tau)  = \frac{1}{2} \left( e^{-z_1} {\rm erfc}(\frac{z_1-z_2-\tau}{2 \sqrt{\tau}}) 
+ e^{-z_2} {\rm erfc}(\frac{z_2-z_1-\tau}{2 \sqrt{\tau}})  \right) 
\ee 
where we used that ${\rm erfc}(x)+{\rm erfc}(-x)=2$, since one has
${\rm erfc}(x)=1-{\rm erf}(x)$ and ${\rm erf}(-x)=-{\rm erf}(x)$.
From this one obtains the explicit form \eqref{phitauexplicit} 
for $\phi_\tau(z)$ given in the text. It can also be 
obtained by noting that
\be 
\phi_\tau(z) = 1+ e^{-z} - g_\tau(z+\tau) 
\ee 
and using \eqref{gtaua}.
\\

{\it Symmetry property}. The function $g_\tau(z)$ obeys an interesting identity. Consider the form \eqref{Iztau}. From it, it is immediate to see that 
$\partial_\tau g_\tau(z+\tau) = \frac{-1}{\sqrt{4 \pi \tau}} e^{- \frac{(z+\tau)^2}{4 \tau}}$. Taking
into account the boundary conditions, we obtain 
\be 
g_\tau(z+\tau)  = \int_{\tau}^\infty \frac{dt}{\sqrt{4 \pi t}} e^{- \frac{(t+z)^2}{4 t} } \label{indentityapp} 
\ee
On this expression, using that $\frac{(t+z)^2}{4 t}  - \frac{(t-z)^2}{4 t} = z$ we 
obtain
\be 
g_\tau(-z+\tau) = \int_{\tau}^\infty \frac{dt}{\sqrt{4 \pi t}} e^{- \frac{(t-z)^2}{4 t} } 
= e^z \int_{\tau}^\infty \frac{dt}{\sqrt{4 \pi t}} e^{- \frac{(t+z)^2}{4 t}}  
= e^z g_\tau(z+\tau)
\ee 
From this we have
\be 
\phi_\tau(-z)=1 + e^z - g_\tau(-z+\tau) = 1 + e^z - e^z g_\tau(z+\tau) = e^z (1 + e^{-z} - g_\tau(z+\tau) ) = e^z 
\phi_\tau(z)
\ee 
which is the symmetry property \eqref{symm0} discussed in the text. Note that
this symmetry is equivalent to the fact that $\Phi(z_1,z_2;\tau)$ is symmetric
in $z_1,z_2$ as can be seen on its explicit form \eqref{Phiexplicit}.
\\

{\it Limit $\tau \to 0$}.
For small time difference $\tau \to 0$ one has
\be  
\lim_{\tau \to 0} g_\tau(z+\tau)  = \int_{0}^\infty \frac{dt}{\sqrt{4 \pi t}} e^{- \frac{(t+z)^2}{4 t} }
= e^{-z} \theta(z) + \theta(-z) 
\ee
Hence 
\be 
\lim_{\tau \to 0} \phi_\tau(z)  = \theta(z) + e^{-z} \theta(-z) \quad , \quad 
\lim_{\tau \to 0} \Phi(z_1,z_2,\tau) = e^{- \min(z_1,z_2)}
\ee
as it should since for $\tau \to 0$, $X(t_2) \to X(t_1)$.

\section{Some details of calculations for $n=2$} 
\label{app:n2}

Let us compute the exponential moments associated to the joint CDF 
\eqref{Q<<} in the text. For the calculations we consider $z_1$ and $z_{21}=z_2-z_1$ as independent real variables. 
The derivatives should be replaced as follows
\be  \label{change2} 
\partial_{z_1} \to  \partial_{z_1}- \partial_{z_{21}} \quad , \quad \partial_{z_2} \to \partial_{z_{21}} 
\ee 
Let us consider the expectation value of exponentials (note that we abusively denote by the same letter the random variable and a real integration variable)
\bea  
&& \langle e^{- a z_1 - b z_2} \rangle = 
\int dz_1 dz_2 e^{- a z_1 - b z_2}  \partial_{z_1} \partial_{z_2} Q_{<<}(z_1,z_2)  \\
&& = \int dz_{21} e^{-b z_{21}} \partial_{z_{21}} 
\int dz_1 e^{- (a+b) z_1}  (\partial_{z_1} - \partial_{z_{21}} ) \, e^{ - e^{-z_1} \phi_\tau(z_{21}) } \\
&& = \int dz_{21} e^{-b z_{21}} \partial_{z_{21}}  (\phi_\tau(z_{21})+ \phi_\tau'(z_{21}))
\int dz_1 e^{- (a+b+1) z_1}  \, e^{ - e^{-z_1} \phi_\tau(z_{21}) } \\
&& = \Gamma(1+a+b) \int dz  e^{-b z} \partial_{z} 
\left(  (1 + \frac{ \phi_\tau'(z)}{\phi_\tau(z)}) 
\frac{1}{\phi_\tau(z)^{a+b}} \right) \label{final} 
\eea  
where we have set $z_{21}=z$ and performed the integration over $z_1$ 
using that $\int dz_1 e^{- A z_1} e^{- p e^{-z_1}} = p^{-A} \Gamma[A]$. 
This is equivalent to \eqref{gener} in the text, with $a+b=s$. 

Let us examine the asymptotic behavior of the terms which appear in the
integral \eqref{final}. One has, using \eqref{asymptoticsphitau} 
\bea   
&&  \!\!\!\!\!\!\!\!\! 1 + \frac{ \phi_\tau'(z)}{\phi_\tau(z)} = 1 + \frac{- e^{-z} + \psi'_\tau(z) }{ e^{-z} +\psi_\tau(z)} =
\frac{\psi_\tau(z)+ \psi'_\tau(z) }{ e^{-z} +\psi_\tau(z)}  \simeq e^z \psi'_\tau(z) 
= - e^{-z} e^{- \frac{ (z+ \tau)^2}{4 \tau}} (\frac{\tau^{1/2}}{z \sqrt{\pi}} + O(z^{-2})) 
~ , ~ z \to - \infty \nn \\
&&  \!\!\!\!\!\!\!\!\!  1 + \frac{ \phi_\tau'(z)}{\phi_\tau(z)} \simeq 1 +  \psi'_\tau(z)   = 
 1 - e^{- \frac{ (z+ \tau)^2}{4 \tau}} (\frac{\tau^{1/2}}{z \sqrt{\pi}} + O(z^{-2})) \quad , \quad z \to + \infty 
\eea   
and 
\bea 
&& \frac{1}{\phi_\tau(z)^{a+b}}  \simeq e^{(a+b) z}  \quad , \quad z \to - \infty  
\quad , \quad  \frac{1}{\phi_\tau(z)^{a+b}}  \simeq 1 \quad , \quad z \to + \infty  
\eea 
Using these asymptotics, one checks that setting $b=0$ in \eqref{final} the integrand is a total derivative and
one obtains 
\be
 \langle e^{- a z_1} \rangle = 
\Gamma(1+a) \int dz \partial_{z} 
\left(  (1 + \frac{ \phi_\tau'(z)}{\phi_\tau(z) }) 
\frac{1}{\phi_\tau(z)^{a}} \right) = \Gamma(1+a) 
\ee
as required since the PDF of $z_1$ is the Gumbel distribution. 
Similarly, setting $a=0$ one obtains 
\be
 \langle e^{- b z_2} \rangle = 
\Gamma(1+b) I_b \quad , \quad I_b = \int dz e^{-b z} \partial_{z} 
\left(  (1 + \frac{ \phi_\tau'(z)}{\phi_\tau(z) }) 
\frac{1}{\phi_\tau(z)^{b}} \right)
\ee
In fact one can show that $I_b=1$, which leads to $\langle e^{- b z_2} \rangle = 
\Gamma(1+b)$ as required since the PDF of $z_2$ is also the Gumbel distribution.
This is indeed a consequence of the symmetry \eqref{symm0} 
which also implies
\be 
\frac{e^{-b z}}{ \phi_\tau(z)^b } (1+ \frac{ \phi_\tau'}{\phi_\tau}(z) ) = - \frac{ \phi_\tau'}{\phi_\tau}(-z) 
\frac{1}{ \phi_\tau(-z)^b } 
\ee 
Hence for $b>0$ integrating by part, using the symmetry and changing $z \to - z$
\be 
I_b = b \int dz e^{-b z} 
 (1 + \frac{ \phi_\tau'(z)}{\phi_\tau(z) }) 
\frac{1}{\phi_\tau(z)^{b}}  =-  b \int dz \frac{ \phi_\tau'}{\phi_\tau}(z) 
\frac{1}{ \phi_\tau(z)^b } = [ \frac{1}{ \phi_\tau(z)^b }  ]_{-\infty}^{+\infty} = 1 
\ee 

One can further simplify the expression of the exponential moments. Indeed
for $b>0$, using the above asymptotics one see that one can integrate by part and get
\be 
 \langle e^{- a z_1 - b z_2} \rangle =  b \, \Gamma(1+a+b) \int dz e^{-b z} 
  (1 + \frac{ \phi_\tau'(z)}{\phi_\tau(z)}) 
\frac{1}{\phi_\tau(z)^{a+b}} 
\ee 
For $a,b>0$ one can perform another integration by part and obtain
\be 
 \langle e^{- a z_1 - b z_2} \rangle =  \frac{a b}{a+b} \, \Gamma(1+a+b) \int dz \frac{e^{-b z}}{\phi_\tau(z)^{a+b}} \label{simplerexp} 
\ee 
Although this closed expression is easy to integrate numerically, 
it is tricky to obtain the moments from it, and we refer to the
discussion in the text. 

Finally, for 3 times one performs the change
$\partial_{z_1} \to  \partial_{z_1}- \partial_{z_{21}}$,
$\partial_{z_2} \to  \partial_{z_{21}}- \partial_{z_{32}}$,
$\partial_{z_3} \to \partial_{z_{32}}$ which leads to 
\eqref{gener3time} and \eqref{PDF3}.

\section{Multi-time order statistics: combinatorics} 
\label{app:combinatorics} 

To display more conveniently the combinatorics associated to the multi-time order
statistics, it is useful to consider the case where each
particle $i=1,\dots,N$ is described by a distinct one time PDF, which in this Appendix we denote as 
$p_i(x)$, and a CDF $P_i(x<X)$, and a two time CDF denoted
$P_i(x<X, x'<X')$. For two-time we denote with prime the quantities at time $t'>t$. 

As a warmup let us recall the PDF and CDF of the maximum at one time 
\be
q(X_1)= \sum_i p_i(X_1) \prod_{j \neq i} P_j(x<X_1) = \partial_{X_1} Q(X_1) \quad , \quad Q(X_1) = \prod_\ell P_\ell(x<X_1) 
\ee
and the joint PDF of the maximum and the second maximum at one time
\be
q(X_1,X_2)= \theta_{X_2<X_1}  \sum_{i \neq j}  p_i(X_1) p_j(X_2) 
\prod_{k \neq i,j} P_k(x<X_2)  
\ee
It can also be retrieved from a joint "CDF", since for $X_2<X_1$ one has 
\be 
{\rm Prob}(X^{(1)}(t) >X_1, X^{(2)}(t) <X_2) = \sum_i P_i(x > X_1) \prod_{k \neq i} P_k(x<X_2) 
\ee 
and taking $- \partial_{X_1} \partial_{X_2}$ it recovers the above expression for $q(X_1,X_2)$. 
It turns out that a generalization of this "CDF" is convenient to obtain the two-time distribution for
the maximum and second maximum.

As a first exercise, let us now write the PDF of the maximum at two times,
and in a second stage check consistency with the result given in the text for the CDF. 
There are
two possibilities, either particle $i$ is the rightmost for both times,
of it is the rightmost only at the first time, but at the second time
particle $j$ has become the rightmost. This leads to the expression of the joint PDF
\bea \label{2max} 
 q(X_1,X'_1) &=& \sum_i p_i(X_1,X'_1) \prod_{k \neq i} P_k(x<X_1,x'<X'_1) 
\\
& +&  \sum_{i \neq j} P_i(X_1,x'<X'_1) P_j(x<X_1,X'_1)  \prod_{k \neq i,j} P_k(x<X_1,x'<X'_1) \nn \\
&=& \partial_{X_1}  \partial_{X'_1} Q(X_1,X'_1) \quad , \quad 
Q(X_1,X'_1)  = \prod_\ell P_\ell(x<X_1,x'<X'_1)
\eea 
Let us now consider the case of identical particles ($p_i=p$ independent of $i$ and so on) 
and estimate \eqref{2max} at large $N$, introducing the associated 
variables $z_1$, $z'_1$ and $\tau_{2,1}=\tau$ as before.
The two terms in \eqref{2max} have factors $N$ and $N(N-1) \simeq N^2$ respectively.
One uses the limits
\bea \label{equiv2} 
&& N P(x>X_1,x'>X_1') \simeq g(z_1,z_1';\tau) = e^{-z_1} + e^{-z_1'} - \Phi(z_1,z_1';\tau)  \\
&& P(x < X_1,x' < X_1')^N \simeq e^{- \Phi(z_1,z_1';\tau) }
\eea
where $g$ and $\Phi$ are defined in \eqref{g_2def} and given explicitly in
\eqref{gexplicit}, \eqref{Phiexplicit}. By two differentiation of the first line, this leads to 
\be
N p(X_1,X'_1) dX_1 dX'_1 \simeq -  \partial_{z_1} \partial_{z'_1} \Phi(z_1,z'_1;\tau) dz_1 dz'_1  
\ee 
To evaluate the mixed PDF/CDF $p(X_1,x'<X'_1)$ we first differentiate the
first line of \eqref{equiv2} w.r.t. $X_1$ which gives 
\be
N p(X_1,x'>X'_1) dX_1 \simeq - \partial_{z_1} g(z_1,z'_1;\tau) dz_1 
\ee
which can be rewritten as
\be
N (p(X_1) - p(X_1,x'<X'_1) ) dX_1 = - \partial_{z_1} g(z_1,z'_1;\tau) dz_1 
\ee
Hence using that $N p(X_1) dX_1 \simeq e^{-z_1} dz_1$ one obtains
\be 
N p(X_1,x'<X'_1) dX_1 = e^{z_1} + \partial_{z_1} g(z_1,z'_1;\tau) = - \partial_{z_1} \Phi(z_1,z'_1;\tau)  dz_1
\ee 
Putting all together we obtain from \eqref{equiv2} the joint PDF of the maximum
at two times
\bea 
&& q(X_1,X'_1) dX_1 dX'_1 \simeq 
\left( -  \partial_{z_1} \partial_{z'_1} \Phi(z_1,z'_1;\tau)  + \partial_{z_1} \Phi(z_1,z'_1,;\tau) 
\partial_{z'_1} \Phi(z_1,z'_1;\tau)  \right) e^{- \Phi(z_1,z_1';\tau) } dz_1 dz'_1 \\
&& = \partial_{z_1} \partial_{z'_1}  e^{- \Phi(z_1,z_1';\tau) } dz_1 dz'_1 
\eea 
in agreement with the result in the text for the joint CDF. This is an equivalent
derivation to the one given in Appendix \eqref{app:derivationPDF} in the case of $n=2$.
\\

\begin{figure}[t]
  \centerline{\includegraphics[width=0.6\columnwidth,height=8cm,keepaspectratio]{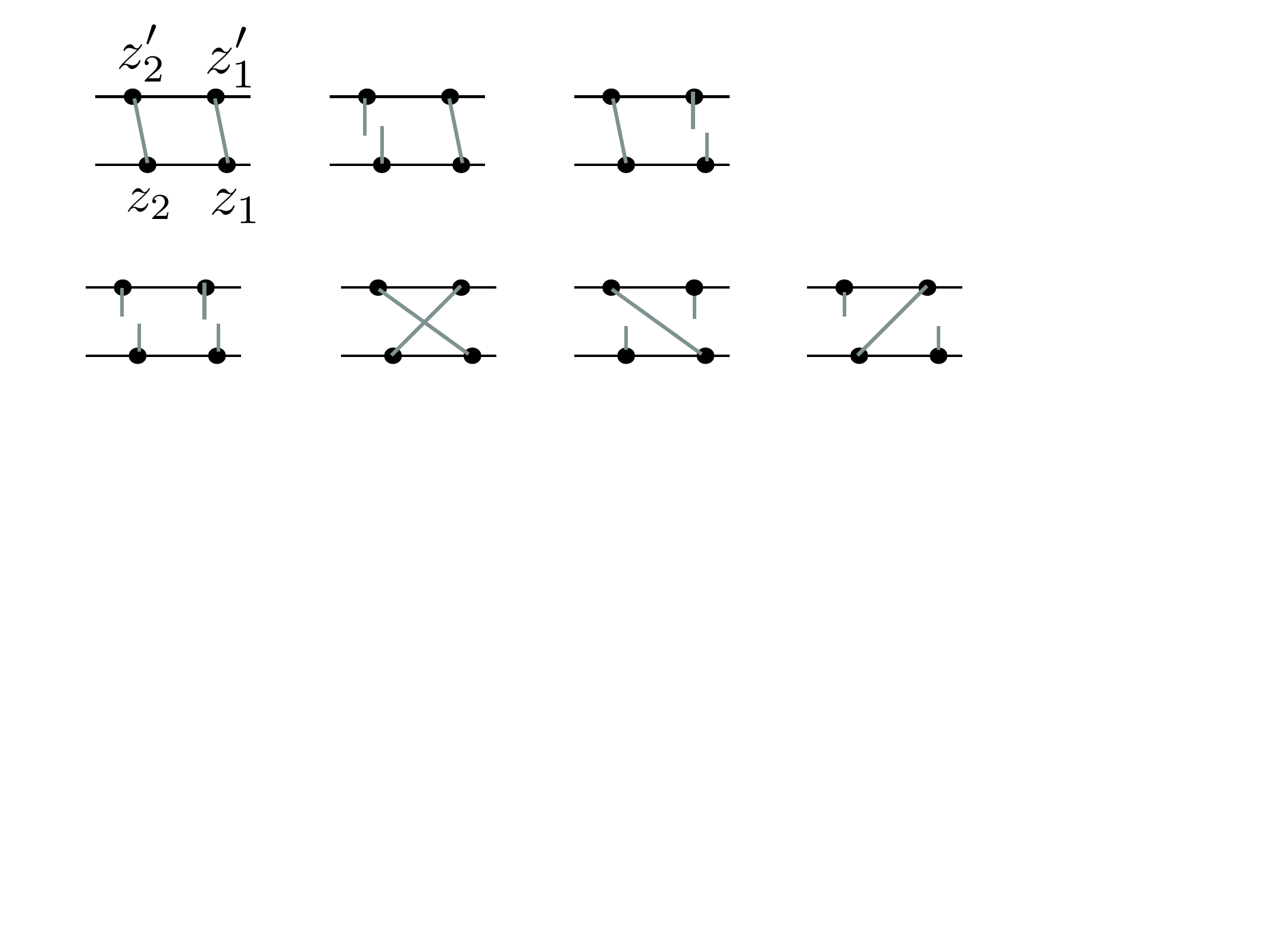}}
    \vspace{-4.2cm}
      \caption{Interpretation of the different terms in \eqref{big} in the order in which they appear, left to right and top to bottom. Only the maximum and second maximum are shown at each time.
      In the second diagram the particle which realizes
      the maximum at $t$ also realizes the maximum at $t'$, but the 
      particle which realizes the second maximum at $t$ is neither maximum
      nor second maximum at $t'$, and so on. }
  \label{fig:fig6}
\end{figure}

We can now turn to the joint PDF $q(X_1,X_2,X'_1,X'_2)$ of (i) the maximum (variable $X_1$)
and second maximum ($X_2$) at time $t$ (ii) the maximum (variable $X'_1$)
and second maximum ($X'_2$) at time $t'$. There are several possible cases, depending 
on which particle realizes the maximum or second maximum at each of the two times.
We can indicate them schematically as
\bea 
&& (X_1,X'_1,X_2,X'_2) = (i,i,j,j) \quad , \quad j \neq i \\
&& (X_1,X'_1,X_2,X'_2) = (i,i,j,j') \quad , \quad j \neq i  , j \neq j' , j' \neq i \\
&& (X_1,X'_1,X_2,X'_2) = (i,i',j,j) \quad , \quad i \neq j  , i' \neq i , j \neq i'  \\
&& (X_1,X'_1,X_2,X'_2) = (i,i',j,j')  \quad , \quad   i \neq i'  , j \neq j' ,  i \neq j  , i' \neq j' 
\eea 
Note that in the last case $i=j'$ is possible and so is $i'=j$ and so is both at the same time.
So in total there are seven terms which read (the terms are illustrated in Fig. \ref{fig:fig6})
\bea \label{big} 
&& q(X_1,X_2,X'_1,X'_2) = 
 \sum_{i \neq j} p_i(X_1,X'_1) p_j(X_2,X'_2) \prod_{r \neq i,j} P_r(x<X_2,x'<X'_2)  \\
&& + \sum_{i \neq j, i \neq j' , j \neq j'}  p_i(X_1,X'_1) p_j(X_2,x'<X'_2) p_{j'}(x<X_2,X'_2)  \prod_{r \neq i,j,j'} P_r(x<X_2,x'<X'_2) \nn \\
&& + \sum_{i \neq i', i \neq j , i' \neq j}  
p_i(X_1,x'<X'_2) p_{i'}(x<X_2,X'_1) p_j(X_2,X'_2)  \prod_{r \neq i,i',j} P_r(x<X_2,x'<X'_2) \nn  \\
&& + \sum_{i,j,i',j' \text{all distinct}} 
p_i(X_1,x'<X'_2) p_{i'}(x<X_2,X'_1) p_j(X_2,x'<X'_2) p_{j'}(x<X_2,X'_2)  
\prod_{r \neq i,i',j,j'} P_r(x<X_2,x'<X'_2)  \nn \\
&& +  \sum_{i \neq j}  p_i(X_1,X'_2) p_j(X_2,X'_1) \prod_{r \neq i,j} P_r(x<X_2,x'<X'_2) \nn \\
&&  + \sum_{i \neq i', i \neq j , i' \neq j} p_i(X_1,X'_2) p_{i'}(x<X_2,X'_1) p_{j'}(X_2,x'< X'_2) 
\prod_{r \neq i,j,j'} P_r(x<X_2,x'<X'_2) \nn \\
&&  + \sum_{i \neq i', i \neq j , i' \neq j} p_i(X_1,x'<X'_2) p_{i'}(X_2,X'_1) p_{j'}(x<X_2,X_2') 
\prod_{r \neq i,j,j'} P_r(x<X_2,x'<X'_2) \nn
\eea 
We can now estimate each term in the large $N$ limit (for identical particles) 
introducing the variables $z_1,z_2,z'_1,z'_2$,
and using the same rules for the asymptotics of the various PDF, CDF and mixed PDF/CDF
as explained above. 
One obtains, for each term of \eqref{big} in the same order (we abusively denote by the
same letter $q$ the two joint PDF) 
\bea \label{verybig1} 
&& q(X_1,X_2,X'_1,X'_2) dX_1 dX_2 dX'_1 dX'_2 \simeq
q(z_1,z_2,z'_1,z'_2) dz_1 dz_2 dz'_1 dz'_2  \\
&& \label{verybig2} q(z_1,z_2,z'_1,z'_2) =
\bigg( 
 \partial_{z_1} \partial_{z'_1} \Phi(z_1,z'_1) \partial_{z_2} \partial_{z'_2} \Phi(z_2,z'_2)   - \partial_{z_1} \partial_{z'_1} \Phi(z_1,z'_1)  \partial_{z_2}  \Phi(z_2,z'_2) \partial_{z'_2} \Phi(z_2,z'_2)   \\
 && - \partial_{z_1} \Phi(z_1,z'_2)  \partial_{z'_1} \Phi(z_2,z'_1)  \partial_{z_2} \partial_{z'_2} \Phi(z_2,z'_2)  + \partial_{z_1} \Phi(z_1,z'_2)  \partial_{z'_1} \Phi(z_2,z'_1)   \partial_{z_2}  \Phi(z_2,z'_2) \partial_{z'_2} \Phi(z_2,z'_2) \nn  \\
 && + \partial_{z_1} \partial_{z'_2} \Phi(z_1,z'_2) \partial_{z_2} \partial_{z'_1} \Phi(z_2,z'_1)   - \partial_{z_1} \partial_{z'_2} \Phi(z_1,z'_2)  \partial_{z'_1} \Phi(z_2,z'_1) \partial_{z_2} \Phi(z_2,z'_2)   \nn \\
 && - \partial_{z_1} \Phi(z_1,z'_2)  \partial_{z_2} \partial_{z'_1} \Phi(z_2,z'_1) 
\partial_{z'_2} \Phi(z_2,z'_2) \bigg) e^{\Phi(z_2,z'_2)}   \nn 
\eea  
where for clarity we have made the time argument implicit, i.e. $\Phi(z,z') \equiv \Phi(z,z',\tau)$.
Recall that each term gives the respective probabilities of how the particle realizing the maximum and second maximum change from time $t$ to time $t'$. For instance the first term corresponds to
events where both the rightmost particle and the second rightmost at $t$ 
have remained so at $t'$, and so on.

It turns out that the rather bulky expression \eqref{verybig1},\eqref{verybig2} is a total derivative, i.e.
one can check that, for $z_1>z_2$, $z'_1>z'_2$, 
\be 
q(z_1,z_2,z'_1,z'_2)  = \partial_{z_1} \partial_{z'_1} \partial_{z_2} \partial_{z'_2} 
(- \Phi(z_1,z'_1) + \Phi(z_1, z'_2) \Phi(z_2,z'_1) )  e^{\Phi(z_2,z'_2)} 
\ee 
which is the result \eqref{q4} displayed in the text, where it is also derived
by a different method using counting statistics. 

\section{Multi-time, multi-space counting statistics}
\label{app:outliers} 

One can generalize the arguments in Section \ref{sec:multicounting}. 
Let us illustrate for $n=3$ times and an arbitrary number of points
$k$, $k'$ and $k''$. Here we denote
$t'-t=t \frac{\tau}{\log N}$ and 
$t''-t'=t \frac{\tau'}{\log N}$, with $\tau,\tau'=O(1)$.
Consider the three sequences of points $\{ X_j \}_{j=1,\dots,k}$,
$\{ X'_{j'} \}_{j'=1,\dots,k'}$, $\{ X''_{j''} \}_{j''=1,\dots,k''}$,
each being in decreasing order. For each time the real axis
is the union of $k+1$ (and then $k'+1$ and $k''+1$) contiguous intervals, e.g. at time $t$ these are $[X_j,X_{j-1}]$, $j=1,k+1$ with
by convention $X_0=+\infty$ and $X_{k+1}=-\infty$, and similarly for $t'$ and $t''$.
Let $n_{j,j',j''}$, for $j=1,\dots,k+1$, $j'=1,\dots,k'+1$, $j''=1,\dots,k''+1$, 
be the numbers of particles which are in $[X_j,X_{j-1}]$ at $t$ and in $[X'_{j'},X'_{j'-1}]$ at $t'$  and in $[X''_{j''},X''_{j''-1}]$ at $t''$. The set of these numbers obey a multinomial distribution.
In the large $N$ limit, in the (multi-time) edge regime, 
defined such that the corresponding variables $z_j,z'_{j'},z''_{j''}$ are all
of order $O(1)$, all of these numbers are of order $O(1)$ with the exception of
$n_{k+1,k'+1,k''+1} \simeq N$. Then this reduced set of numbers
are independent Poisson variables, each with mean parameter $\lambda_{j,j',j''}$.
These parameters can be related to the functions defined in this paper 
as follows 
\bea 
&& \lambda_{j,j',j''} = \langle (\theta_{x>X_j}- \theta_{x>X_{j-1}}) (\theta_{x'>X'_{j'}}- \theta_{x'>X'_{j'-1}})
(\theta_{x''>X'''_{j''}}- \theta_{x''>X'''_{j''-1}})  \rangle \\
&& \simeq \sum_{\ell=0,1} \sum_{\ell'=0,1} \sum_{\ell''=0,1} (-1)^{\ell+\ell'+\ell''} 
g_3(z_{j-\ell},z'_{j'-\ell'},z''_{j''-\ell''};\tau,\tau') 
\eea 
where $g_3$ was defined in \eqref{defg3}. We recall
the convention $z_0=z'_0=z''_0=+\infty$ and 
$z_{k+1}=z'_{k'+1}=z''_{k''+1}=-\infty$. As mentionned in Section \ref{sec:first},
$g_3$ vanishes when any of the $z$ argument is taken to $+\infty$, and 
reduces to $g_2=g$ when any of the $z$ argument is taken to $-\infty$,
more precisely one has 
$g_3(-\infty,z',z'',\tau,\tau')=g(z',z'',\tau')$, 
$g_3(z,-\infty,z'',\tau,\tau')=g(z,z'',\tau+\tau')$,
$g_3(z,z',-\infty,\tau,\tau')=g(z,z',\tau)$, 
and similarly for $g_2=g$ which reduces to $g_1(z)=e^{-z}$. One can 
check that the sum of all the $\lambda_{j,j',j''}$ (over all indices, not including $(j,j',j'')=(k+1,k'+1,k''+1)$)
equals $\Phi(z_k,z'_{k'},z''_{k''};\tau,\tau')$, yielding the normalization
factor $e^{- \Phi(z_k,z'_{k'},z''_{k''};\tau,\tau')}$ for the
multiple independent Poisson distribution of the $n_{j,j',j''}$. 

{\it Two times, three first maxima}. Let us first return to the case of $2$ times. The probability that
the maximum is in $[X_1,+\infty[$ and the second 
maximum is in $]-\infty,X_2]$ at $t$, and 
the maximum is in $[X'_1,+\infty[$ and the second 
maximum is in $]-\infty,X'_2]$ at $t'$
was given in \eqref{cum12max}, and reads, translated in the present notations 
\be 
(\lambda_{11} + \lambda_{13} \lambda_{31}) e^{- \Phi(z_2,z'_{2};\tau)}
\ee 
which is a sum over the two permutations of 2 elements. Upon
taking the derivatives $\partial_{z_1} \partial_{z_2} \partial_{z'_1} \partial_{z'_2}$
yields the two time joint PDF of the maximum and second maximum.

\begin{figure}[t]
  \centerline{\includegraphics[width=0.6\columnwidth,height=8cm,keepaspectratio]{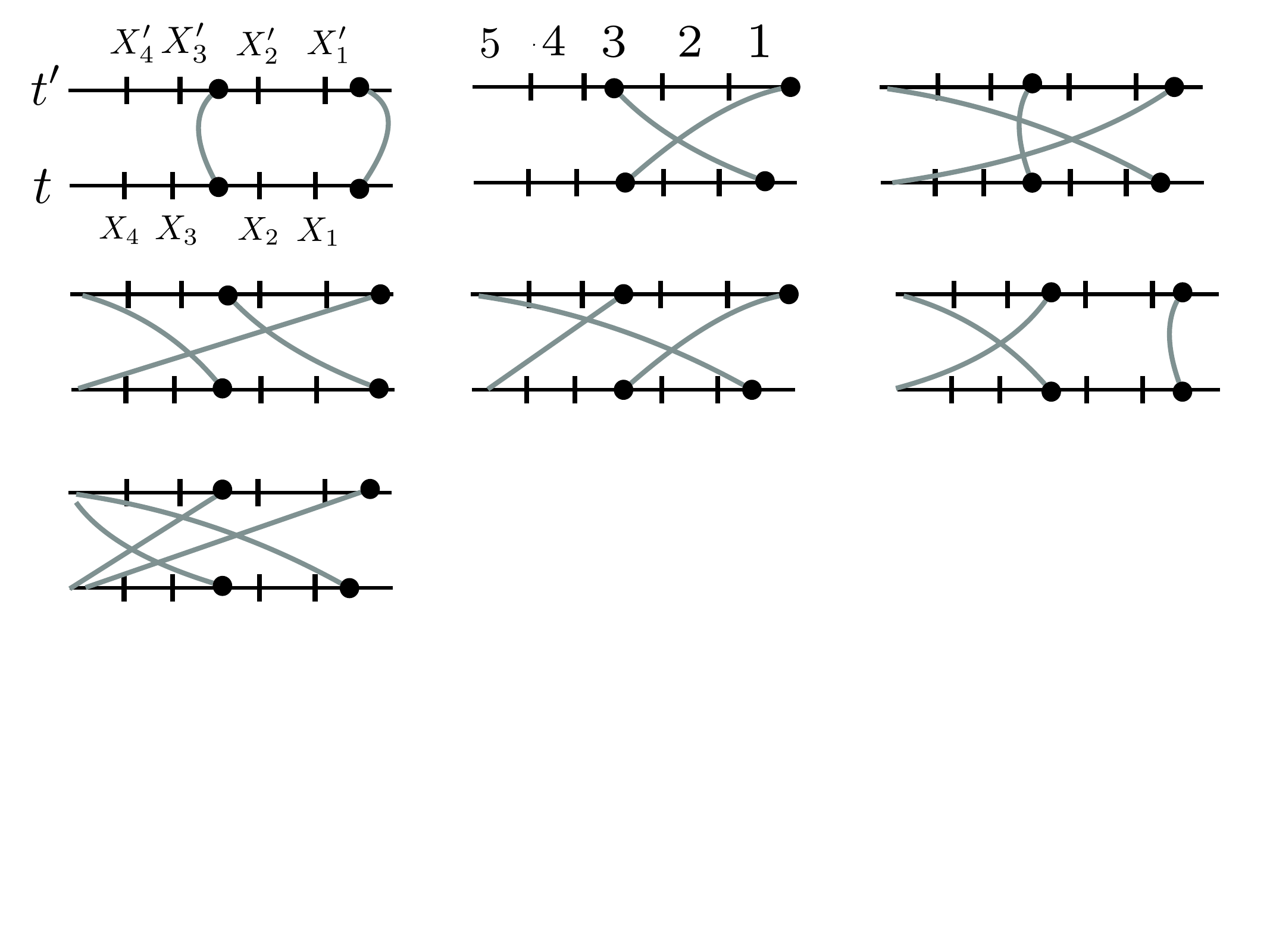}}
    \vspace{-2.5cm}
      \caption{Interpretation of the different terms in \eqref{list1} in the order in which they appear, left to right and top to bottom. Only the maximum and second maximum and their trajectories are shown at each time.
      }
  \label{fig:fig7}
\end{figure}

This can be generalized. For instance consider $n=2$ and $k=k'=4$
and the following joint probability
\be \label{cum123max} 
 {\cal P} = {\rm Prob}( X^{(1)}(t) >X_1, X^{(2)}(t) \in [X_3,X_2] , X^{(3)}(t) < X_4,
X^{(1)}(t') >X'_1, X^{(2)}(t') \in [X'_3,X'_2] , X^{(3)}(t') < X'_4) 
\ee 
From ${\cal P}$ one can obtain by differentiation the two time joint PDF of the maximum,
second maximum and third maximum. The various cases are shown in Fig. \eqref{fig:fig7}
and one obtains 
\bea \label{list1} 
{\cal P} \simeq (\lambda_{11} \lambda_{33} +  \lambda_{13} \lambda_{31} 
+ \lambda_{15} \lambda_{51} \lambda_{33} + \lambda_{13} \lambda_{35} \lambda_{51}  
+ \lambda_{15} \lambda_{53} \lambda_{31}  + \lambda_{11} \lambda_{35} \lambda_{53}  
+ \lambda_{15} \lambda_{51}  \lambda_{35} \lambda_{53}  ) e^{- \Phi(z_3,z'_3,\tau)} 
\eea 
which, apart from the last term, is a sum over the six permutations of three elements. Note that the intervals $[X_2,X_1]$
and $[X_4,X_3]$ remain empty. The two time 3-order statistics PDF is
obtained as (in the $z$ variables) 
\bea 
q(z_1,z_2,z_3,z'_1,z'_2,z'_3) = \partial_{z_1}\partial_{z_2}\partial_{z_3} 
\partial_{z'_1}\partial_{z'_2}\partial_{z'_3} {\cal P} 
\eea 

\begin{figure}[h]
  \centerline{\includegraphics[width=0.6\columnwidth,height=8cm,keepaspectratio]{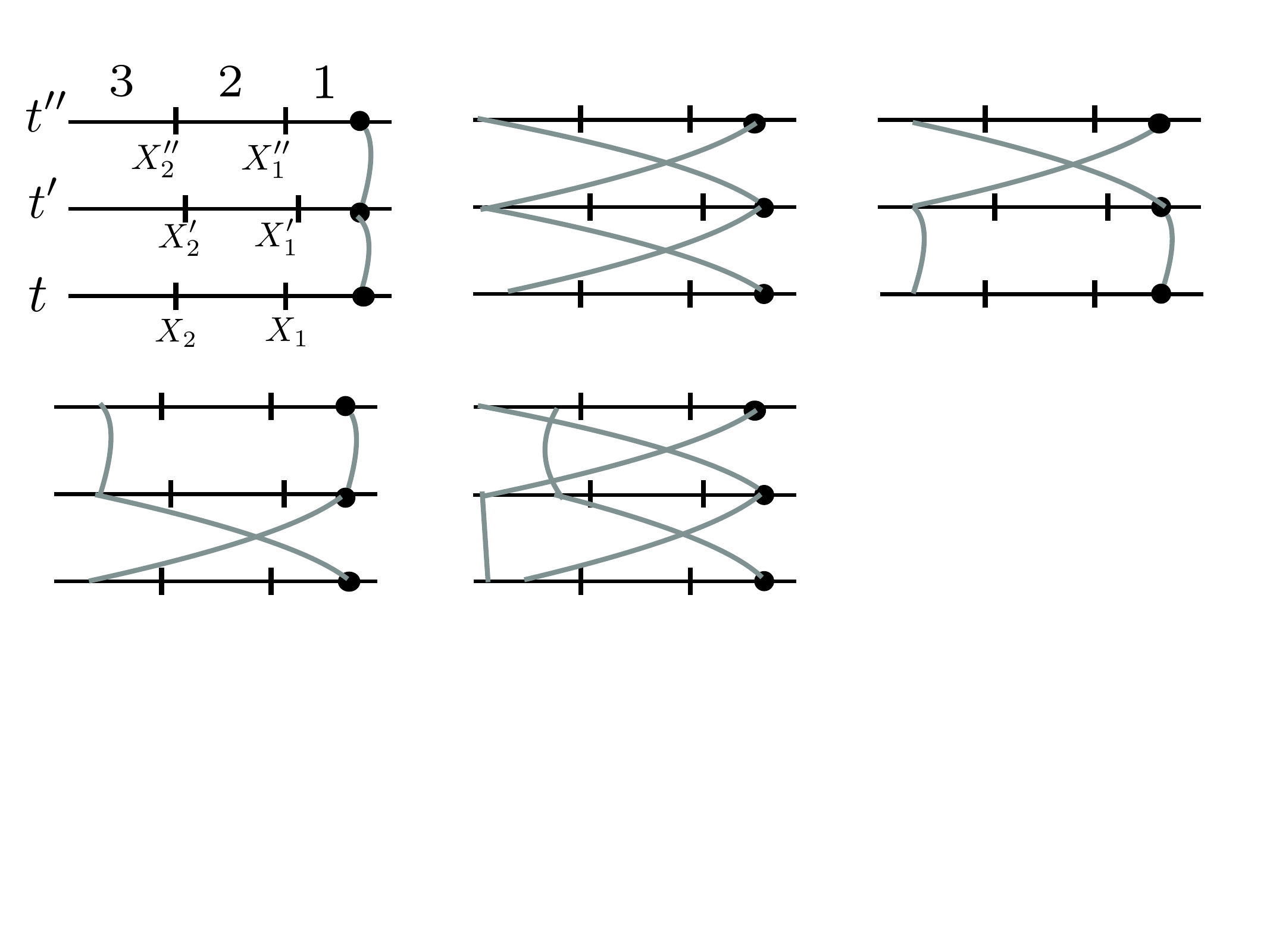}}
    \vspace{-2.5cm}
      \caption{Interpretation of the different terms in \eqref{list2} in the order in which they appear, left to right and top to bottom. Only the maximum and its trajectory is shown at each time.
      }
  \label{fig:fig8}
\end{figure}

{\it Three times, two first maxima}. One can also obtain the three time joint PDF of the maximum and
second maximum. From Fig. \ref{fig:fig8}
we see that, for $z_1>z_2$, $z'_1>z'_2$, $z''_1>z''_2$,
\bea \label{list2} 
&& q(z_1,z_2,z'_1,z'_2,z''_1,z''_2) = \\
&&  \partial_{z_1}\partial_{z_2}\partial_{z'_1} 
\partial_{z'_2}\partial_{z''_1}\partial_{z''_2} 
\big( (\lambda_{111} +  \lambda_{131} \lambda_{313} +
\lambda_{113} \lambda_{331}
+ \lambda_{133} \lambda_{311} + \lambda_{133} \lambda_{331} \lambda_{313}
 ) e^{- \Phi(z_2,z'_2,z''_2;\tau,\tau')}  \big) \nn
\eea 
which has $2!^2$ terms (two successive permutations of 2 elements). 
To be specific we give
\bea 
&& \lambda_{111}= g_3(z_1,z'_1,z''_1;\tau,\tau') \\
&& \lambda_{113}= g(z_1,z'_1,\tau) - g_3(z_1,z'_1,z''_2;\tau,\tau') \\
&& \lambda_{331}= e^{-z''_1} - g_2(z_2,z''_1;\tau+\tau') 
- g(z'_2,z''_1,\tau')  + g_3(z_2,z'_2,z''_1;\tau,\tau') \\
&& \lambda_{133} = e^{-z_1} - g(z_1,z'_2;\tau) - g(z_1,z''_2;\tau+\tau') +
g_3(z_1,z'_2,z''_2;\tau,\tau') \\
&& \lambda_{311} = g(z'_1,z''_1,\tau') - g_3(z_2,z'_1,z''_1;\tau,\tau') \\
&& \lambda_{131} = g(z_1,z''_1,\tau+\tau') - g_3(z_1,z'_2,z''_1;\tau,\tau') \\
&& \lambda_{313} = e^{-z'_1} - g(z_2,z'_1;\tau) - g(z'_1,z''_2;\tau') +
g_3(z_2,z'_1,z''_2;\tau,\tau') 
\eea 

\section{Continuum time observables: calculations} \label{app:continuous}

\subsection{First passage time} \label{app:FPT} 

Let us recall the following. 
Let $B(t)$ the standard Brownian motion (i.e. with $B(0)=0$) 
and $W(t)=\sqrt{D} B(t)+\mu$ a Brownian with drift $\mu$
and diffusion coefficient $D$. Let us denote $T^{\mu,D}_z$ the first passage time
of $W(t)$ at level $z$. For $z>0$
the PDF of $T=T_z^{\mu,D}$ is 
\be 
p_z(T)  = \frac{z}{\sqrt{2 \pi D} T^{3/2} } e^{- \frac{(z-\mu T)^2}{2 D T} }  + \delta_{+\infty}(T) (1-e^{2 \mu z/D})
\theta(-\mu) \label{FPTdistrib} 
\ee 
One has upon integration
\be \label{FPTCDF} 
{\rm Prob}(T_z^{\mu,D} > t) = \frac{1}{2} ( {\rm erfc}( \frac{\mu t - z}{\sqrt{2 D t}}) 
- e^{2 \mu z/D}  {\rm erfc}( \frac{\mu t + z}{\sqrt{2 D t}})  )
\ee 
For $\mu=0$ it simplifies into
\be \label{mu0} 
{\rm Prob}(T_z^{0,D} > t) = {\rm erf}(\frac{z}{\sqrt{2 D t}}) 
\ee 
Eq. \eqref{FPTCDF} is also the probability that a Brownian 
survives up to time $t$ in presence of an absorbing wall at $x=z$
and can thus also be obtained from the image method.
Indeed one has, denoting $x$ the position of the Brownian at time $t$,
\be 
{\rm Prob}(T_z^{\mu,D} > t) =  \int_{-\infty}^{z}  dx ( 
 \frac{e^{-(x-\mu t)^2/(2 D t)}}{\sqrt{2 \pi D t}} - e^{2 \mu z/D} \frac{e^{-(x-2 z -\mu t)^2/(2 D t)}}{\sqrt{2 \pi D t}} ) 
\ee 
which upon integration recovers \eqref{FPTCDF}.

\subsection{Probability that the maximum remains below a straight line} 
\label{app:checks} 

Let us consider the following observable for the maximum process $X(t)=X^{(1)}(t)$
of $N$ standard Brownian motions started at the origin
\be 
{\rm Prob}(X(t) < X_1 + v (t-t_1) ~,~ \forall t \in [t_1,t_2]) = {\rm Prob}(x(t) < X_1 + v (t-t_1) ~,~  \forall t \in [t_1,t_2])^N
\ee
Now one has
\be 
{\rm Prob}(x(t) < X_1 +v (t-t_1) ~,~ \forall t \in [t_1,t_2]) = \int_{-\infty}^{X_1} dx_1 \frac{e^{-x_1^2/(2 t_1)}}{\sqrt{2 \pi t_1}} 
 {\rm Prob}(T^{-v}_{X_1-x_1} > t_2-t_1)
\ee 
where here we must set $D=1$. Indeed, asking that a standard Brownian motion starting at $x_1$ at time $t_1$, 
remains below $X_1+v(t-t_1)$ until $t_2$ is equivalent to asking
that a standard drifted Brownian of drift $-v$ remains below $X_1$ until $t_2$,
which is also equivalent to asking that the first passage time
at level $X_1$ of a drifted Brownian of drift $-v$ started at $x_1$ at time $t_1$ is 
larger than $t_2-t_1$. Using \eqref{FPTCDF} one obtains
\bea 
&& {\cal P} := {\rm Prob}(x(t) < X_1 +v (t-t_1) ~,~ \forall t \in [t_1,t_2]) \\
&& =  \int_{-\infty}^{X_1} dx_1 \frac{e^{-x_1^2/(2 t_1)}}{\sqrt{2 \pi t_1}} \frac{1}{2} 
\bigg( 
 {\rm erfc}( \frac{-v (t_2-t_1) - (X_1-x_1)}{\sqrt{2  (t_2-t_1)}}) 
- e^{- 2 (X_1-x_1) v}  {\rm erfc}( \frac{-v (t_2-t_1) + (X_1-x_1)}{\sqrt{2  (t_2-t_1)}}  \bigg) \nn
\eea 
We will now scale $X$ near the edge, and scale the time window as well,
 i.e. choose
\be 
X_1= \sqrt{2 t_1 \log N} (1+ \frac{z_1+c_N}{2 \log N}) \quad , \quad \frac{t_2-t_1}{t_1} = \frac{\tau_{2,1}}{\log N} 
\ee
and we will need to scale the slope $v$ as
\be
v =  w \sqrt{ \frac{\log N}{2 t_1}}  \quad , \quad w=O(1) 
\ee 
In that region ${\cal P}$ is close to unity, with 
$1-{\cal P}= O(1/N)$ and
\be 
{\rm Prob}(X(t) < X_1 + v (t-t_1) ~,~ \forall t \in [t_1,t_2])  = {\cal P}^N \simeq e^{- N (1- {\cal P})} 
\ee
Using that the edge coordinates satisfy
\be 
N p_{t_1}(x_1) dx_1= e^{-y_1} dy_1  \quad , \quad 
X_1-x_1 =  \frac{\sqrt{2 t_1}}{2 \sqrt{\log N}} (z_1-y_1) 
\ee
We can write
\bea 
&& N(1-{\cal P})= N 
 \int_{-\infty}^{+\infty}  dx_1 \frac{e^{-x_1^2/(2 t_1)}}{\sqrt{2 \pi t_1}}  \\
 && \times \bigg( 1- \theta(X_1-x_1) \frac{1}{2}
\big( 
 {\rm erfc}( \frac{-v (t_2-t_1) - (X_1-x_1)}{\sqrt{2  (t_2-t_1)}}) 
- e^{- 2 (X_1-x_1) v}  {\rm erfc}( \frac{-v (t_2-t_1) + (X_1-x_1)}{\sqrt{2  (t_2-t_1)}}  \big) \bigg) \nn \\
&& \simeq  \int dy_1 e^{-y_1} (1 - \theta(z_1-y_1) \frac{1}{2}
\big( 
 {\rm erfc}( \frac{- w \tau_{2,1} - (z_1-y_1)}{2 \sqrt{\tau_{2,1}}}) 
- e^{-  (z_1-y_1) w}  
 {\rm erfc}( \frac{- w \tau_{2,1} + (z_1-y_1)}{2 \sqrt{\tau_{2,1}}}) \big) \bigg) \nn \\
 && = \int dy e^{-y_1} (1- \theta(z_1-y_1)  {\rm Prob}({\sf T}^{-w}_{z_1-y_1} > \tau_{2,1})) 
  = e^{-z_1} \int dy e^{y} (1- \theta(y)  {\rm Prob}({\sf T}^{-w}_{y} > \tau_{2,1})) 
\eea
where in the last line we have set $y_1=z_1-y$ and 
where ${\sf T}^{-w}_{z} = T^{-w,D=2}_z$ is the first passage time
for a Brownian with drift $-w$ and diffusion coefficient $D=2$, with
\be 
 {\rm Prob}({\sf T}^{-w}_{y} > \tau) 
 = \frac{1}{2}
\big( 
 {\rm erfc}( \frac{- w \tau - y}{2 \sqrt{\tau}}) 
- e^{-  y w}  
 {\rm erfc}( \frac{- w \tau + y}{2 \sqrt{\tau}}) \big) \bigg) 
\ee 

In summary we find that 
\be
 {\rm Prob}(X(t) < X_1 + v (t-t_1) ~,~ \forall t \in [t_1,t_2]) ~ \simeq ~ e^{- e^{-z_1} 
 \Psi_{w}(\tau_{2,1}) } 
\ee
where 
\bea 
&& \Psi_w(\tau) = \int dy e^{y} (1- \theta(y)  {\rm Prob}({\sf T}^{-w}_{y} > \tau)) 
  = 1 -  \int_{y>0}  dy \, e^{y} \, {\rm Prob}({\sf T}^{-w}_{y} < \tau)  
\eea 
An explicit calculation and one finds, for $w \neq 1$
\be \label{explicit} 
\Psi_w(\tau) = \frac{2-w}{1-w} \left( 
-\frac{1}{2} e^{\tau-\tau w}
   \text{erfc}\left(\frac{1}{2}
   \sqrt{\tau}
   (w-2)\right)+\left(\frac{1}{w-2}+
   \frac{1}{2}\right)
   \text{erfc}\left(\frac{\sqrt{\tau}
   w}{2}\right)+1
\right) 
\ee 
with $\Psi_w(0) =1$. For $w=1$ it gives \eqref{Psi1} in the text, and for 
$w=0$ one has
\be
\Psi_0(\tau) =
e^\tau \left(\text{erfc}\left(\sqrt{\tau}\right)-2\right)+2 
\ee 

Now under the rescaling described here and the definition of $z(\tau)$ 
in the text one finds that 
\be 
 {\rm Prob}(X(t) < X_1 + v (t-t_1) ~,~ \forall t \in [t_1,t_2]) 
 \simeq 
 {\rm Prob}(z(\tau) + \tau < z_1 + w \tau) 
\ee 
which shows the result \eqref{resultT} conjectured in the text.


\subsection{Running maximum and arrival time of the first particle: one time} \label{app:detailsrunning1} 

Here we derive the one-time distributions by a more detailed calculation. 
For the standard Brownian motion, we see from \eqref{mu0} that Eq. \eqref{P1def} 
becomes
\be 
{\rm Prob}(r(t_1)<R_1) = {\rm Prob}( T_{R_1} > t_1) = {\cal P}_1 =  {\rm erf}(\frac{R_1}{\sqrt{2  t_1}}) 
\ee 
This can be either interpreted as the CDF for the running maximum at fixed $t_1$
or for the "CDF" of the first passage time at fixed $R_1$. Similarly for $N$ Brownian one has
\be 
{\rm Prob}(R(t_1) <R_1) = {\rm Prob}( T^{\min}_{R_1} > t_1) \label{same} 
\ee 
can be interpreted either as the CDF of $R(t_1)=\max_i r_i(t_1)$ at fixed $t_1$,
or the "CDF" of the arrival time of the first particle,
$ T^{\min}_{R_1} = \min_i T^{(i)}_{R_1}$ at fixed $R_1$. At large $N$ \eqref{same} is estimated as
\be 
\simeq e^{- N (1- {\cal P}_1) } = e^{- N (1- {\rm erf}(\frac{R_1}{\sqrt{2  t_1}}))} 
\simeq e^{- N \frac{\sqrt{2 t_1}}{\sqrt{\pi} R_1} e^{- \frac{R_1^2}{2 t_1}} } = e^{-e^{-z}} \label{283} 
\ee 
where 
\be \frac{R_1^2}{2 t_1} = \log N + z - \log \frac{\sqrt{\pi} R_1}{\sqrt{2 t_1}}  \simeq \log N + z+  c'_N \quad , \quad c'_N = - \frac{1}{2} \log(\pi \log N) = c_N + \log 2 
\ee 
If we are interested in the running maximum at fixed $t_1$ we obtain from this estimate
\be 
R(t_1)=R_1 = \sqrt{2 t_1 \log N} ( 1 + \frac{z_1 + c_N}{2 \log N}) ) \quad , \quad z_1 = z+ \log 2
\ee 
in agreement with the text, where $z$ is Gumbel distributed (by definition from \eqref{283}), and the shift of $\log 2$
agrees with the one obtained in \eqref{running1}. On the other hand,
if one is interested in the arrival time of the first particle one obtains, from
the same estimate
\be 
T^{\rm \min}_{R_1} = t_1= \frac{R_1^2}{2 \log N} ( 1 - \frac{z + c'_N}{\log N}) ) 
\ee 
where $z$ is Gumbel distributed. One may ask why is the 
arrival time of the first particle also distributed with (minus) Gumbel,
since the distribution of the first passage time is very different from a Gaussian, see 
\eqref{FPTdistrib}. To see that immediately one can consider
that $\min_i T^{(i)}_R=1/(\max_i U_i)$ where $U=1/T_R$
has a distribution with an exponential tail which clearly
belongs to Gumbel class.
\\

\subsection{Running maximum and arrival time of the first particle: two time}
 \label{app:2timerunning} 

We give here more details of the calculation of the 
two-time joint CDF of the running maximum depicted in the text. 
Let us start from the exact expression for ${\cal P}$ in \eqref{calP0}. 
In the large $N$ limit, with the scaling \eqref{sc},
using similar estimates as in Appendix \ref{app:derivationPDF},
we obtain 
\bea \label{manip} 
&& N(1- {\cal P}) =N  \int dx_1 \int dx_2 \bigg(  \frac{e^{- \frac{x_1^2}{2 t_1}}}{\sqrt{2 \pi t_1}} 
\frac{e^{- \frac{(x_2-x_1)^2}{2 (t_2-t_1)}}}{\sqrt{2 \pi (t_2-t_1)}} \\
&& \times 
\bigg( 1 - \theta(R_1-x_1) \theta(R_2-x_2) (1- e^{- \frac{2 R_1(R_1-x_1)}{t_1} }) 
(1- e^{- \frac{2 (R_2-x_1)(R_2-x_2)}{t_2-t_1} }) \bigg) \nn \\
&& \simeq
\int dy_1 \int dy_2  \frac{e^{-\frac{(y_2-y_1+\tau)^2}{4 \tau}}}{\sqrt{4 \pi \tau} }
(e^{-y_1} - \theta(z_1-y_1) \theta(z_2-y_2) (e^{-y_1} - e^{-  (2 z_1-y_1)} ) (1- e^{- \frac{(z_2+\tau-y_1)(z_2-y_2)}{\tau} } ) \nn
\eea 
where we have changed variables denoting $x_i = \sqrt{2 t_i \log N} (1+ \frac{y_i + c_N}{2 \log N})$.
As for the functions $\Phi$ we can give some interpretation to this formula in terms of the Brownian motion
with unit negative drift and diffusion coefficient $D=2$, with however an important difference. 
The factor $(e^{-y_1} - e^{-  (2 z_1-y_1)}) \theta(z_1-y_1)$ is the stationary measure in the
presence of an absorbing wall at $z_1$. The other factor can be written as
\be 
\theta(z_2-y_2) \left( 
\frac{e^{-\frac{(y_2-y_1+\tau)^2}{4 \tau}}}{\sqrt{4 \pi \tau} }
- \frac{e^{-\frac{(y_2-(2 (z_2+\tau)-y_1)+\tau)^2}{4 \tau}}}{\sqrt{4 \pi \tau} } \right) 
\ee 
which vanishes at $y_2=z_2$ but is not exactly the propagator in presence of a fixed absorbing wall, since
the wall is effectively moving. Performing the change of variable 
$y_1 \to z_1-y_1$ and $y_2 \to z_2-y_2$ in the last line of \eqref{manip} one 
obtains 
\bea
{\rm Prob}(R(t_1)<R_1, R(t_2) < R_2) \simeq 
e^{- N(1- {\cal P}) } \simeq e^{ - e^{-z_1} \gamma_\tau(z_{21}) } 
\eea 
with the function $\gamma_\tau(z)$ displayed in \eqref{gammatau0} in the text. 
To obtain the explicit expression \eqref{gammaexplicit} one splits
the integral into $y_1>0$ and $y_1<0$ and use that 
$\int_{y_1<0}  \int dy_2 e^{y_1} G(z-y_{2,1},\tau) =1$.
The part $y_1>0,y_2<0$ is found to equal $\phi_\tau(z)-1$.
The remaining part $y_1>0,y_2>0$, which is positive, can be integrated explicitly.

\subsection{Multi-time CDF for the running maximum} \label{app:multi-running} 

The two-time calculation of the previous section can easily be extended to
any number of times $n$. Here we just give the result. One finds,
under the scaling \eqref{sc} and $t_j-t_{j-1}=t_1 \tau_{j,j-1}/\log N$
\bea
{\rm Prob}(R(t_1)<R_1,\dots,R(t_n)<R_n)
= {\rm Prob}(T^{\rm min}_{R_1} > t_1 ,\dots,T^{\rm min}_{R_n} > t_n )
 \simeq e^{- \Gamma(z_1,\dots,z_n;\tau_{2,1},\dots,\tau_{n,n-1}) } \label{runningn} 
\eea 
where  
\bea \label{Gamman} 
&& \Gamma(z_1,\dots,z_n;\tau_{2,1},\dots,\tau_{n,n-1}) =
\int dy_1 \int dy_2 \dots dy_n  \, e^{-y_1} 
G(y_{2,1},\tau_{2,1}) \dots G(y_{n,n-1},\tau_{n,n-1}) \\
&& \times 
(1- \prod_{i=1}^n \theta(z_i-y_i) (1- e^{-  2 (z_1-y_1)} ) 
(1- e^{- \frac{(z_2+\tau-y_1)(z_2-y_2)}{\tau} } ) 
\dots 
(1- e^{- \frac{(z_n+\tau_{n,n-1}-y_n)(z_n-y_n)}{\tau_{n,n-1}} } ) \nn
\eea 
Similarly, the result \eqref{runningn}, \eqref{Gamman} can be read as a result for the
multi-time joint "CDF" of the arrival times $T^{\rm min}_{R_i}$
of the first particle at $R_i$, which allows to
obtain the joint distribution of $z_1$, and of 
the scaled delay times $\tau_{2,1},\dots,\tau_{n,n-1}$,
defined as in \eqref{tauz1} in the text.


\begin{thebibliography}{12}
\bibitem{footnoteconvergence} 
e.g. for the initial condition
considered $\hat F(x,0)=+\infty$ for $x<0$ and $F(x,0)=0$ for $x<0$.
Also $\hat F  = K e^{-z}$ would be an equally good stationary condition,
corresponding to a translation in $z$. 
This is not surprising since here some details of the initial condition
have been erased.
\end{thebibliography}
\end{document}